%% file: main.tex
\documentclass[twocolumn, twocolappendix]{aastex701}

\usepackage{amsmath}
\usepackage{gensymb}
\usepackage{placeins}
\usepackage{longtable}
\usepackage{comment}

\newcommand{\CHECK}[1]{{#1}}

\newcommand{\cyone}[1]{{#1}}
\newcommand{\cytwo}[1]{{#1}}
\newcommand{\cythree}[1]{{#1}}
\newcommand{\cyfour}[1]{{#1}}
\newcommand{\cyfive}[1]{{#1}}
\newcommand{\cyfiveeqn}[1]{{#1}}

\begin{document}

\reportnum{DES-2025-0909}
\reportnum{FERMILAB-PUB-25-0573-LDRD-PPD}

\title{DELVE Milky Way Satellite \cyfive{Galaxy} Census I: \\ Satellite Population and Survey Selection Function \cyfive{in DES, DELVE, and Pan-STARRS}}

\input{authors}




\begin{abstract}
The properties of Milky Way satellite galaxies have important implications for galaxy formation, reionization, and the fundamental physics of dark matter. However, the population of Milky Way satellites includes the faintest known galaxies, and current observations are incomplete. To understand the impact of observational selection effects on the known satellite population, we perform rigorous, quantitative estimates of the Milky Way satellite galaxy detection efficiency in three wide-field survey datasets:  the Dark Energy Survey Year 6, the DECam Local Volume Exploration Data Release 3, and  the Pan-STARRS1 Data Release 1. Together, these surveys cover $\sim$13,600 deg$^2$ to $g \sim 24.0$ and $\sim$27,700 deg$^2$ to  $g \sim 22.5$, spanning $\sim$91\% of the high-Galactic-latitude sky ($|b| \geq 15\degree$). We apply multiple detection algorithms over the combined footprint and recover {49} known satellites above a strict census detection threshold. To  characterize the sensitivity of our census, we run our detection algorithms on a large set of simulated galaxies injected into the survey data, which allows us to develop models that predict the detectability of satellites as a function of their properties. We then fit an empirical model to our data and infer the luminosity function, radial distribution, and size-luminosity relation of Milky Way satellite galaxies. Our empirical model predicts a total of $265^{+79}_{-47}$ satellite galaxies with $-20 \leq M_V \leq 0$, half-light radii of $15 \leq r_{1/2} ,(\rm pc) \leq 3000$, and galactocentric distances of $10 \leq D_{\rm GC} (\rm kpc) \leq 300$. We also identify a mild anisotropy in the angular distribution of the observed galaxies, at a significance of {$\sim$$2\sigma$}, which can be attributed to the clustering of satellites associated with the LMC.
\end{abstract}



\section{Introduction} \label{sec:intro}
The standard cosmological model comprised of a cosmological constant and cold dark matter ($\Lambda$CDM) predicts that the Milky Way is surrounded by a dark matter halo that hosts thousands of dark matter subhalos, many of which host luminous satellite galaxies \citep{Press_Schechter:1974, White:1978, Blumenthal:1984, Kauffmann:1993, Kravtsov:2010}.  These satellite galaxies span a wide range of sizes and luminosities, with stellar masses ranging from $\sim$$10^2 M_\odot$ to $\sim$$10^9 M_\odot$ (see \citealt{Simon:2019}, \citealt{Doliva-Dolinsky:2025}, and \citealt{Pace:2024} for recent reviews). 

\begin{figure*}[ht]
\centering
\includegraphics[width=\linewidth]{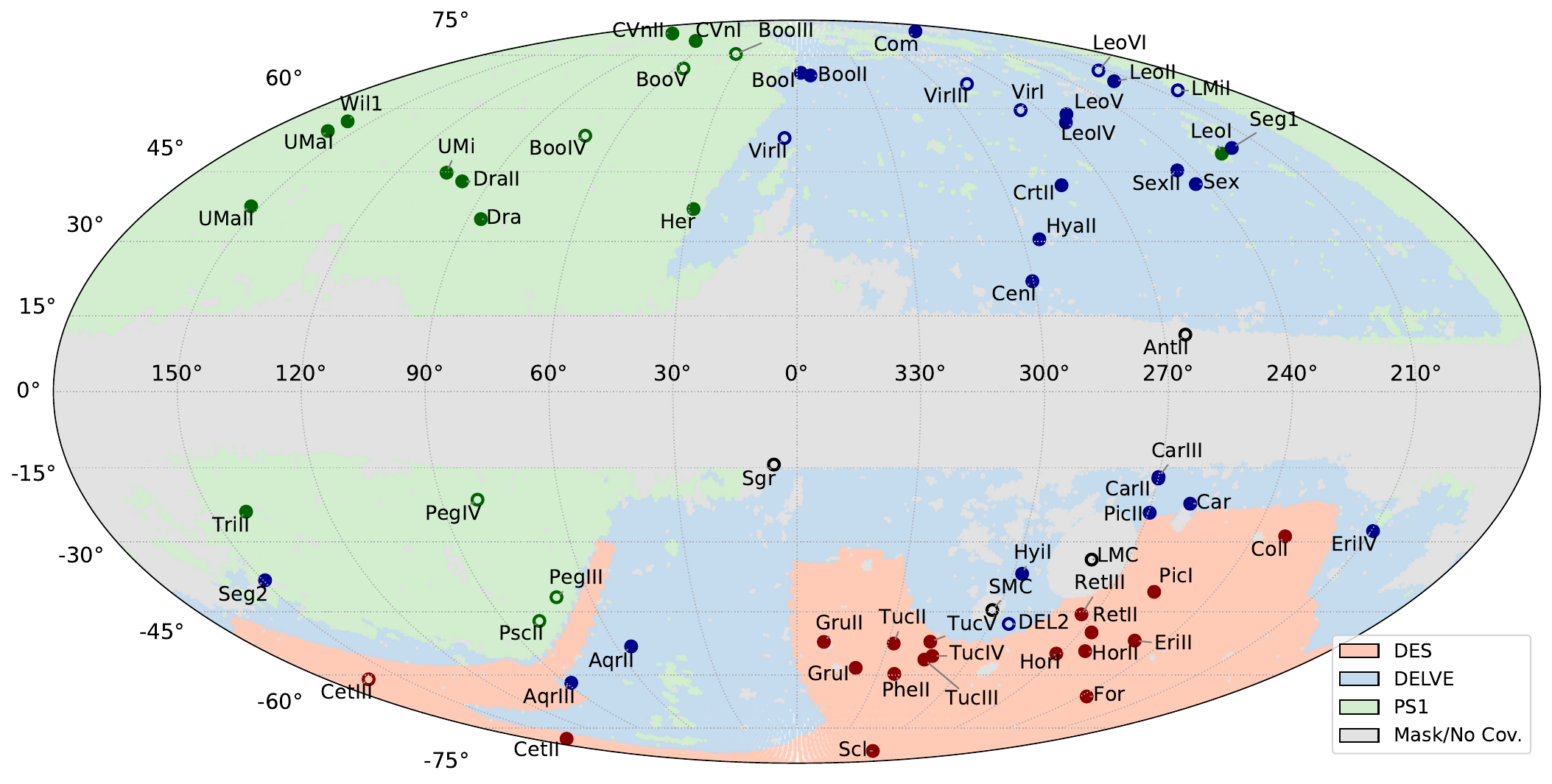}
%\vspace{-3cm}
\caption{Galactocentric locations of Milky Way satellites  with $r_{1/2} > 15$ pc. \cytwo{In total, our census recovered 49 out of 62 known Milky Way satellites within our survey footprint.} Filled circles indicate systems recovered by our analysis, while open circles represent unrecovered systems. Each satellite is colored according to the survey region that it is located in. The red, blue, and green regions show the coverage areas of DES Y6, DELVE DR3, and PS1 DR1 surveys, respectively. The gray region marks areas excluded from our census, either because they are at low Galactic latitudes, are near other stellar systems (e.g., globular clusters), or lack sufficient survey coverage. \cythree{We note that four relatively bright satellite (LMC, SMC, Sagittarius, and Antlia II) are outside of our survey footprint.\label{fig:census_dwarfs} }}
\end{figure*}
\setcounter{footnote}{0} 
\cythree{The Milky Way’s satellite dwarf galaxies can be distinguished from other bound stellar systems, such as globular clusters, by their large
velocity dispersion relative to their baryonic content \citep{Willman:2012}. In the context of the $\Lambda$CDM model, this is interpreted as a dark matter mass that is hundreds of times greater than their stellar mass \citep{Wolf:2010}. In addition to a large velocity dispersion, the dark-matter-dominated nature of dwarf galaxies can also be confirmed spectroscopically by measuring a large spread in the metallicities of member stars, which suggests a dark matter halo that is massive enough to retain supernova ejecta and support multiple generations of star formation} \citep{2013ApJ...770...16K, Simon:2019}. 

Due to their large dark matter content, relative proximity, and small sizes, the Milky Way satellite dwarf galaxies have played an outsized role in understanding dark matter 
\citep[e.g.,][and references therein]{Bullock:2017, Simon:2019, Sales:2022}.  For example, the fundamental properties of dark matter such as its particle mass and interaction cross section can greatly impact the luminosity function, mass density profiles, and kinematics of the Milky Way satellites (see \citealt{Bullock:2017} for a review). Furthermore, their lack of gas and other astrophysical sources of high-energy particles  makes Milky Way satellite galaxies excellent targets for indirect gamma-ray searches for signals from dark matter annihilation or decay \citep[e.g.,][and references therein]{Strigari:2018}. On the other hand, the faintest Milky Way satellites also represent the extreme end of galaxy formation. Their properties, in particular the luminosity function, are sensitive to physical processes such as reionization \citep[e.g.,][]{Bullock:2000, 2002MNRAS.333..177B, Manwadkar:2022, Ahvazi:2024}, molecular hydrogen cooling \citep[e.g.,][]{Ahvazi:2024}, and dark matter–baryon streaming \citep[e.g.,][]{Nadler:2025}, allowing us to probe the importance of these mechanisms in galaxy formation.

Over the past few decades, the satellites of the Milky Way have attracted attention due to the apparent discrepancy between the number of observed satellite galaxies and the orders-of-magnitude larger population of dark matter subhalos predicted by $\Lambda$CDM simulations, a tension historically referred to as the ``Missing Satellites Problem'' \citep{1999ApJ...522...82K, 1999ApJ...524L..19M}.
However, from the outset it was pointed out that this apparent problem could be resolved through observational incompleteness and the physics governing the formation of the faintest galaxies \citep[e.g.,][]{1999ApJ...522...82K, 1999ApJ...524L..19M, Bullock:2000}. 
Over the last two decades, it has been demonstrated that improved treatments of observational selection effects and the development of more detailed galaxy--halo connection models bring $\Lambda$CDM simulations and observations of Milky Way satellites into agreement, at least to the limit of current observations \citep[e.g.,][]{Jethwa:2018, Kim:2018, Newton:2018,  MWCensus2, Sales:2022, 2022MNRAS.515.3685S,  2025MNRAS.540.1107S}.
Furthermore, these $\Lambda$CDM simulations predict that a large fraction of Milky Way satellites likely remain undiscovered \citep[e.g.,][]{Tollerud:2008, Hargis:2014, Manwadkar:2022, Tsiane:2025},  \cyfour{further highlighting the need to carefully account for observational selection effects when studying Milky Way satellites.}

Observations of the Milky Way satellite population have advanced by leaps and bounds. While the LMC and SMC have been well-known to naked-eye observers since prehistory, the next Milky Way satellite galaxies were not discovered until the photographic sky surveys of the 20th century \citep{1938BHarO.908....1S, 1938Natur.142..715S, Harrington50, 1955PASP...67...27W, 1977MNRAS.180P..81C, Ibata:1994}. The 11 galaxies discovered before 2003 are collectively known as the ``classical'' dwarf satellites \citep{Mateo:1998, Willman:2010}. Beginning in 2005, an explosion of discoveries of ``ultra-faint'' Milky Way satellites was spurred by the advent of the Sloan Digital Sky Survey \citep[{SDSS:}][]{York:2000gk}, which more than doubled the number of known Milky Way satellites \citep[e.g.,][]{Willman:2005a, Willman:2005b, Belokurov07, Belokurov08, Grillmair:2009, Zucker:2006}. The wide coverage and uniformity of SDSS enabled the first careful, systematic search for Milky Way satellites, providing a foundation for estimating the total satellite population of the Milky Way. \citep{Koposov:2008, Walsh:2009}.

A second wave of discoveries of even fainter satellites occurred with the start of more recent surveys such as the the Panoramic Survey Telescope and Rapid Response System $3\pi$ survey \citep[{Pan-STARRS1,PS1:}][]{Chambers:2016}, the Dark Energy Survey \citep[DES:][]{DES:2016}, and the Hyper Suprime-Cam Subaru Strategic Program \citep[HSC-SSP;][]{Aihara:2018}. This second wave of discoveries caused the number of known satellites to double again to more than 50 systems by 2020 \citep[e.g.,][]{laevens:2015,  Koposov15a, Bechtol:2015, Kim:2015b, Drlica-Wagner:2015, Homma:2016, Homma:2018, Homma:2019}. To leverage these surveys,  we have conducted a Milky Way satellite census using DES Year 3 and PS1 data, yielding strong constraints on the galaxy–halo connection and on alternative dark matter models \citep{MWCensus1, MWCensus2, MWCensus3, MWCensus4, Jethwa:2016}.

Since 2020,  the DECam Local Volume Exploration Survey \citep[DELVE;][]{Drlica-Wagner:2021,Drlica-Wagner:2022}, the Ultraviolet Near Infrared Optical Northern Survey \citep[UNIONS;][]{Gwyn:2025}, and the Kilo-Degree Survey \citep[KiDS;][]{KiDS:2013} have joined other on-going surveys to discover even more faint galaxies around the Milky Way \citep[e.g.,][]{Smith:2023, Smith:2024, Homma:2024, Gatto:2022}. DELVE alone has identified more than a dozen Milky Way satellites  with absolute magnitudes in the range $-5.6 \leq M_V\leq 0.4 $ \citep{Mau:2020, Cerny:2021,Cerny:2021b,Cerny:2023c,Cerny:2023,Cerny:2023b, Cerny:2024, Tan:2025}.
At the writing of this paper, the Local Volume Database \citep[LVDB;][]{Pace:2024} contains $66$ candidate and confirmed Milky Way satellites. Properties of the currently known Milky Way satellites, including both confirmed and candidate systems, are summarized in Table~\ref{table:all_dwarfs}, while their on-sky distribution is shown in Figure~\ref{fig:census_dwarfs}.

In this paper, we leverage recent observational data to conduct the largest systematic census of Milky Way satellite galaxies to date. \cytwo{We combine wide-field imaging from the full six years of DES (DES Y6) and the recent third data release from DELVE (DELVE DR3), supplemented by PS1 DR1 data as analyzed by \citet{MWCensus1}.}  Compared to  previous efforts, the wider and deeper dataset allows us to better account for observational biases in the known galaxy sample. This facilitates better comparisons between the population-level properties of the observed Milky Way satellites (e.g., the satellite luminosity function) and the predictions from simulations and semi-analytic models of galaxy formation.

The primary goal of this census can be described in two parts: (1) to develop a standardized detection pipeline that yields a pure sample of confirmed dwarf galaxies, and (2) to characterize the detection efficiency of this pipeline by constructing a galaxy selection function that can be applied to model galaxy populations, thereby mimicking the same search process. In line with \citet{Willman:2010}, the purpose of this census is to establish a well-defined, uniform dwarf galaxy sample that is complete and pure to the faintest achievable limits. We impose a strict criterion on the purity of the galaxy sample to avoid biasing the observed galaxy number counts by including spurious detections. To ensure such purity, we adopt a high detection threshold, which excludes some of the fainter confirmed and candidate systems that have lower detection significances. 
%when comparing our observed satellite population with predictions from simulations. 
%
% Additionally, this stringent threshold likely omits a subset of as-yet-undiscovered dwarf galaxies with lower significance.

%This is the first paper in the  DELVE Milky Way Satellite Census series, which aims to study the Milky Way satellite population using data from the DELVE DR3 releases. 

The structure of the paper is as follows. We describe the survey data used to produce the census in Section~\ref{sec:data}, while detailing the search algorithms and methods used to find Milky Way satellites in Section~\ref{sec:methods}. The complete census of recovered Milky Way satellites is then presented in Section~\ref{sec:results}. In Section \ref{sec:det_efficiency}, we discuss the observational selection function of the census, which is derived through the injection and recovery of simulated satellites at the catalog level. We use this information in Section~\ref{sec:mwpop} to infer the properties of the total Milky Way satellite population. We conclude our analysis in Section~\ref{sec:discussion}. All associated data products are made publicly available on the DELVE GitHub repository, \footnote{\url{https://github.com/delve-survey/delve_mw_census} \label{github_link}} \cyfive{which is also preserved on Zenodo at \url{https://doi.org/10.5281/zenodo.18383157}.}

\input{Table1}

\begin{figure*}[t]
\centering
\includegraphics[width=0.95\linewidth]{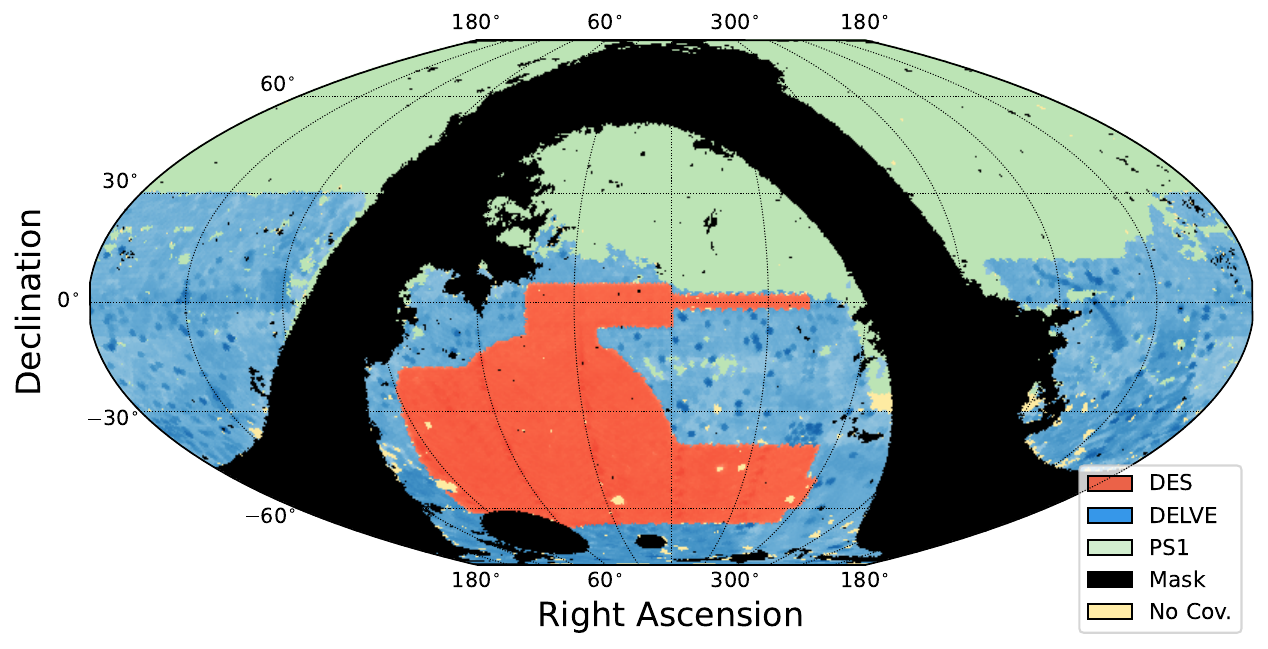}
%\vspace{-3cm}
\caption{Equatorial McBryde--Thomas projection of the coverage area of  DES, DELVE, and PS1 surveys  used in our census. \cytwo{Darker shades indicate regions with deeper magnitude limits, which is particularly apparent in the relatively inhomogeneous DELVE survey.} We also highlight in black the masked regions not considered in our census. This includes regions around the Galactic plane with high interstellar reddening ($E(B-V) > 0.2$) or high stellar density ($\rho_{G<21}>$ \CHECK{8} stars/arcmin$^2$), and regions around other types of stellar systems (such as globular clusters and the Magellanic Clouds). Additionally, we represent in yellow the regions where we do not have coverage from any of our three surveys.\label{Figure:mask} }
\end{figure*}

\section{Census Data \& Footprint} \label{sec:data}
To perform our census, we use catalogs of astronomical objects produced from three different  multi-band optical/near-infrared  wide-field surveys: DES Y6, DELVE DR3, and PS1 DR1, which have 10$\sigma$ median depths of $g \sim 24.7$, $g \sim24.2$, and $g \sim 22.5$, respectively. The catalogs include all classes of sources detected in the imaging, including member stars of the target Milky Way satellites, foreground Milky Way stars, and distant background galaxies.
In Sections~\ref{sec:DES}, \ref{sec:DELVE}, and \ref{sec:PS1},  we briefly discuss the properties of the surveys and how high-quality stellar objects are selected for our analysis.  We also discuss in Section \ref{sec:mask} the geometric mask used to remove problematic regions, such as those near the Galactic Plane, where our search algorithms are not expected to perform well. In total, the combined survey data cover $\sim$\CHECK{27,700}\,deg$^2$, corresponding to $\sim$67\% of the entire celestial sphere, \cythree{$\sim$91\% of the high-Galactic-latitude sky ($|b| \geq 15\degree$)},  and $\sim$98\%
of the unmasked region. A summary of the survey footprint and the geometric mask is shown in Figure \ref{Figure:mask}.

\subsection{DES Y6} 
\label{sec:DES}

DES is a ground-based survey performed using the Dark Energy Camera \citep[{DECam:}][]{flaugher_2015_decam} on the NSF's Víctor M.\ Blanco 4-meter Telescope at the Cerro Tololo Inter-American Observatory (CTIO) in Chile (PropID: 2012B-0001). The DES Wide Field survey covers $\sim$5,000\,deg$^2$ of the southern Galactic cap in five broad-band filters ($grizY$) collected over the span of six years, and is optimized for cosmological analyses with supernovae, weak gravitational lensing and galaxy clustering \citep{DES:2005, DES:2016}.  In this analysis, we use the DES Y6 Gold catalog \cyone{of astronomical objects} detailed in \citet{Bechtol:2025}. DES Y6 Gold is derived from the publicly available DES Data Release 2 \citep{Abbott:2021}, and features improved photometry and object classification. Further details of the DES Data Management (DESDM) image reduction and coaddition pipeline used to process the DES images can also be found in \citet{Morganson:2018}. We note that the full six years of DES Y6 observations provide deeper data than the Y3 release used in \citet{MWCensus1}, reaching a  10 $\sigma$ depth of 
$g\sim 24.7$ compared to $g\sim24.3$.

We use the \texttt{PSF\_MAG\_APER\_8} magnitude measurements from the DES Y6 Gold catalog, which were obtained by fitting individual-epoch Point Source Function (PSF) models to each object and have been  been normalized to the \texttt{MAG\_APER\_8} system as described by \citet{Bechtol:2025}. We use the dereddened  measurements (i.e., with the  \texttt{\_CORRECTED} suffix), which were obtained by applying the interstellar extinction correction $A_b = R_b \times E(B - V)$ where $R_g$ = 3.186, $R_r$ = 2.140, $R_i$ = 1.569, and $R_z$ = 1.196 \citep{Abbott:2021}. The $E(B - V)$ values are obtained from \citet{Schlegel1998} reddening maps with the calibration adjustment suggested by \citet{Schlafly:2011}.

To ensure that we obtain a high-quality sample of objects, we have excluded objects with  \texttt{GOLD\_FLAGS} $> 0$. The \texttt{GOLD\_FLAGS}  exclude objects that are subject to data processing issues and objects with unphysical or unusual measurements \citep{Bechtol:2025}. We search for resolved Milky Way satellites by looking for overdensities in the stellar distribution.  To increase the effectiveness of our search, we remove contaminating background galaxies from our object sample. This is done using the  \texttt{EXT\_XGB} flag, which classifies the objects in DES Y6 into different morphological classes using the \texttt{XGBoost} algorithm \citep{Bechtol:2025}. For our analysis, we only use likely stars with $0\leq\texttt{EXT\_XGB}\leq2$.

We assess the completeness of the high-quality sample of stars by comparing our sample with the catalogs of stars from the Deep/UltraDeep fields of the HSC-SSP Public Data Release 3 (PDR3), which reach a 5$\sigma$ depth of $g \sim 27.4$ \citep{Aihara:2022}.\footnote{We can get a rough estimate of the 10$\sigma$ depth from  5$\sigma$ depth using Pogson's equation: $m_{{\rm lim,} 10 \sigma} = m_{{\rm lim,} 5 \sigma} - 1.25\log_{10}(4)$ \label{sn_conversion}}  We find an overlap of  $\sim$15\,deg$^2$ between DES Y6 and the HSC-SSP Deep and UltraDeep fields. Figure \ref{fig:completeness} shows the stellar completeness as functions of DES $r$-band magnitude  compared to the HSC stellar catalog. We find that our sample achieved $\gtrsim$90\% completeness relative to HSC-SSP down to a 10 $\sigma$ DES magnitude limit of \CHECK{$r\sim23.5$}. For our completeness analysis, we use $r$-band measurements rather than the $g$-band values typically reported for survey depth. This choice reflects the fact that $g$-band photometry is not used for object detection in the catalogs; instead, detections are performed on the combined $r+i+z$ coadd \citep{Morganson:2018, Abbott:2021}.

Finally, we only consider objects located in the DES Y6 Gold footprint. The footprint is expressed as a  \texttt{HEALPix} map with resolution of $\texttt{nside} = 4096$ and includes regions satisfying two conditions: (1) At least two exposures per band in each of the $g, r, i, z$ bands, and (2) $f_{griz} > 0.5$, where $f_{griz}$ is the fraction of the \texttt{HEALPix} pixels that has simultaneous coverage in all four bands, \cytwo{computed using a higher-resolution map with $\texttt{nside} = 16384$.}  
Regions with incomplete coverage are excluded because our  estimates of the background stellar density are inaccurate in those regions, leading to a higher rate of false positive detections. We require coverage in the $g$, $r$, $i$, and $z$ bands because the satellite detection algorithms rely on photometry from the $g$, $r$, and $i$ bands, while catalog object detection was performed on the $r$ + $i$ + $z$ detection coadd \citep{Abbott:2021}. Restricting our analysis to regions in the DES Y6 Gold footprint that are not excluded by our geometric mask (Section~\ref{sec:mask}), we end up with a total area of $\sim$\CHECK{4,900}\,deg$^2$. Using the DECam Survey Property Maps (\texttt{decasu})\footnote{\url{https://github.com/erykoff/decasu}} tool, we estimate the 10$\sigma$ point-source depth for DES Y6 to be \CHECK{$g \sim 24.7,\ r \sim 24.4,\ i \sim 23.8$}, with a standard deviation of 0.2 in each of the three bands. %\texttt{decasu}  estimates the depth of the surveys through  a random forest classifier that was trained on the  combined survey characteristics, such as coverage, seeing, and sky brightness.}

\subsection{DELVE DR3} 
\label{sec:DELVE}
%%%%%%Leo6%%%%%%
DELVE is a DECam community survey program (PropID: 2019A-0305) that has assembled contiguous imaging of a large portion of the high-Galactic-latitude southern sky in the $g, r, i,$ and $z$ bands outside the DES footprint \citep{Drlica-Wagner:2021, Drlica-Wagner:2022}. The forthcoming DELVE DR3 (\citealt{Tan:2025}; Drlica-Wagner et al.\ in prep.)\footnote{\url{https://datalab.noirlab.edu/data/delve}} combines data from more than \CHECK{150} nights of dedicated DELVE observing with  public archival DECam data. This includes data from the DECam Legacy Survey \citep[DECaLS;][]{Dey:2019}, the DECam eROSITA Survey \citep[DeROSITAS;][]{Zenteno:2025}, and  numerous community programs.\footnote{We note that the official DELVE DR3 release includes the DES Y6 catalogs. However, we treat the DES Y6 data separately and use DELVE DR3 to refer to the non-DES portion of DELVE DR3.}  The images in the DELVE dataset were processed using the DESDM pipeline \citep{Morganson:2018}, with the image de-trending and coaddition pipeline closely following DES Y6 to ensure consistency.

\cyone{Compared to DES Y6, DELVE is less homogeneous (see Figure~\ref{Figure:mask}) due to the fact that it is an amalgamation of many different observing programs. Despite this, DELVE data have been used to discover 14 new Milky Way satellites \citep{Mau:2020, Cerny:2021,Cerny:2021b,Cerny:2023c,Cerny:2023,Cerny:2023b, Cerny:2024, Tan:2025}. In addition, the dataset has been used to produce a weak-lensing shape catalog for cosmic shear analyses that are robust to survey inhomogeneities \citep{Anbajagane:2025a, Anbajagane:2025b, Anbajagane:2025c, Anbajagane:2025d, Anbajagane:2025e}.  In this section, we summarize the details of DELVE DR3 that are relevant to the Milky Way satellite search, and we refer the reader to \citet{Tan:2025} and \CHECK{Drlica-Wagner et al.\ (in prep.)} for more details.}

\begin{figure}[t]
\centering
\includegraphics[width=\linewidth]{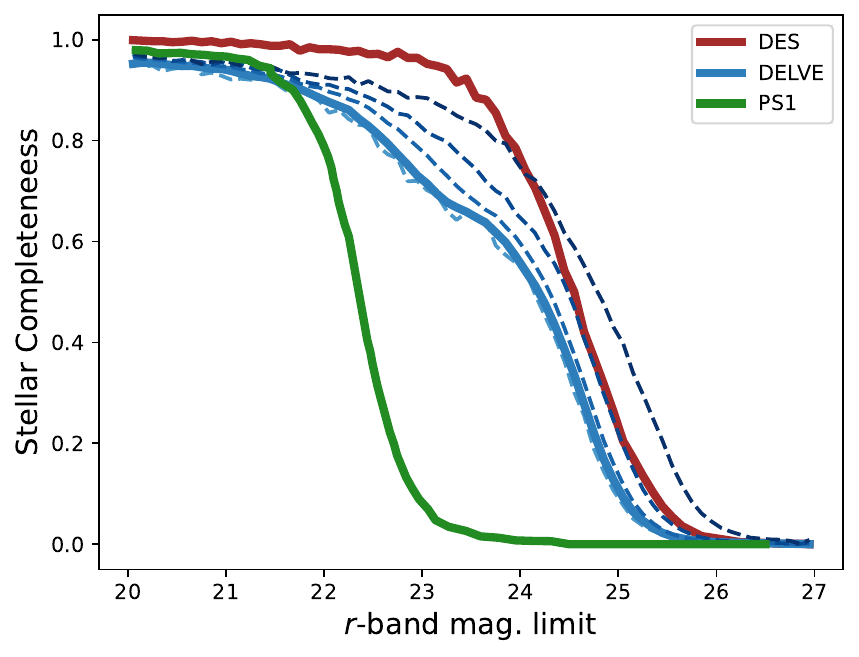}
%\vspace{-3cm}
\caption{Stellar completeness of the DES Y6, DELVE DR3, and PS1 DR1 stellar catalogs relative to the HSC-SSP PDR3 stellar catalog as a function of $r$-band magnitude limit.
For the inhomogeneous DELVE survey, stellar completeness is illustrated at the median $r$-band depth ($m_{\mathrm{lim},r} = 23.5$) with a solid line, and at additional depths $m_{\mathrm{lim}}\sim\{23.0, 24.0, 24.5, 25.0\}$ using dashed lines, \cytwo{with darker lines indicate deeper limiting magnitudes.} \label{fig:completeness}}
\end{figure}

\cyone{Following the procedure described for DES Y6 in Section~\ref{sec:DES}, we use the \texttt{PSF\_MAG\_CORRECTED} measurements from DELVE DR3. We also exclude objects with \texttt{GOLD\_FLAGS>0}, which follows the same definition as in DES Y6. However, for the star--galaxy separation, we use the \texttt{EXT\_FITVD} flag, which classifies objects based on their multi-epoch \texttt{fitvd} photometric quantities, specifically the bulge and disk model fit (BDF) measurements. We specifically use likely stars with  $0\leq\texttt{EXT\_FITVD}\leq2$ for our \texttt{ugali} search and $0\leq\texttt{EXT\_FITVD}\leq1$ for our more noise-sensitive \texttt{simple} search. \cytwo{Due to the different star–galaxy classification methods used in the DELVE region (\texttt{EXT\_FITVD}) and the DES region (\texttt{EXT\_XGB}), we observe significantly different stellar completeness levels, even at similar depths (Figure \ref{fig:completeness}). We note that in general the  \texttt{EXT\_FITVD} classification  is less complete but more pure compared to the \texttt{EXT\_XGB} classification \citep{Bechtol:2025}, which makes it more suitable for the more inhomogeneous DELVE survey.}  More information of the \texttt{EXT\_FITVD} classifier can be found in Appendix A of \citet{Bechtol:2025}.}

We also assess the completeness of DELVE DR3 by comparing it to star catalogs from HSC-SSP. However, due to the wide variation in depth within the DELVE DR3 data, we separate the DELVE regions based on their $r$-band magnitude limit and compare each subset to the HSC-SSP Wide data, which reaches a 5$\sigma$ depth of $g \sim 26.5$ \citep{Aihara:2022}.$^{\ref{sn_conversion}}$ We use the HSC-SSP Wide fields, since their broad sky coverage overlaps with multiple DELVE regions of varying depth, with a total overlapping area of $\sim$560\,deg$^2$. We find that the completeness estimates differ only slightly when comparing results based on the HSC-SSP Wide fields to those using the HSC-SSP Deep and UltraDeep fields for the same region. Figure~\ref{fig:completeness} shows the completeness of DELVE DR3 in regions with different magnitude limits, where deeper regions exhibit much higher completeness. These deeper regions thus yield slightly higher detection efficiency; details on how we account for variations in detection significance across the DELVE footprint due to depth differences can be found in Section~\ref{sec:selfunc_ML}.

\cyfive{We note the presence of a small hump in the DELVE detection efficiency at $r \sim 23$ in regions with shallower $r$-band depth. Many of these shallower regions are covered by only a few exposures, making them more susceptible to variations in observing conditions such as poor seeing. Since the magnitude limit and the star--galaxy separation efficiency depend differently on the PSF FWHM, two regions with the same magnitude limit but different PSF FWHM can have different stellar completeness (e.g., see Figure 15 in \citealt{Slater:2020}). Averaging over regions with different observational properties causes the stellar completeness function to differ from the standard sigmoidal shape.}

Similar to our analysis of DES Y6, we only consider objects that are located in the DELVE DR3 footprint, which is defined to be regions with at least 1 exposure per band in each of the  $g,r,i,z$ band and $f_{griz} > 0.5$. Additionally, to create a more uniform survey footprint, we manually remove small, discontinuous regions from the DELVE coverage.  To prevent double-counting, we also remove DELVE DR3 regions that have overlap with the deeper DES Y6 regions, which results in an effective unmasked survey area of $\sim$\CHECK{12,000} deg$^2$.  Using \texttt{decasu}, we estimate the median 10$\sigma$ point-source depth for DELVE DR3 to be \CHECK{$g\sim 24.2, r\sim 23.7, i\sim23.2$}, with a larger standard deviation of 0.5 across the three bands. \cytwo{For the $r$-band depth used in the stellar completeness analysis, we find that approximately 11\%, 45\%, 29\%, 11\%, and 3\% of the area have limiting magnitudes of $r_{\mathrm{lim}} \sim$ 23.0, 23.5, 24.0, 24.5, and 25.0, respectively.}

%\begin{figure*}
%    \centering
%    \includegraphics[width=\linewidth]{figures/test_data.png}
%    \caption{Diagnostic Plot}
%    \label{fig:diagnostic}
%\end{figure*}

\subsection{PS1 DR1}
\label{sec:PS1}
To cover regions that were not observed by DES or DELVE, our census also included data from the 3$\pi$ Survey performed with the PS1 Gigapixel Camera 1, on the 1.8-m PS1 telescope at Haleakala Observatories in Hawai`i. Due to the similarity between our analysis and that of \citet{MWCensus1}, we choose not to perform a new Milky Way satellite search in the PS1 region, and instead we use the recovered satellite sample and selection function from \citet{MWCensus1}.\footnote{We explored analyzing PS1 DR2; however, we did not find any significant improvement in sensitivity.} To construct the PS1 footprint, we consider regions that are: 1) in the PS1 survey footprint defined in \citet{MWCensus1}, 2) not covered by the deeper DES Y6 and DELVE DR3 surveys \cyone{and 3) not covered by our geometric mask (Section \ref{sec:mask}), resulting in a total area of \CHECK{$\sim$10,900} deg$^2$}. The 10$\sigma$ detection limit of PS1 DR1 is estimated to be $g$ =22.5 and $r$ =22.4.  More details about the PS1 DR1 data, such as the photometry and the star--galaxy separation, can be found in \citet{MWCensus1}.

%As we  we did not observe a significant improvement in detection effiency in PS1  DR2,

\subsection{Geometric Mask}
\label{sec:mask}

\cyone{As with \citet{MWCensus1}, we apply a  geometric mask to exclude regions of the survey footprint where our search algorithms are expected to produce a large number of false positives. While we remove candidates detected within the masked regions, these regions may still be used by the search algorithms for background estimation.}

Around the Galactic plane, we mask regions with high interstellar extinction, defined as $E(B-V) > 0.2$ \citep{Schlegel1998}, as well as areas with a high density of Milky Way stars. The latter are defined as regions
where the density of \textit{Gaia} DR2 \citep{Gaia:2018} sources with $G < 21$ exceeds \CHECK{8} arcmin$^{-2}$  for $G < 21$ and Galactic latitude is $|b| < 15\degree$. \cyone{These regions are excluded because their photometry and stellar density can vary significantly over small scales. Such small-scale variations are poorly captured by the background estimation of our search algorithms, which assume a constant background level, leading to numerous spurious detections.} Furthermore, isochrone-based search methods are not particularly effective in regions with high densities of foreground Milky Way stars. In fact, dwarf galaxies discovered near the Galactic plane have typically been identified using alternative methods. For example, Sagittarius was discovered through the distinct kinematics of its member stars relative to foreground stars \citep{Ibata:1994}, while Antlia II was identified using a combination of astrometry, photometry, and variable star detections \citep{Torrealba:2019b}. For the same reason, we also mask circular regions with radii of $9\degree$ and $3\degree$ around the LMC and SMC, respectively, corresponding to approximately three times their half-light radii \citep{Choi:2018, Munoz:2018}.

We also mask regions around resolved stellar systems that are not Milky Way satellite dwarf galaxies and regions known to produce spurious hotspots.  This mask includes Milky Way globular clusters \citep{Harris:1996, Sitek:2017, Pace:2024}, open clusters \citep{Paunzen:2008}, regions around bright stars \citep{Hoffleit:1991}, \cytwo{and nearby galaxies (outside of the viral radius of the Milky Way) with resolved members stars \citep{Corwin:2004, Nilson:1973, Webbink:1985, Kharchenko:2013, Bica:2008}}.  

For our main analysis, we also mask regions around ambiguous compact ultra-faint systems with half-light radii $r_{1/2} < 15$ pc, such as DELVE 1 and Ursa Major III/UNIONS 1 \citep{Mau:2020, Smith:2024}.  These compact systems lie outside the parameter space considered in our census, as further discussed in Section~\ref{sec:results}. However, we study their detection efficiency in Appendix \ref{Appendix:ambigous} using the same dataset.
For the injection--recovery tests used to characterize the detection efficiency of the census (Section~\ref{sec:sims}), we also mask the known satellites from Table \ref{table:all_dwarfs}.

Our geometric mask is expressed as a $\texttt{nside} = 4096$ \texttt{HEALPix}  map as shown in Figure \ref{Figure:mask}. For extended objects with size information, we apply a circular mask with a radius that matches the object's half-light radius (with a minimum radius of 0.05\degree). For stars and objects with no size information available, we set the radius of the circular mask to 0.1\degree. \cyone{The \texttt{HEALPix} mask, which includes both the geometric mask and survey footprint for DES, DELVE, and PS1, is available in the GitHub repository.$^{\ref{github_link}}$}

%Finally, our masked regions also exclude regions that are not in the survey footprint defined as in Section \ref{sec:data}.  After applying the mask, we end up with a combined area of \CHECK{27,723} deg$^2$ \TODO{reorganized area} using the DES, DELVE and PS1 survey datasets.

\section{Satellite Search Methods} \label{sec:methods}
To find Milky Way satellites in our wide-field survey data, we employ two search algorithms, \texttt{ugali}\footnote{\url{https://github.com/DarkEnergySurvey/ugali}} and \texttt{simple}\footnote{ \url{https://github.com/DarkEnergySurvey/simple} }, described in Sections \ref{sec:ugali} and \ref{sec:simple}, respectively. Both algorithms detect Milky Way satellites as overdensities of resolved stars that follow a distinct locus in color–magnitude space, but \texttt{ugali} uses a more rigorous maximum-likelihood approach, while \texttt{simple} finds local density peaks in the stellar density field. Although the two algorithms show similar performance in detecting real systems, false positives from one are often not shared by the other. Therefore, by requiring a candidate to be independently detected by both algorithms, we can significantly reduce the number of false positives. A more detailed discussion of the detection criteria that we set for a candidate to be included in our census  is provided in Section~\ref{sec:det_criteria}. 

\begin{figure*}[t]
\centering
\includegraphics[width=\linewidth]{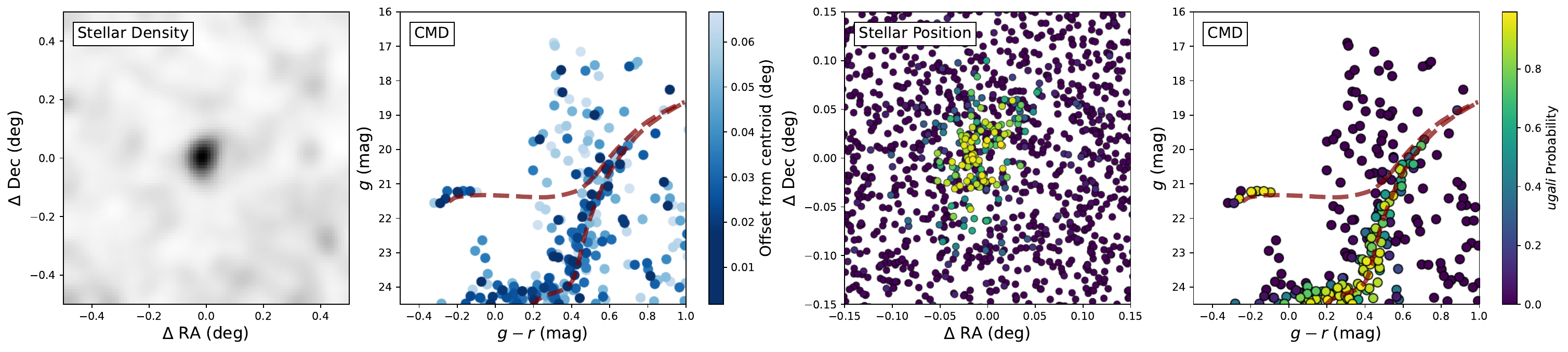}
%\vspace{-3cm}
\caption{ \cytwo{Diagnostic plots for one of the census galaxies, Hydra II, constructed using DELVE DR3 data.
Left: Smoothed, isochrone-filtered stellar density map centered on Hydra II. 
Center Left: Color–magnitude diagram (CMD) of stars near the system, with points colored by their angular distance from the Hydra II centroid. A best-fit \texttt{PARSEC} isochrone is overlaid as a dashed maroon line.
Center Right: Spatial distribution of stars, colored by their \texttt{ugali} membership probability. 
Right: CMD of stars, colored by their \texttt{ugali} membership probability. 
For this illustration, we adopt an elliptical Plummer profile using the best-fit structural parameters as the \texttt{ugali} galaxy model. However, in our actual search, we use a radially symmetric Plummer profile, with galaxy parameters drawn from a predefined grid.
} \label{fig:example_system}}
\end{figure*}

\subsection{Likelihood-based search algorithm: ugali} 
\label{sec:ugali}

The first search algorithm used in the census employs a likelihood-based approach implemented in the Ultra–faint GAlaxy LIkelihood toolkit, \texttt{ugali} \citep{Bechtol:2015, MWCensus1}. This approach identifies Milky Way satellite candidates by comparing the likelihood of two models: (1) includes only a uniform distribution of background sources,\footnote{We collectively refer to  foreground stars and misclassified background galaxies as ``background'' sources.} while (2) adds a Milky Way satellite galaxy. The log-likelihood function is given by
\begin{equation}
    \log\mathcal{L}(\mathcal{D}| \lambda,\theta) = -f\lambda-\sum_{i \in \rm stars}\log(1-p(\mathcal{D}_i| \lambda,\theta)) ,
\end{equation}
where the richness, $\lambda$, represents the total number of member stars of the galaxy with masses greater than $0.1M_\odot$, $f$ represents the fraction of member stars that are within the survey's spatial footprint and magnitude limits, and are thus observable in our data.
The summation represents all the stars in the catalog within $r<0.5 \degree$ of the candidate system. 

The term $p(\mathcal{D}_i| \lambda,\theta)$ represents the probability that star $i$, given its data $\mathcal{D}_i$, is a member of the dwarf galaxy with richness, $\lambda$, and structural parameters, $\theta$, as opposed to the uniform background. 
This probability is given by
\begin{equation}
    p(\mathcal{D}_i| \lambda,\theta) = \frac{\lambda u(\mathcal{D}_i|\theta)}{\lambda u(\mathcal{D}_i|\theta) + b_i} ,
\end{equation}
where $u(\mathcal{D}_i| \theta)$ is the normalized probability that the astrometric and photometric properties of the star are consistent with the satellite galaxy, and $b_i$ is the expected density function for the background source population (determined from a circular annulus surrounding each candidate at $0.5\degree < r_{\rm ann} < 2.0\degree$).

We assume that  $u(\mathcal{D}_i|\theta)$ can be separated into a spatial component and a color-magnitude component, $u(\mathcal{D}_i|\theta) = u(\mathcal{D}_{s,i}|\theta_s) u(\mathcal{D}_{c,i}|\theta_c)$. For the spatial component, we consider the celestial coordinates of the stars, \( \mathcal{D}_{s,i} = \{ \rm{RA}_i, \rm{Dec}_i \} \), and assume that the probability density function (PDF) of candidate member stars is radially symmetric and follows a Plummer profile \citep{Plummer:1911}. The profile is defined by the following parameters: centroid coordinates and half-light radius, \( \theta_s = \{ {\rm RA}, {\rm Dec}, r_h \} \). For the color--magnitude component, we consider the $g,r$-band magnitude and magnitude error of the stars  $\mathcal{D}_{c,i} = \{ g_i, \sigma_{g,i}, r_i, \sigma_{r,i}\}$. We  built our PDF in color--magnitude space ($g$, $g-r$) using old, metal-poor \cytwo{\texttt{PARSEC v1.2S}} isochrones \citep{Bressan:2012, Chen:2014, Tang:2014, Chen:2015} with distance modulus, age, and metallicity, $\theta_c = \{(m-M)_0, \tau, Z\}$, that have been weighted with  a \citet{Chabrier:2001} initial mass function. \cytwo{For our analysis, we use a composite isochrone consisting of four isochrones with the following galaxy parameters: $ Z = \{0.0001, 0.0002\}$ ([Fe/H]$\sim$$\{-2.2, -1.9\}$),  $ \tau = \{$10 Gyr, 12 Gyr$\}$. }  \cytwo{As an example, the \texttt{ugali} membership probability of the stars around Hydra II is shown in Figure \ref{fig:example_system}.  } We also perform our \texttt{ugali} search using \( g,i \) band pairs in place of \( g,r \).

Hotspots are identified when the model that includes the satellite galaxy provides a significantly larger log-likelihood than the background-only model. 
To search for Milky Way satellite candidates, we evaluate the likelihood over a spatial grid of \texttt{HEALPix} pixels ($\texttt{nside} = 4096$; spatial resolution of $\sim\,0.08 \arcmin$). 
For each point, we scan through a grid of half-light radii,  $r_h= {1.2\arcmin, 4.2\arcmin, 9.0\arcmin}$ and distance moduli ranging  from $16 \leq (m - M)_0 < 23$ in steps of 0.5\,mag (corresponding to heliocentric distances of 16\,kpc$\ \leq D < \ $400\,kpc). At each grid point, we find the combination of parameters that maximizes the likelihood. We then quantify the statistical significance of a hotspot using a Test Statistic (TS) based on the likelihood ratio between the model that includes the satellite and the uniform-background-only model such that
\begin{equation}
    \rm{TS} = 2 [\log\mathcal{L}(\mathcal{D}| \lambda= \hat{\lambda}, \theta=\hat{\theta}) - log\mathcal{L}(\mathcal{D}| \lambda= 0) ],
\end{equation}
where $\hat{\lambda}$ and $\hat{\theta}$ are the values of the stellar richness and satellite parameters, respectively, that maximize the likelihood. We then find isolated peaks (i.e., contiguous regions exceeding a ${\rm TS} > 10$) in the likelihood maps to obtain a list of hotspots.

\subsection{Stellar-overdensity search algorithm: simple} 
\label{sec:simple}
The second search algorithm used in our census is \texttt{simple}. The algorithm has been successfully used to discover more than twenty Milky Way satellite to date \citep[e.g.,][]{Bechtol:2015, Drlica-Wagner:2015, Mau:2020,Cerny:2021, Cerny:2023c, Tan:2025}. However, it also outputs a larger number of false positives compared to \texttt{ugali} due to its comparatively simple implementation. 

The algorithm identifies candidates by comparing the stellar density in a region of interest to the background source density. To enhance the contrast, it further applies a simple isochrone filter to remove stars that are unlikely to be associated with an old, metal-poor stellar system. To perform the isochrone filter cut, we use a \texttt{PARSEC} isochrone with metallicity $Z= 0.0001$ ([Fe/H] $  \sim \cyfive{-}2.2)$ and age of $\tau = 12$ Gyr. We perform the search multiple times across different distance moduli of \CHECK{$16 \leq (m - M)_0 < 23$} in steps of 0.5\,mag, and select stars which have color differences with the template isochrone of $\Delta(g-r)<\sqrt{0.1^2+\sigma^2_g+\sigma^2_r}$ where $\sigma^2_{g,r}$ are the $g$- and $r$-band uncertainties. 

To search for Milky Way satellites in the data, we first partition the footprint into $\texttt{nside} = 32$ \texttt{HEALPix} pixels. For each \texttt{HEALPix} pixel,  we smooth the filtered stellar density field with a Gaussian kernel with $\sigma = 2^\prime$ and identify local density peaks  by iteratively raising a density threshold until there are fewer than 10 disconnected regions above the threshold value. For each of the local density peaks, we compute the Poisson significance of the observed stellar counts within the aperture given the local density field, 
\begin{equation}
    \rm{SIG} = \rm{ISF}_\mathcal{N} \left[ \rm{SF}_{\mathcal{P}(\lambda=N_b)} (N_{\rm{obs}})\right],
\end{equation} 
where $\rm{SF}_{\mathcal{P}(\lambda=N_b)} $ is the survival function of a Poisson distribution with the counts estimated from the background and  $\rm{ISF}_\mathcal{N} $ is the inverse survival function of a Gaussian distribution with $\mu=0$ and $\sigma=1$. We iterate through circular apertures with radii between 0.6$\arcmin$ to 18$\arcmin$  in steps of 0.6$\arcmin$ and choose the radius that maximizes the detection significance. The local field density is estimated from an annulus between 18$\arcmin$ and 30$\arcmin$ surrounding the peak, accounting for the survey coverage.

\subsection{Detection Criteria}
\label{sec:det_criteria}

\begin{figure}[t]
\centering
\includegraphics[width=\linewidth]{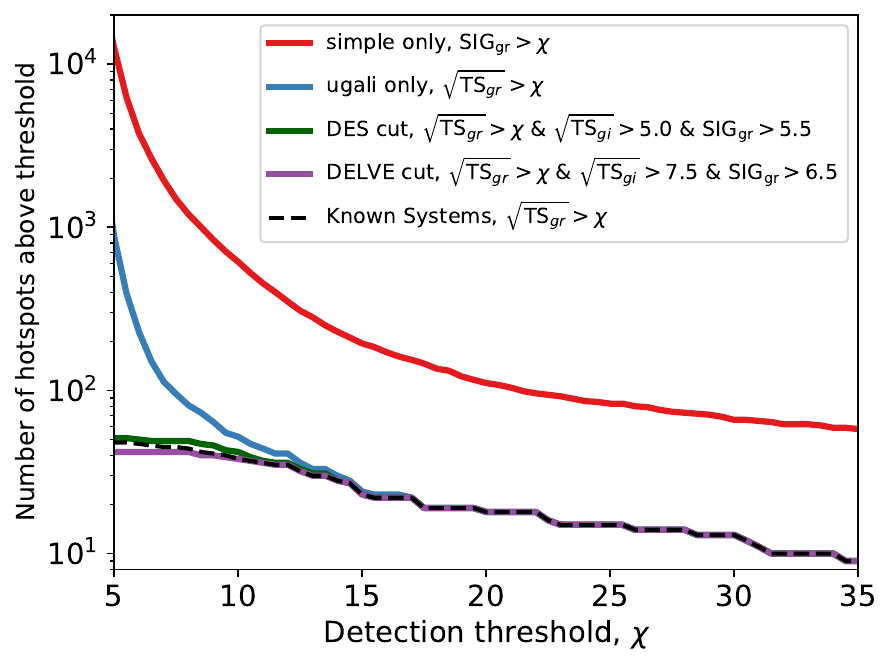}
\caption{Distribution of ``hotspots'' identified by the search algorithms in the DES and DELVE data. The $x$-axis represents the detection threshold value, $\chi$. The red line shows hotspots returned by \texttt{simple} that pass the SIG$_{gr}>\chi$ threshold, while the blue line shows those returned by \texttt{ugali} that meet the $\sqrt{\mathrm {TS}_{gr}}>\chi$ threshold. The green and purple lines represent subsets of \texttt{ugali} hotspots that meet additional detection significance thresholds on $\sqrt{\mathrm {TS}_{gi}}$ (based on $g-i$ bands) and SIG$_{gr}$ (from \texttt{simple}),  both of which are required for inclusion in the census within the DES and DELVE regions, respectively. The dashed black lines indicate the number of previously known dwarf galaxy systems in our DES and DELVE footprint that meet the \texttt{ugali} $\sqrt{\mathrm{TS}_{gr}}>\chi$ threshold.\label{fig:junk_number}}
\end{figure}

%As mentioned previously, running  over the entire census footprint at a detection threshold of $\sqrt{TS} $= 5 ($\sigma=5$) results in  $\sim$1,000 ($\sim$10,000) hotspots including many false positives. 

When running our search algorithms (\texttt{ugali}/\texttt{simple}) on the entire census footprint, we obtain thousands of ``hotspots''  (i.e., locations where the detection significance exceeds the detection threshold), with the majority of these hotspots having relatively low significance (see Figure~\ref{fig:junk_number}). While many known dwarf galaxies correspond to high-significance hotspots, the nature of the remaining hotspots is less clear, with many likely being false positives caused by survey artifacts or inaccuracies in the survey coverage maps. 

As outlined in Section~\ref{sec:intro}, our analysis requires that the census of Milky Way satellite galaxies be pure---i.e., consist exclusively of real satellites. As shown in Figure~\ref{fig:junk_number}, increasing the \texttt{ugali} detection threshold raises the fraction of hotspots corresponding to real systems, eventually reaching 100\% purity. Thus, by adopting a sufficiently high threshold, we can obtain a pure Milky Way satellite sample. At lower thresholds, we can further suppress false positives by restricting the sample to hotspots detected by both \texttt{ugali} and \texttt{simple} and, for \texttt{ugali}, by requiring detections in both the $g,r$ and $g,i$ band pairs. Using the detection significance and identity of candidates recovered by \texttt{ugali}/\texttt{simple} as a guide, we adopt a detection threshold that excludes all unidentified hotspots likely to be false positives, while maximizing the recovery of faint confirmed systems.

For DES Y6, we identify hotspots with \texttt{ugali} significance in the  $g$ and $r$ bands of $\sqrt{\mathrm{TS}_{gr}}>5.0$. We then evaluate the significance of these hotspots with \texttt{ugali} using the $g$ and $i$ bands, as well as with \texttt{simple} using the $g$ and $r$ bands.  Only hotspots that meet all three of the following criteria are included:  $\sqrt{\mathrm{TS}_{gr}}>5.0$, $\sqrt{\mathrm{TS}_{gi}}>5.0$, and SIG$_{gr}>5.5$. For the more inhomogeneous DELVE DR3 data (see Section \ref{sec:DELVE}), we required a higher detection threshold of $\sqrt{\mathrm{TS}_{gr}}>8.0$, $\sqrt{\mathrm{TS}_{gi}}>7.5$, and SIG$_{gr}>6.0$. For the PS1 region, we adopt the detection criteria from \citet{MWCensus1}: $\sqrt{\mathrm{TS}_{gr}}>6.0$ and SIG$_{gr}>6.0$, with no detection threshold for $\sqrt{\mathrm{TS}_{gi}}$.

To assess the suitability of our detection criteria, we visually inspect several hotspots that fall just below these thresholds. We find that some of the hotspots are obvious survey artifacts, while the nature of others are sufficiently ambiguous that we cannot conclusively determine their nature without additional follow-up observations. If we target completeness rather than purity, we would need to lower our detection threshold to $\sqrt{\mathrm{TS}_{gr}}>4.5$ and  SIG$_{gr}>4.5$ in order to included every known dwarf galaxy in the DES and DELVE regions. This selection would result in $\sim$150 unidentified candidates in the DES footprint and $\sim$350 in DELVE.

\section{Census Satellite Population }
\label{sec:results}

We identify 17 and 21 confirmed and candidate Milky Way satellites in the DES and DELVE footprint, respectively. These systems pass our detection criteria (Section~\ref{sec:det_criteria}) and can be unambiguously cross-matched with our compilation of Milky Way satellites (Table~\ref{table:all_dwarfs}; \citealt{Pace:2024}) to within $0.1\degree$.

The classification of some recently discovered old, metal-poor halo systems, such as Eridanus III, DELVE 1 and Ursa Major III/UNIONS 1, as either dark-matter-dominated dwarf galaxies or baryon-dominated star clusters remains highly debated \citep{Simon:2024, Smith:2024, Errani:2024, Cerny:2026arXiv260217652C}. Thus, to reduce the number of ambiguous systems which might be misclassified as star clusters in our sample, we only considered systems with physical half-light radii $r_{1/2}>15$\,pc. This threshold corresponds to the size of Virgo II, the smallest system commonly regarded as a likely dwarf galaxy upon its discovery \citep{Cerny:2023c}. Although some of the recovered satellites above the size cut have not yet been spectroscopically confirmed as dwarf galaxies, none are believed to be particularly contentious. We therefore adopt the simplifying assumption that any system passing the size cut is considered a Milky Way satellite galaxy. We also refer the reader to Appendix~\ref{Appendix:ambigous} for the detection efficiencies of the compact ambiguous systems below our size cut.

\cyfive{Furthermore, we only considered systems with heliocentric distances between 16\,kpc and 400\,kpc, and treat systems outside this distance range as undetectable in our search. We impose a lower distance limit of 16\,kpc, corresponding to the minimum distance at which our detection methods are optimized. The larger upper limit of 400\,kpc is set as roughly half the distance from the Milky Way to M31 \citep{Stanek:1998} and is comparable to common estimates of the Milky Way virial radius \citep{Dehnen:2006, Garrison-Kimmel:2014, Ou:2024}. Systems outside this distance range include the ambiguous Ursa Major III/UNIONS 1 and Kim 3, with a heliocentric distance of less than 16\,kpc \citep{Smith:2024, Kim:2016b}, as well as more distant Local Group galaxies such as Phoenix I, Leo T, Leo K, and Leo M, all of which lie at heliocentric distances greater than 400\,kpc \citep{Battaglia:2012, Higgs:2021, McQuinn:2024}.}

Within the DES Y6 region, we recover 17 of the 18 known Milky Way satellites. The significance for both search methods is presented in Table \ref{table:des_search} in Appendix~\ref{appendix:detection_efficiencies}. This sample matches the population identified in the previous search using the DES Y3 data by \citet{MWCensus1}, although our search recovers candidates at higher significance due to the deeper DES Y6 data. The only known system that we failed to recover is Cetus III, a faint ($M_V = -3.5$) and distant ($D = 251$\,kpc) system discovered in deeper HSC-SSP data \citep{Homma:2018}. \cytwo{While we measure a higher significance for Cetus III compared to the previous analysis by \citet{MWCensus1}, $\sqrt{\mathrm{TS}_{gr}}=4.8$, $\sqrt{\mathrm{TS}_{gi}}=4.1$, SIG$_{gr}=4.9$, it still narrowly misses our detection threshold of $\sqrt{\mathrm{TS}_{gr}}=5.0$, $\sqrt{\mathrm{TS}_{gr}}=5.0$,  SIG$_{gr}=5.5$.} %\cyfive{As in \citet{MWCensus1}, we did not recover in the deeper DES Y6 data the candidates DES J2038-4609 (Indus II) and DES J0225+0304, originally reported by \citet{Drlica-Wagner:2015} and \citet{Luque:2017}, respectively.}

For the DELVE DR3 region, we recovered 21 out of the 27 known Milky Way satellites (see Table \ref{table:delve_search} in Appendix~\ref{appendix:detection_efficiencies}). The slightly lower recovery rate is expected given the shallower depth of DELVE DR3 relative to DES Y6, and the stricter detection threshold applied to mitigate the higher incidence of imaging artifacts. Four of the unrecovered systems (DELVE~2, Leo~VI, Leo Minor~I, and Virgo~II) were originally discovered in DELVE data, and they were re-detected here with a lower significance than we required for inclusion in our systematic census \citep{Cerny:2021b, Cerny:2023c, Tan:2025}. Furthermore, follow-up DECam data that was obtained to confirm the discovery of Leo~VI and Leo Minor~I was not included in the DELVE DR3 processing. Two of the remaining systems, Virgo~I and Virgo~III, were discovered in much deeper HSC-SSP data. 

One system we would like to highlight is Carina~III, which was not identified by our \texttt{ugali} peak finder as an isolated hotspot. 
This is because Carina~III resides only 18$\arcmin$ from its brighter neighbor Carina~II \citep{Torrealba:2018}, which causes \texttt{ugali} to interpret them as a single hotspot. Nevertheless, by evaluating the likelihood at the known position of Carina~III, we recover a strong detection with a \texttt{ugali} significance of $\sqrt{\mathrm{TS}_{gr}} = 21.6$. The system is also flagged by \texttt{simple} as a possible independent candidate, with a high significance of $\sigma_{gr} = 7.1$. We choose to include Carina~III in our census despite the lack of a unique \texttt{ugali} detection for two reasons. First, once a brighter system such as Carina~II is discovered, visual inspection of the surrounding region would naturally reveal nearby companions like Carina~III. Second, our injection tests (Section~\ref{sec:det_efficiency}) do not account for the case of two systems in such close proximity, and would have included this system. We further note that Carina~III's close angular separation from Carina~II (0.3\degree) is observationally rare, with the second closest pair of known dwarf galaxies having a separation of 1.5\degree.

We reuse the search of PS1 DR1 performed by \citet{MWCensus1} to expand our coverage of the northern celestial hemisphere.
In the region where we rely on the PS1 data, \citet{MWCensus1} recovered \CHECK{11} out of the \CHECK{17} known Milky Way satellites (Table~\ref{table:ps1_search} in Appendix~\ref{appendix:detection_efficiencies}).
This comparatively lower recovery rate is expected due to the significantly shallower depth of PS1 DR1, $g\sim22.5$. Two of the unrecovered satellites (Pisces~II and Pegasus~III) were found using data from SDSS, but were only confirmed through deeper follow-up imaging with the 4-m Mayall Telescope and DECam, respectively \citep{Belokurov:2010, Kim:2015a}. Three of the satellites were discovered using data from surveys that were significantly deeper than PS1 DR1. Bo\"{o}tes~IV was found in the HSC-SSP \citep{Homma:2019}, Pegasus~IV was identified in DELVE \citep{Cerny:2023}, and Bo\"{o}tes~V was independently discovered in both DELVE and UNIONS \citep{Cerny:2023c, Smith:2023}.  Finally, Bo\"{o}tes~III is a diffuse object (with an elliptical half-light radius of $a_h=40\arcmin$) first identified in filtered stellar density maps from SDSS DR5 \citep{Grillmair:2009}. Bo\"{o}tes~III has proven difficult to detect with automated search algorithms due to its diffuse nature and complex morphology \citep{Koposov:2008, Walsh:2009, MWCensus1}. We note that several of the unrecovered satellites (Bo\"{o}tes~III, Bo\"{o}tes~V, Pegasus~III, Pegasus~IV, and Pisces~II) have coverage in DELVE DR2, but lie within the PS1 footprint of our census rather than the DELVE footprint. This is because the deeper, coadd image-based DELVE DR3 only includes contiguous regions with coverage in all four $griz$ bands, resulting in a smaller footprint compared to DELVE DR2.

We note that four known, relatively massive satellites fall outside our survey footprint (see Figure~\ref{fig:census_dwarfs}). The LMC and SMC are excluded due to their high stellar densities, which interfere with the background estimation of our search algorithm. Our search algorithms are also not designed to search for  satellites this luminous. Sagittarius and Antlia II are also excluded, but in this case because they lie within our Galactic plane mask. 

To incorporate the brightest satellites into a Milky Way population analysis, even if they lie outside the survey footprint, we can modify the selection function by assuming that all bright satellites ($M_* \gtrsim 10^7\ M_\odot$) have already been discovered. This assumption is motivated by the fact that the last bright Milky Way satellite discovered was Sagittarius in 1994 \citep{Ibata:1994}, despite extensive modern surveys. To implement this, we assume that any Milky Way satellite with an absolute magnitude of $M_V < -12.5$ is always detected regardless of their location or distance.  Under this assumption, we extend our census to include the LMC, SMC, and Sagittarius, but exclude Antlia~II. We note when running our search algorithms on Antlia~II using DELVE data, we get a detection significance of $\sqrt{\mathrm{TS}_{gr}}=9.2$,  and  SIG$_{gr}=5.5$, despite its location at low Galactic latitude.

In summary, by combining data from DES Y6, DELVE DR3, and PS1 DR1, we recover \CHECK{49} of the \CHECK{62} known Milky Way satellites within our census footprint, representing the largest uniformly selected sample of Milky Way dwarf satellites obtained to date. If we additionally assume that all bright satellites ($M_V < -12.5$) have been discovered, we include three further systems, raising the total to \CHECK{52} out of \CHECK{66} known satellites.  Through the design of the detection thresholds, this analysis did not identify any new candidates that were not already previously identified as Milky Way satellite. We leave the investigation of these less prominent Milky Way satellite candidates to future work.

\section{Census Detection Efficiency}
\label{sec:det_efficiency}
In this section, we describe how we estimate the detection efficiency of our census (i.e., the detection probability of a satellite as a function of its physical properties and location), which allows us to infer the total Milky Way satellite population based on the subset of satellites recovered in our census. We first simulate Milky Way satellites with a wide range of properties and inject them at the catalog-level into our survey data  (Section~\ref{sec:sims}). We then run the same search methods that we applied to the real data (Section \ref{sec:methods}) to determine our efficiency for recovering simulated satellites as a function of their physical properties. We express our detection efficiency using  two different methods: 1) a simple analytic approximation based on the $50\%$ detection contour as a function of the satellite properties
 (Section~\ref{sec:selfunc_analytical}) and 2) a machine-learning-based classifier (Section~\ref{sec:selfunc_ML}). We note that the detection efficiency estimation in this analysis is performed only for the DECam-based DES Y6 and DELVE surveys. For the PS1 region, we adopt the detection efficiency from \citet{MWCensus1}.

\subsection{Satellite Simulations}
\label{sec:sims}

To simulate Milky Way satellites for our census, we generate mock catalogs of their member stars. We randomly sample a wide range of Milky Way satellite sizes, distances, stellar masses, and photometric properties as shown in Table \ref{table:galsims_params}. The simulated population is intended to cover a wide range of the possible parameter space for Milky Way satellites to determine changes in detection efficiency and it is not intended to represent the actual satellite distribution.

\begin{deluxetable}{lccc}
\tabletypesize{\scriptsize}
\tablewidth{0pt} 
%\tablenum{1}
\tablecaption{The parameter range of the simulated satellite used in estimating the census satellite selection function.\label{table:galsims_params} }
\tablehead{
Parameters & Range &  Unit & Sampling  } 
\startdata 
%%%%%%%%%%%%%%%%%%%%%%%%%%%%%%%%%%%
Stellar Mass & $[10,10^6]$  &$ M_\odot $&  log \\ 
Heliocentric Distance &[10,1000] & kpc & log \\
2D half-light radius & [1,2000] & pc & log\\
Ellipticity & [0.1,0.8] & - & linear \\
Position Angle & [0,180] & deg & linear \\
Age & \{10,12,13.5\}& Gyrs& discrete \\
Metallicity & \{0.0001, 0.0002\}& - & discrete \\
%%%%%%%%%%%%%%%%%%%%%%%%%%%%%%%%%%%%%%%%%%%%%%%
\enddata
\end{deluxetable}

\begin{figure*}[ht]
\centering
\includegraphics[width=0.9\linewidth]{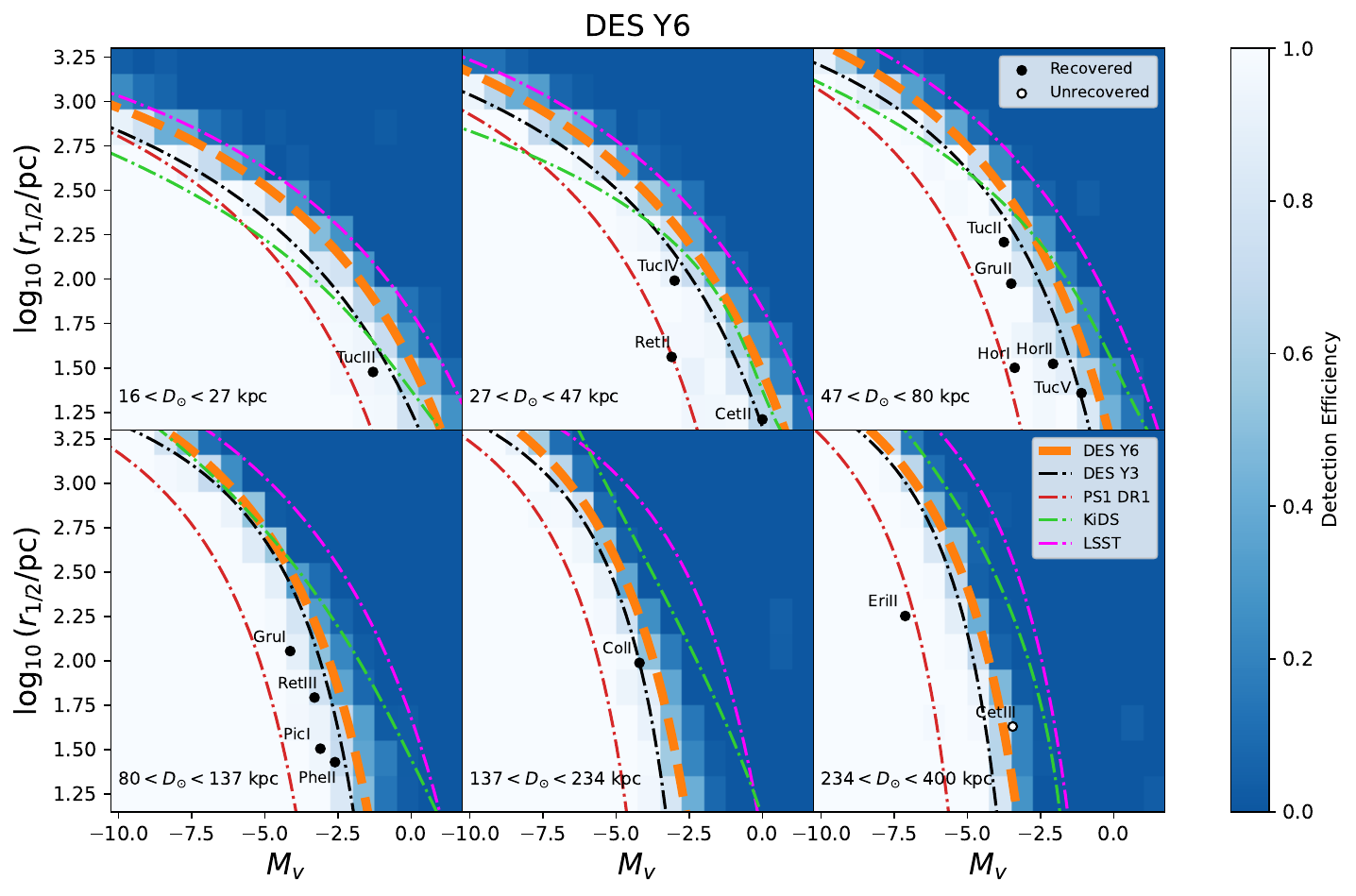}
%\vspace{-3cm}
\caption{{Detection efficiency of our census in the DES Y6 region}. The detection efficiency of a satellite is given as a function of its absolute magnitude, $M_V$, size, $r_{1/2}$ and its heliocentric distance, $D_\odot$, and ranges from 0 (always undetectable) to 1 (always detectable). 
The orange dashed line shows the 50\% detectability contour for the DES Y6 survey, while the black and red dash-dotted lines show the  50\% detectability contours for the DES Y3 and PS1 surveys from \citet{MWCensus1}. The green  line indicates the 50\% detectability limit for the Kilo-Degree Survey (KiDS), based on catalog-level simulations \citep{Zhang:2025}, while the purple line shows the corresponding limit for simulated LSST data, assuming perfect star/galaxy classification \citep{Tsiane:2025}. {We also overlay known satellites within the DES region, with filled circles indicating systems that pass our detection threshold, and open circles for  systems that do not. }\label{fig:des_effiencency}}
\end{figure*}

\begin{figure*}[ht]
\centering
\includegraphics[width=0.9\linewidth]{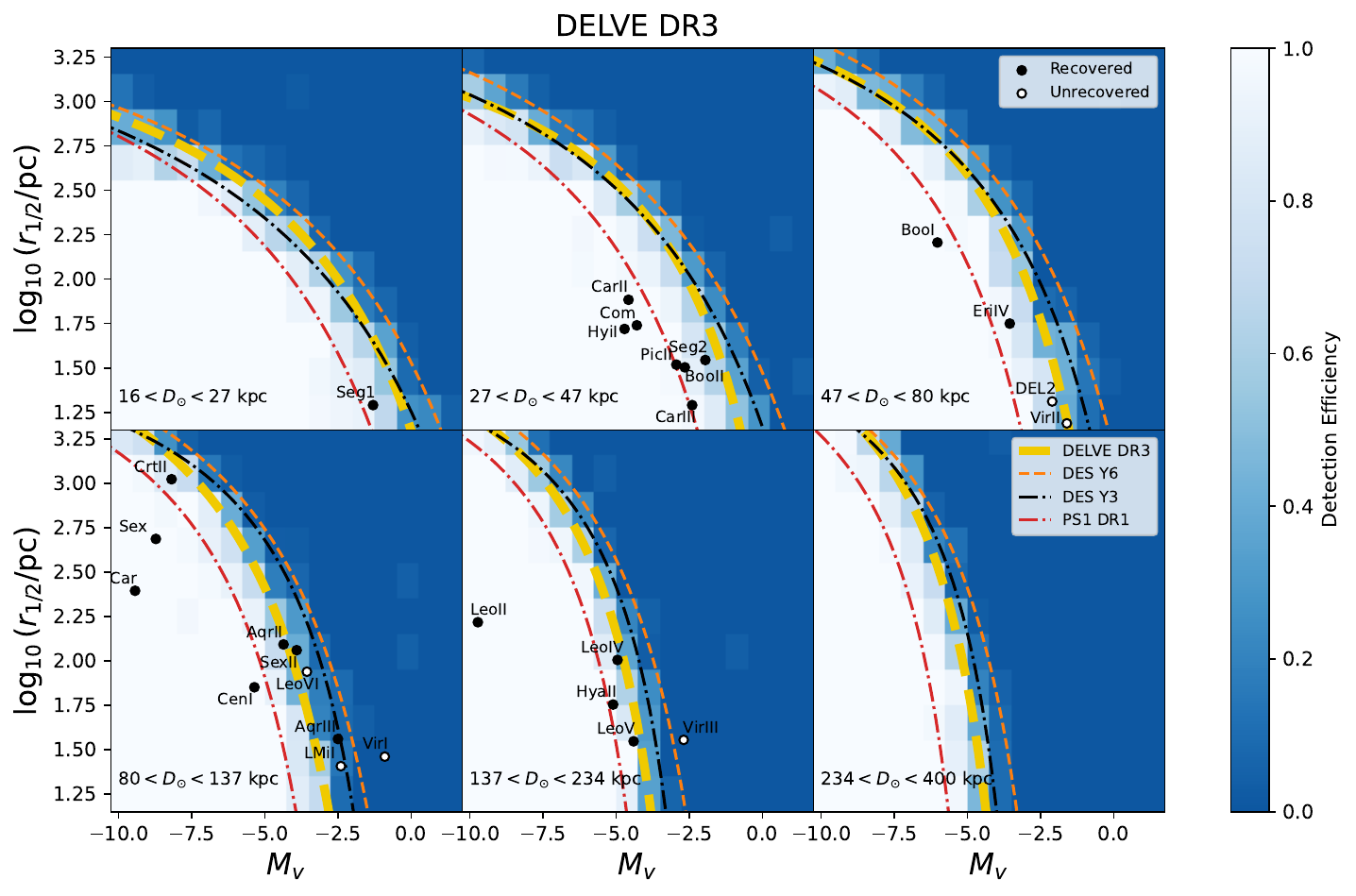}
%\vspace{-3cm}
\caption{{Detection efficiency of our census in the DELVE DR3 region}. The
yellow dashed line shows the 50\% detectability contour for the DELVE DR3 survey. As in Figure \ref{fig:des_effiencency}, the orange dashed line represents the 50\% detectability contours for DES Y6 survey, while black and red dash-dotted lines represent the contours for the DES Y3 and PS1 surveys, respectively. \cytwo{We also overlay known satellites within the DELVE region, with filled circles indicating systems that pass our detection threshold, and open circles for  systems that do not. }\label{fig:delve_effiencency} }
\end{figure*}

The photometry of the systems was simulated from \texttt{PARSEC} isochrones over a range of ages and metallicities (Table \ref{table:galsims_params}). We populate the member stars by sampling from the \citet{Chabrier:2001} initial mass function, with the lower mass bound set at 0.08 $M_\odot$ (the hydrogen-burning limit). Using the \texttt{PARSEC} isochrones, we convert the  stellar mass into absolute magnitudes in DECam $g$, $r$ and $i$ bands. We then obtain the apparent magnitudes by adding the distance modulus of the simulated satellite and apply interstellar extinction using the same maps described in Section \ref{sec:data}. For a simulated star with the magnitude $m$,  we  apply a magnitude uncertainty, $\sigma_m$, based on 
\begin{equation}
    \sigma_m = 0.01 + e^{\gamma_0(m_{\rm lim}-m)+\gamma_1}  + \gamma_2 ,
\end{equation}
where $m_{\rm lim}$ is the depth of the survey at the location of the star and $\gamma_0, \gamma_1, \gamma_2$ are constants determined from the survey data.\footnote{For both DES Y6 and DELVE DR3, $\gamma_0, \gamma_1, \gamma_2$ are given by $[-0.9133, -2.0625, 0.0003]$, $[-0.9127, -2.1588, 0.0005]$, and $[-0.9044, -2.2738, 0.0005]$ for $g$, $r$, and $i$ band, respectively.} \cytwo{To incorporate these uncertainties into the magnitude measurements, we resample the flux using a Gaussian distribution with a standard deviation set by $\sigma_m$, and then convert the flux back to magnitudes.}  We also use the \texttt{PARSEC} isochrones to obtain the absolute magnitudes of the satellites from the stellar mass.

We sample the spatial distribution of the member stars independently from their photometry, using a  2D elliptical Plummer profile parameterized by the 2D half-light radius along the semi-major axis, $a_{1/2}$, its ellipticity, $\epsilon$, and its position angle, P.A. \footnote{The azimuthally averaged half-light radius, $r_{1/2}$, reported in Table~\ref{table:all_dwarfs} and used throughout this paper, is related to the semi-major axis of an ellipse containing half of the light, $a_{1/2}$, by $r_{1/2} = a_{1/2} \sqrt{1-\epsilon}$.}. \cyfive{We draw these structural parameters from the ranges listed in Table~\ref{table:galsims_params}}. As discussed in Section \ref{sec:data}, our survey catalogs are incomplete at the faint end due to star/galaxy classification inefficiency and detection incompleteness. Thus, our analysis does not include all the stars present in the satellites. To account for this, we applied the stellar completeness functions for each survey (Figure~\ref{fig:completeness}) to probabilistically select a fraction of stars to inject into the data based on their magnitudes.

We injected $10^5$ simulated Milky Way satellites into each of the DES Y6 and DELVE DR3 \cyfive{at the catalog-level}, and we ran the detection algorithms described in Section~\ref{sec:methods} on both datasets.  When attempting to recover simulated satellites, we made several modifications to the search pipeline to reduce computational time. For example, instead of running our search over the entire survey, we fixed the spatial location and distance modulus to the search grid values closest to the true position and distance of the simulated satellite. \citet{MWCensus1} found that freeing search parameters only changed the detection significance by at most a few percent. Additionally, we assume that bright simulated satellites that have $>1000$ stars detected with $g \leq 22$ and surface brightnesses of $\mu <23.5$\,mag/arcsec$^2$ are always detected. We then record the \texttt{ugali} and \texttt{simple} detection significance for each satellite, along with whether it would have passed the detection threshold used in our census.

%Assumtions of peak finder 

\subsection{Selection Function: Detectability Contour}
\label{sec:selfunc_analytical}

\begin{deluxetable*}{|c | c | c c  c | c c c  |}
\tabletypesize{\footnotesize}
\tablecaption{\cyfive{Best-fit parameters of the 50\% satellite detectability contour for both the DES Y6 and DELVE DR3 surveys (see Equation~\ref{table:p50fit}). Distances are given in kiloparsecs, and the geometric mean is used to represent the average distance of each distance bin. } \label{table:p50fit}}
\tablehead{ 
Distance &  Distance Bins &  $A_{0, {\rm DES}}$ & $M_{V,0,{\rm DES}}$ & $\log_{10}\left(r_{1/2, 0}\right)_{\rm DES}$  & $A_{0, {\rm DELVE}}$ & $M_{V,0,{\rm DELVE}}$ & $\log_{10}\left(r_{1/2, 0}\right)_{\rm DELVE}$  }
\startdata
20.8 &  [16, 27] & 20.3 & 7.9 & 4.1 & 15.4 & 5.6 & 3.9 \\ 
35.6 &  [27, 47] & 23.9 & 7.9 & 4.5 & 10.2 &  3.1  &  3.8  \\ 
61.3 &  [47, 80] & 17.7 & 5.1 &  4.5  & 10.7 & 2.1 & 4.1 \\ 
104.7 &  [80, 137] & 13.0 & 2.4 & 4.5  & 12.1 & 0.8 & 4.5 \\ 
179.0 &  [137, 234] & 10.8 & 0.6 & 4.5 & 7.9  & -1.3 & 4.3 \\ 
305.9 &  [224, 400] & 9.4  &  -0.5 &  4.5 & 7.9 & -2.0 & 4.5 \\ 
\enddata
\end{deluxetable*}

\cyfive{Following \citet{Koposov:2008}}, \citet{Walsh:2009} and \citet{MWCensus1}, we bin our simulated satellites based on their heliocentric distance, $D_\odot$, absolute magnitude, $M_V$, and  half-light radius, $r_{1/2}$, and show the average probability of detection for each bin in parameter space for DES Y6 (Figure \ref{fig:des_effiencency}) and DELVE DR3 (Figure \ref{fig:delve_effiencency}). The binning is performed across the entire DES and DELVE footprints, averaging the detection efficiency over regions with varying  depths.

Similar to what was found by \citet{Koposov:2008}, for a fixed distance and size, we find that there is a sharp dropoff in detection efficiency below a certain absolute $V$-band magnitude, $M_V$. Therefore, to quantify the detection efficiency of the census, it is useful to define a contour to delineate the parameters of satellites with 50\% detection efficiency, $P_{\rm det}(D_\odot, M_V , r_{1/2}) = 0.5$. For a fixed heliocentric distance, $D_\odot$, we parameterize the 50\% detection probability contour in the $r_{1/2}$ vs.\ $M_V$ parameter space with

\begin{align}
  \cyfiveeqn{ \log_{10}\left(\frac{r_{1/2}}{\rm{1~pc}}\right)= \frac{A_0}{M_V-M_{V,0} } + \log_{10}\left(r_{1/2, 0}\right)}
  %  M_V =& A \log_{10}\left(\frac{r_{1/2}}{\rm{1~pc}}\right)^3 + B \log_{10}\left(\frac{r_{1/2}}{\rm{1~pc}}\right)^2 +\\ \nonumber
   % &C \log_{10}\left(\frac{r_{1/2}}{\rm{1~pc}}\right) + E,
\end{align}
\cyfive{where $A_0$, $M_{V,0}$ and $\log_{10}\left(r_{1/2, 0}\right)$} are distance- and survey-dependent constants that we fit to the binned data. Table \ref{table:p50fit} shows the best-fit values for \cyfive{$A_0$, $M_{V,0}$ and $\log_{10}\left(r_{1/2, 0}\right)$} for both DES Y6 and DELVE DR3.

We also overlay the best-fit 50\% detectability contour for the DES Y6 data as a dashed orange line in Figure~\ref{fig:des_effiencency}, and for the DELVE DR3 data as a dashed yellow line in Figure~\ref{fig:delve_effiencency}. For compassion, we also overlay the 50\% detectability contours for DES Y3 and PS1 DR1 from \citet{MWCensus1}. We find that the deeper DES Y6 survey has a higher detection efficiency than the DES Y3 survey, while our DELVE DR3 search has a detection efficiency that lies between the DES Y3 and PS1 DR1 detection. While the DELVE DR3 survey has a similar depth to DES Y3 ($g\sim24.3$), we applied a higher detection threshold to reduce the rate of false positives caused by survey inhomogeneities---i.e., ($\sqrt{\mathrm{TS}_{gr}}>8.0$,  $\sqrt{\mathrm{TS}_{gi}}>7.5$ and ${\rm SIG}_{gr}>6.5$) for DELVE DR3 vs.\ ($\sqrt{\mathrm{TS}_{gr}}>6.0$ and  ${\rm SIG}_{gr}>6.0$) for DES Y3. As a result, the detection efficiency of DELVE DR3 is reduced compared to DES Y3. If we instead use the DES Y3 detection threshold for DELVE DR3, we find very similar 50\% detectability contour as DES Y3. We note that with this lower threshold, we would be able to recover four additional satellites within the DELVE footprint (DELVE~2, Leo~VI, Leo Minor~I, and Virgo~II).  

Additionally, in Figure~\ref{fig:des_effiencency}, we overlay the 50\% detectability contour for the  Kilo-Degree Survey (with a $5\sigma$ limiting magnitude of $g \sim 24.96$$^{\ref{sn_conversion}}$;  We note that the 50\% detectability contour reported by \citealt{Zhang:2025} was derived using different satellite detection thresholds that placed less emphasis on sample purity. We also overlay the forcasted sensitivity of the upcoming Vera C.\ Rubin Observatory’s Legacy Survey of Space and Time (LSST; \citealt{Ivezic:2019}), based on the LSST Dark Energy Survey Data Challenge 2 \citep{2021ApJS..253...31L}, which has a $5\sigma$ limiting magnitude of $g \sim 27.0$  \citep{Tsiane:2025}.$^{\ref{sn_conversion}}$ \cyfive{For the 10-year LSST forecast, we assume perfect star–galaxy classification, motivated by the expectation that star–galaxy separation methods will have improved substantially relative to current implementations, with approaches that combine ground-based data and space-based imaging being particularly promising. In addition, alternative detection algorithms such as \texttt{ugali} have been shown to be less sensitive to galaxy contamination than the \texttt{simple} algorithms adopted in \citet{Tsiane:2025}, further reducing the impact of contamination.}

As noted in \citet{Zhang:2025}, our catalog-level injections may yield slightly optimistic results, whereas the more realistic image-level injections could lead to lower detection efficiencies, particularly for compact objects where blending effects become significant.
%However, in the range of satellite sizes that is relevant for our Milky Way satellit galaxy census ($r_{1/2} > 15$\,pc), \citet{Zhang:2025} showed that the correction to the sensitivity was ``small''. 
However, the largest deviations seen in \citet{Zhang:2025} were found for very compact satellites, which are not considered in our census of Milky Way satellite galaxies. \cyfive{As shown in Appendix \ref{appendix:balrog}, where we perform a limited set of image-level simulations, we  find that blending does not significantly affect the simulated satellites considered in our analysis, for which we impose a size and distance limit ($r_h \geq 15$ pc and $D_\odot \leq 400$ kpc).}

\subsection{Selection Function: Machine-Learning Classifier}
\label{sec:selfunc_ML}

\cytwo{While the 50\% detectability contour is a useful metric to compare detection efficiencies, it does not fully capture the  detectability gradient in the intermediate-efficiency regime where most known satellites reside.} Thus, to better encapsulate the information from our simulation, we trained a gradient-boosted decision tree classifier to predict the detectability of Milky Way satellites based on their properties.

For our classifier, we built a feature vector from the properties of the simulated satellite and the local stellar density around the satellite: $\vec{X} = \{ M_V, r_{1/2}, D_\odot, \rho(g<22)\}$ where $\rho(g<22)$ is an estimate of the density of (foreground) Milky Way stars with $g<22$ in units of arcmin$^{-2}$. We then seek to predict the relationship between $\vec{X}$ and a set of labels, $\vec{Y}$, where $Y_i \in \{0,1\}$ with 1 indicating that the satellite is detected (based on the criteria in Section \ref{sec:det_criteria}) and 0 representing an undetected satellite. For a given $\vec{X} = \{ M_V, r_{1/2}, D_\odot, \rho(g<22)\}$, our classifier outputs the probability that such a satellite would be detected. 

Given the significant variation in depth across DELVE DR3, we also considered including a local estimate of the survey depth (as captured by the 10$\sigma$ limiting magnitude) around the satellite, $m_{lim}$, as an additional feature. However, we found that including the depth in the feature set changes the detection probability by at most a few percent and does not significantly improve the performance of the classifier while increasing its complexity. Thus, we choose not to include this information in our final classifier.

\begin{figure}
    \centering
    \includegraphics[width=1\linewidth]{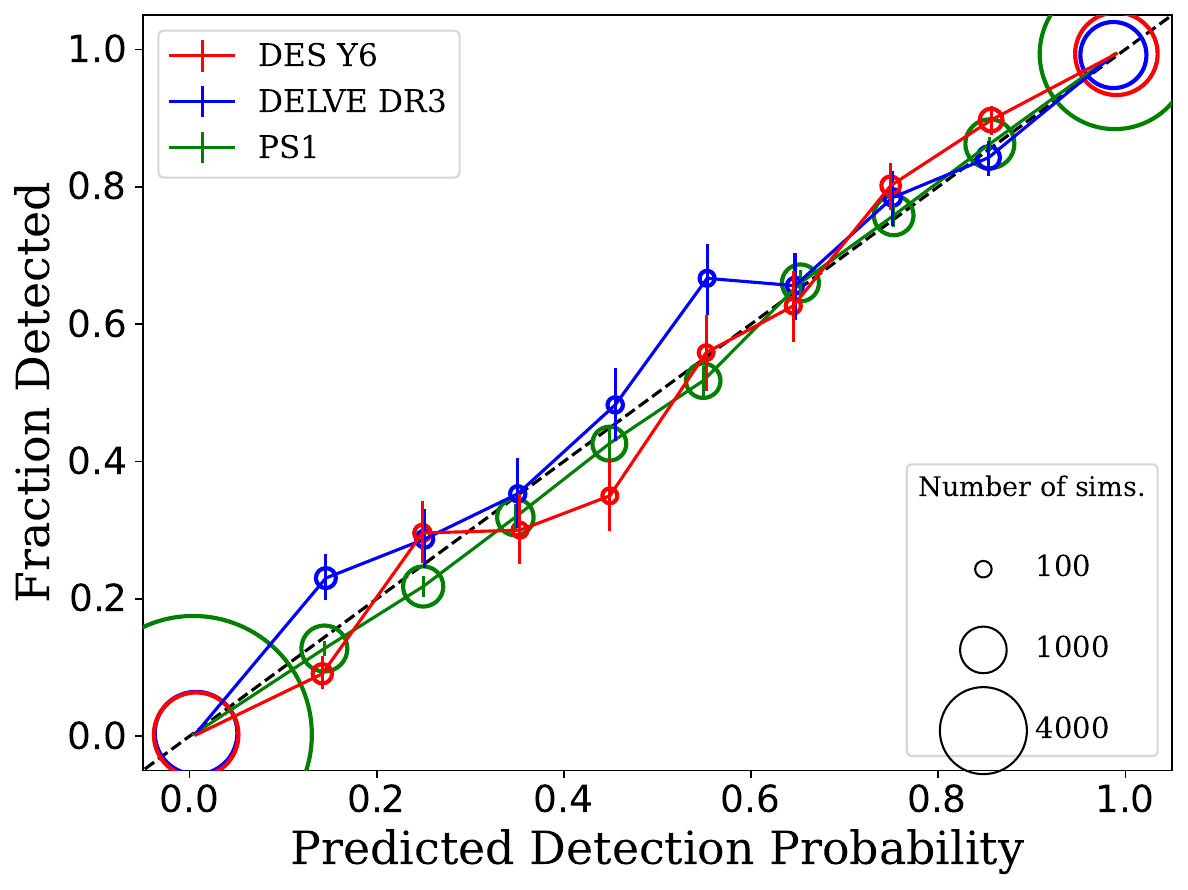}
    \caption{Fraction of simulated satellites that passed our detection criteria vs.\ the detection probability predicted by our \texttt{XGBoost}-based classifier for DES Y6 and DELVE DR3. \cyfive{The size of the circle face markers represent the number of simulated systems} in each predicted detection probability bin, with darker shades indicating bins containing more systems. Most of the simulated systems fall into either the 0\% or 100\% detection fraction bins. The dashed black one-to-one  line represents perfect performance for our classifier. \cyfive{We also include the PS1 performance from \citet{MWCensus1}, which uses more simulations owing to the smaller number of member stars per satellite due to the shallower PS1 imaging}. \label{fig:ml_eval}}
\end{figure}

We first split our sample of simulated satellites into training and test sets, which contain $\sim90\%$ and $\sim10\%$ of the total sample, respectively.  We trained the classifier using \texttt{XGBoost} \citep{Chen:2016} and \texttt{scikit-learn} \citep{Pedregosa:2011} following the procedure described in \citet{MWCensus1}. To evaluate the robustness of our machine learning classification, we run our trained classifier on our test sample. \CHECK{Since the classifier outputs a detection probability between 0 and 1, to create a realization of the detected satellite population, we draw a uniform random number in the interval $[0, 1]$ and classify the system as detected if it is less than the predicted probability.}

We found that, for detected simulated systems, the DES and DELVE classifiers have a true positive rate of {97\%} and {94\%}, respectively. For undetected systems, the DES and DELVE classifiers have a true negative rate of {97\%} and {97\%}, respectively. Since the majority of simulated systems reside in regions of parameter space where they are always detected or undetected (in contrast to the observed population), we also specifically evaluated the robustness of our classifier in the region of intermediate detectability. This is done by binning the simulated satellites based on their predicted detection probabilities and comparing them to the fraction of satellites that actually pass our detection threshold in each bin. As shown in Figure~\ref{fig:ml_eval}, our classifier also accurately predicts the detection probability in intermediate regions.

The Milky Way satellite detectability classifiers for DES Y6 and DELVE DR3 are available publicly.$^{\ref{github_link}}$ They can be used to obtain the detection probability of Milky Way satellites as a function of their physical size, $r_{1/2}$, absolute magnitude, $M_V$, heliocentric distance, $D_\odot$, and sky location (RA, Dec). The sky position is used to retrieve information about the local stellar density, census survey footprint and geometric mask. The detection probabilities can thus be used to predict the number of satellites that would be observed given an underlying satellite population (e.g., from a galaxy formation model or numerical simulations). For the PS1 DR1 regions, we use the trained classifiers from \citet{MWCensus1}, with an updated mask to reflect changes in the PS1 region used in our census.

\section{Estimates of the total Milky Way Satellite Galaxy Population}
\label{sec:mwpop}
%\subsection{tion}

To infer the properties of the underlying Milky Way satellite population, which includes both observed and undetected satellites, we combined the results of our census (i.e., the recovered Milky Way satellites and the observational selection function) with a parametric model describing the luminosity, size, and distance distributions of Milky Way satellites. To facilitate comparison with other studies, we provide the posteriors of our empirical model parameters and the resulting satellite population properties (such as the luminosity function), properly accounting for survey selection effects.$^{\ref{github_link}}$

\begin{deluxetable*}{llccc}
\tabletypesize{\scriptsize}
\tablewidth{0pt} 
%\tablenum{1}
\tablecaption{\cyone{Best-fit values for the empirical satellite model parameters, $\theta$, for the Milky Way compared to those for the M31 satellites estimated by \citet{Doliva-Dolinsky:2023}. \label{table:model_variables}} }
\tablehead{
\colhead{Parameter} & \colhead{Description}& 
\colhead{Equation}& \colhead{MW Value}  & \colhead{M31 Value}  } 
\startdata 
%%%%%%%%%%%%%%%%%%%%%%%%%%%%%%%%%%%
$\beta$& Luminosity Function Power-Law Slope &  \ref{eqn:lum_PL} & $0.16^{+0.02}_{-0.02}$ & $0.20^{+0.04}_{-0.04}$  \\ 
$r_s$& Scale Radius of the  cored-NFW radial profile (kpc) & \ref{eqn:NFW} & $21^{+9}_{-6}$ & ---$^*$  \\ 
$z_{S-L}$&  Median $\log r_{1/2}$ value at $M_V=-6$ for the size-luminosity function & \ref{eqn:size_median} &$2.07^{+0.04}_{-0.04}$ & $2.5^{+0.2}_{-0.1}$  \\ 
$m_{S-L}$&  Slope of the size-luminosity function & \ref{eqn:size_median}  &$-0.12^{+0.01}_{-0.01}$ &  $-0.05^{+0.03}_{-0.02}$  \\ 
$\sigma_{S-L}$& Gaussian scatter in the size-luminosity function&  \ref{eqn:size_scatter}  & $0.24^{+0.03}_{-0.03}$  & $0.33^{+0.06}_{-0.06}$ \\ 
$N$&  Total number of satellites ($-20\leq M_V\leq0$,  $15\leq r_{1/2} (\rm pc) \leq 3000$, $10\leq D_{\rm GC} (\rm kpc)\leq 300$) & \ref{eqn:Ntrue} & $265^{+79}_{-47}$ & ---$^{\dagger} $  \\
%%%%%%%%%%%%%%%%%%%%%%%%%%%%%%%%%%%%%%%%%%%%%%%
\enddata
\tablenotetext{*}{Radial distribution modeled using a 3D power-law profile instead of cored-NFW.}
\tablenotetext{\dagger}{\cytwo{\citet{Doliva-Dolinsky:2023}  estimated $92^{+19}_{-26}$ systems with $-17.5\leq M_V\leq-5.5$,} compared to our Milky Way prediction of $29^{+9}_{-7}$ within the same luminosity, distance, and size ranges.}
\end{deluxetable*}

\subsection{The Empirical Model }
\label{sec:mw_empirical_model}
We follow the probabilistic model developed by \citet{Doliva-Dolinsky:2023} to describe the satellite population of M31. This model assumes that the global distribution of the physical and spatial properties of satellites can be modeled using a combination of simple analytic functions that depend on a set of model parameters, $\theta$. The model consists of three independent components: (1) the luminosity function, $P_{LF}(M_V|\theta)$, (2)~the radial distribution of the satellites, $P_{\rm sp}(D_{GC}| \theta)$, and (3)~the size--luminosity relation, $P_{r}(r_{1/2}|M_v, \theta)$ . The model also assumes a spherically symmetric distribution of satellite galaxies (though see Section~\ref{sec:anisotropy} for discussion of anisotropy). Thus, the probability of sampling a galaxy with the following parameters $\mathcal{D} = \{M_V, r_{1/2}, D_{GC}\}$ from the overall galaxy distribution is given by:
\begin{equation}
    P(\mathcal{D}|\theta)  \propto  P_{LF}(M_V|\theta) P_{\rm sp}(D_{GC}| \theta) P_{r}(r_{1/2}|M_v, \theta).
\end{equation}

Following \citet{Tollerud:2008} and \citet{Doliva-Dolinsky:2023}, we model the luminosity function (i.e., the probability that a galaxy has a given absolute magnitude) as a power law with an exponent $\beta$:
\begin{equation}
\label{eqn:lum_PL}
     P_{LF}(M_V|\beta) \propto10^{-\beta M_V},
\end{equation}
over a magnitude range of $-20\leq M_V\leq0$. \cythree{To incorporate the brightest satellites into our empirical model, we assume that all galaxies with $M_V < -12.5$ are always detected and adopt a 52-galaxy sample that includes the LMC, SMC, and Sagittarius.}

For the radial distribution of Milky Way satellites (i.e., the probability that a galaxy has a given Galactocentric distance), we only consider galaxies with Galactocentric distance in the range of 10\,kpc $\leq D_{GC}\leq$  300\,kpc, and thus exclude the distant dwarf satellite Eridanus II.  We further assume that the distribution follows a spherically symmetric cored Navarro–Frenk–White (NFW) profile \citep{Zhao:1996, Navarro:1997}:
\begin{equation}  
\begin{aligned}
\label{eqn:NFW}
        P_{\rm sp}(D_{GC} | \alpha) &\propto  \int ^{2\pi}_0 \int^\pi_{0} \frac{1}{(D_{GC}+r_s)^3} D_{GC}^2\sin\theta ~ d\theta d\phi \\
         &\propto  \frac{D_{GC}^2}{(D_{GC}+r_s)^3}.
\end{aligned}
\end{equation}
where $r_s$ is the scale radius of the distribution. 
We also tested a generalized NFW  profile, $\rho(D_{GC}) = D_{GC}^{-\gamma}(D_{GC}+r_s)^{\gamma-3}$, but found that the posterior distribution favors a cored NFW profile, corresponding to $\gamma=0$. Following \citet{Doliva-Dolinsky:2023}, we also tested a power-law profile for the radial distribution, but we found a significantly lower likelihood at $\ln(\mathcal{L}_{PL}/\mathcal{L}_{cNFW}) \sim 4.5$.

We model the size--luminosity relation (i.e., probability of a galaxy with a fixed luminosity having a given size) following \citet{Shen:2003} and \citet{Brasseur:2011} by assuming that the mean half-light radius $\langle\log(r_{1/2})\rangle$ follows a linear relation with absolute magnitude, $M_V$:
\begin{equation}
\label{eqn:size_median}
     \langle \log r_{1/2} \rangle (M_V, z_{S-L},m_{S-L}) = z_{S-L} + m_{S-L}(M_V +6.0),
\end{equation}
where $z_{S-L}$ and $m_{S-L}$ are free parameters that sets the offset and slope of the size--luminosity relation. We then assume that the size distribution of satellite galaxies, $\log (r_{1/2})$, follows a Gaussian distribution around the mean, $\langle \log r_{1/2} \rangle$, with scatter, $\sigma_{S-L}$, such that:
\begin{equation}
\label{eqn:size_scatter}
     P_{r}(r_{1/2}|\theta) \propto  \frac{1}{\sqrt{2\pi}\sigma_{S-L}} \exp \left( - \frac{1}{2} \left( \frac{\log r_{1/2} - \langle \log r_{1/2} \rangle}{\sigma_{S-L}} \right)^2 \right).
\end{equation}Here we only consider galaxies with physical sizes of $15\,{\rm pc} \leq r_{1/2} \leq 3000\,{\rm pc}$.

\begin{figure*}
    \centering
    \includegraphics[width=0.95\linewidth]{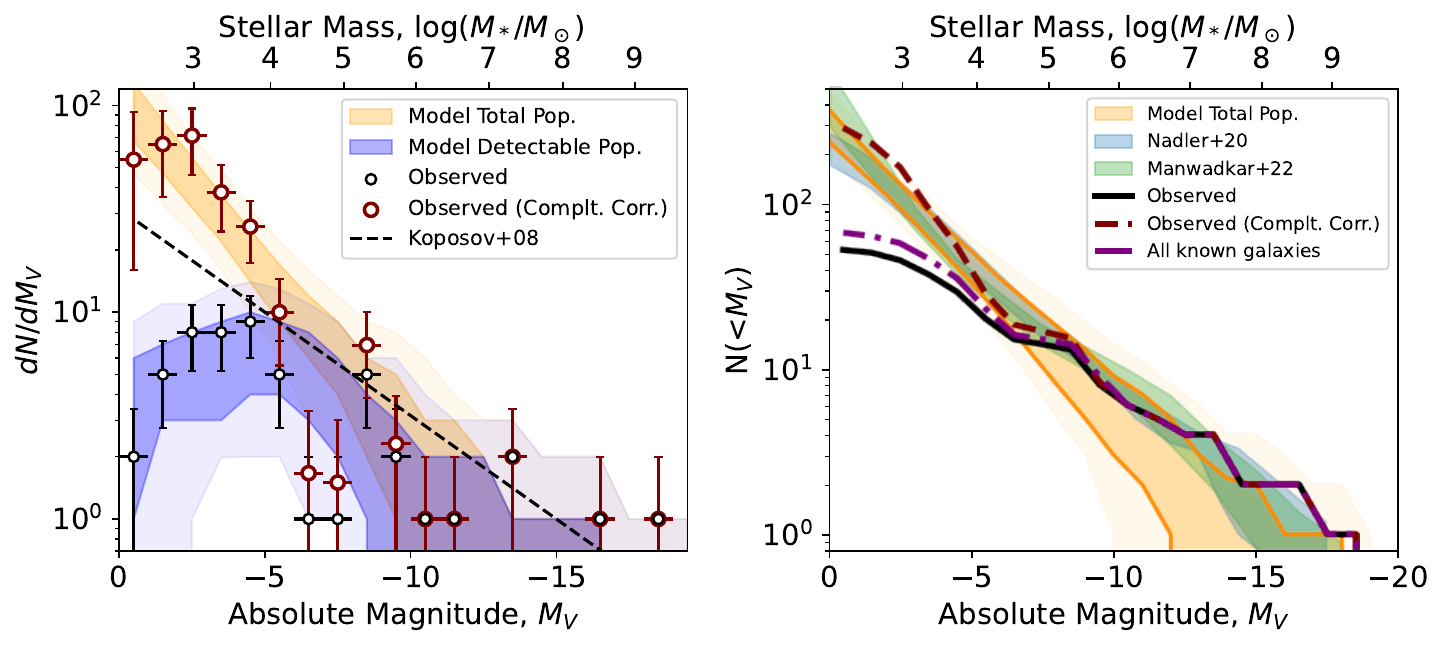}
    \caption{The inferred luminosity function of the total Milky Way satellite population from our empirical model (yellow). The dark (light)  bands correspond to 68\% (95\%) credible intervals. Left: Differential number of satellites within $M_V$ bins. The purple region shows the luminosity function from the predicted  detectable galaxies  in our census. The black points show the galaxies recovered in our census while the maroon points shows the completeness corrected luminosity function. \cytwo{Each magnitude bin has a width of $\Delta M_V = 1$ wide.} Right: Cumulative number of satellites. The 
    black solid line represents galaxies recovered in our census while the dashed maroon line shows the median completeness-corrected luminosity function. The purple dot-dashed line shows the luminosity function of all known galaxies, including systems outside the census footprint. For comparison, we also include predictions from \citet{MWCensus2}  and \citet{Manwadkar:2022} with 68\% credible intervals.}  
    \label{fig:lum_func}
\end{figure*}

To incorporate information about the total number of galaxies around the Milky Way, we then normalize $P(\mathcal{D}\,|\,\theta)$ such that 
\begin{equation}
\label{eqn:Ntrue}
   P_{\rm total}(\mathcal{D}|\theta)  =  \frac{N_{\rm True}}{P_0} P(\mathcal{D}|\theta), 
\end{equation}
where $N_{\rm True}$ is the total number of galaxies and $P_0$ is the normalization constant, $P_0 = \int P(\mathcal{D}|\theta) d\mathcal{D}$. Thus, our model has a total of 6 free parameters, $\theta$, \cyone{which we summarize in Table \ref{table:model_variables}. }

By combining $P_{\rm total}(\mathcal{D}|\theta)$ with the galaxy selection function from Section \ref{sec:det_efficiency}, we are able to obtain the number of observed satellite galaxies predicted by each realization of the model parameters via:
\begin{equation}
    N_{\rm {pred}}  = \int  \tau(\mathcal{D})P_{\rm total}(\mathcal{D}|\theta) d\mathcal{D},
\end{equation}
where $\tau(\mathcal{D})$ is the probability of detecting the galaxy in our census based on its properties. 

\cytwo{To evaluate the likelihood of the observed data, $\mathcal{D_{\rm obs}}= \{M_{V, obs},r_{1/2, obs},D_{GC, obs}\}$, given a set of model parameters, $\theta$, we adopt the unbinned Poisson likelihood formalism (e.g., see Appendix C of \citealt{MWCensus1} and \citealt{Doliva-Dolinsky:2023}). The log-likelihood is given by}
\begin{align}
\label{eqn:unbinned-like}
    \log(\mathcal{L}(\mathcal{D}_{\rm obs}|\theta)) =&  - \int  \tau(\mathcal{D})P_{\rm total}(\mathcal{D}|\theta)  d\mathcal{D}  \nonumber \\
    &+ \sum_{i\in G_{\rm obs }} \log (\tau(\mathcal{D}_i)P_{\rm total}(\mathcal{D}_i|\theta)) \\
    &+ \rm{const,}  \nonumber
\end{align}
where for the first term, which corresponds to $N_{\rm pred}$, we integrate the $\tau(\mathcal{D})P_{\rm total}(\mathcal{D}|\theta)$ over the entire $\{r_{1/2},D_{GC},M_V\}$ parameter space, while for the second term, we evaluate the sum of $\log(\tau(\mathcal{D})P_{\rm total}(\mathcal{D}|\theta))$ over the parameters of recovered galaxies in our census, $G_{\rm obs }$.

We sampled the likelihood from Equation \ref{eqn:unbinned-like} with the Markov Chain Monte Carlo (MCMC) method using \texttt{emcee} \citep{Foreman_Mackey:2013}. We assume uniform \cythree{linear} priors on all model parameters, $\theta$, but imposed that the scatter on satellite galaxy sizes was positive, $\sigma>0$. We note that when evaluating the parameters of our model, we do not account for the uncertainties on the measured parameters of the known galaxies in our census. \cytwo{However, we repeat our analysis randomizing the properties of the observed galaxies in accordance with their measurement uncertainties, and we find that it makes a minimal difference in our results (Appendix \ref{appendix:uncertainties})}.

\subsection{Inferred properties of the Milky Way satellites}
We find that the posterior distribution of the parameters from our empirical model are all well constrained by the limited data from our census and show their best-fit values in Table \ref{table:model_variables} (see also Figure \ref{fig:mcmc_chains} in the Appendix \ref{appendix:uncertainties}). \cyone{To estimate the values of parameters and their uncertainties, we use the peak of the posterior (obtained through a kernel density estimation) and the highest density region
containing 68\% of the posterior,  respectively.}

\cyone{In the left panel of Figure \ref{fig:lum_func}, we present the predicted luminosity function of the total Milky Way satellite population, assuming our empirical model and using parameter values sampled from the posterior distribution. We then apply the detection efficiency and geometric mask from our analysis to predict the detectable satellite population, which can be directly compared against the observed population in our census.}
\cyone{For reference, we also show the binned luminosity function of the observed satellites recovered in our census, along with the luminosity function that has been completeness-corrected. To perform the completeness correction, we generate a synthetic galaxy sample using our empirical model with parameters drawn from the posterior. For each magnitude bin, we then compute the ratio of detectable to total satellites in this sample and apply this ratio to the observed population to estimate the intrinsic luminosity function of the Milky Way satellites.}

\begin{figure}
    \centering
    \includegraphics[width=1\linewidth]{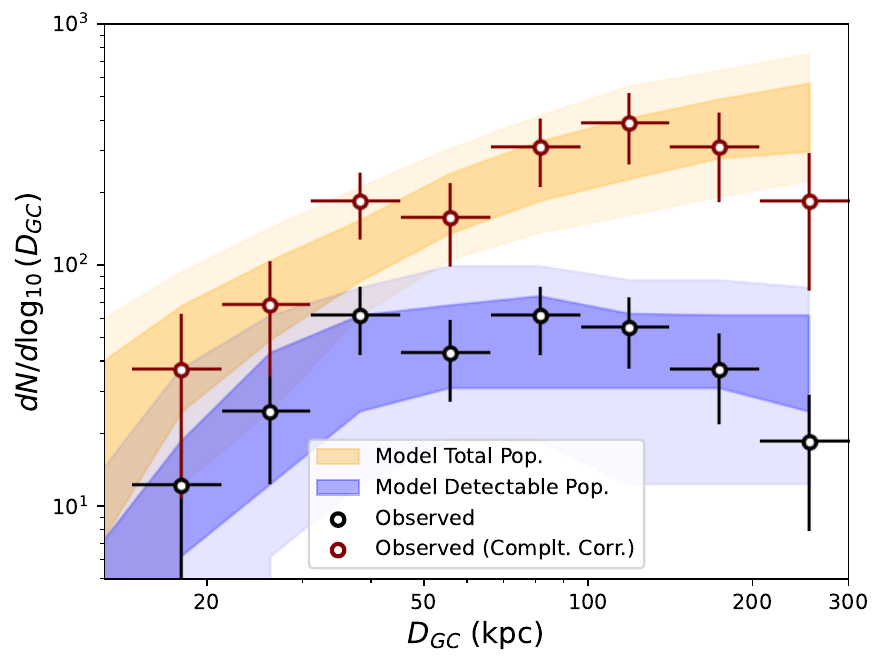}
    \caption{\cyone{{The inferred radial distribution of the total and detectable  Milky Way satellite galaxy population.} \cytwo{The dark (light)  bands correspond to 68\% (95\%) credible intervals}. The black circles show the radial distribution of the  galaxies recovered in our census while the maroon points shows the completeness corrected version. \cytwo{Each distance bin has a width of $\Delta\log_{10}(D_{GC}/{\rm kpc}) \sim 0.164$ wide.} \label{fig:radial_dist}}}
\end{figure}

We generally find good agreement between our empirical model and the observed data. However, not all features in the observed luminosity function are captured by our simplistic power-law-based model. In particular, we find that our steep luminosity function under-predicts the number of MC-size systems. The association of the Milky Way with two massive MCs has long been recognized as unusual in both data \citep[e.g.,][]{Lorrimer:1994, James:2011, Liu:2011} and simulations \citep[e.g.,][]{Boylan-Kolchin:2010, Buscha:2011, 2020MNRAS.497.4311E}. More recently, \citet{Mao:2024} have noted that SAGA galaxies with satellite abundances and masses similar to the Milky Way typically lack satellites as massive as the LMC. \cytwo{They suggest that the Milky Way may be an older, less massive host which experienced a rare, recent LMC/SMC accretion event, resulting in a larger number of very bright satellites than is typical for such galaxies.}  We also observe a deficit of galaxies with absolute magnitude $M_V \sim -7$ compared to the predictions of our empirical model. This feature has been noted by \citet{Bose:2018}, who attribute it to a bimodality in the satellite population driven by the effects of reionization. \cyfour{Furthermore, we see indications of a downturn in the lowest completeness-corrected luminosity bin. It is unclear whether this feature is statistically significant. However, such an effect could be the result of selection effects, such as the 15\,pc physical size cut removing some compact systems that are actually dwarf galaxies. } We note that the empirical model used in this analysis is primarily intended to provide a data-driven estimate of the total number of Milky Way satellites, and we plan to explore more physically motivated models in future work.

% We find that the total luminosity function derived from our Milky Way model is significantly steeper than that derived from SDSS data by \citet{Koposov:2008}, predicting a larger number of satellites at faint magnitudes ($M_V \gtrsim -6$). However, our results are found to be in agreement with more recent estimates of the luminosity function from \citet{Nadler:2019b} and \citet{Manwadkar:2022}.

\cyone{Our models predict that the total number of satellite galaxies around the Milky Way with $-20\leq M_V\leq0$, $15\leq r_{1/2} (\rm pc) \leq 3000$, and $10\leq D_{GC} (\rm kpc)\leq 300$ is $265_{-47}^{+79}$. If we expand the parameter space to include more compact galaxies with $10\leq r_{1/2} (\rm pc) \leq 3000$, our predicted number of galaxies increases to $280_{-50}^{+83}$. 
Compared to other estimates for the Milky Way satellite population, we find that the total luminosity function derived from our Milky Way model is steeper than that derived from SDSS data by \citet{Koposov:2008} and predicts a larger number of faint satellites with $M_V \gtrsim -6$ (left panel of Figure~\ref{fig:lum_func}).
However, our estimates are consistent with the $\sim$270 satellites predicted by \citet{MWCensus1}, the $220\pm50$ satellites predicted by \citet{MWCensus2} and the $440^{+201}_{-147}$ predicted by \citet{Manwadkar:2022}, all of which used the similar parameter space of $ D (\rm kpc)\leq 300$ and  $M_V \leq$  0 (see the right panel of Figure~\ref{fig:lum_func}). However, our result is somewhat higher than the estimate of $124^{+40}_{-27}$ satellites brighter than $M_V$ = 0 from \citet{Newton:2018}, though our estimates are consistent with \citet{Jethwa:2018}, who predict 178--235  satellites with $M_V < -1$ compared to our 159--239 ($\pm 1\sigma$ intervals). \cyfive{We note that our higher predicted number of galaxies compared to \citet{Newton:2018} is likely due to the fact that \citet{Newton:2018} excluded LMC satellites from their analysis, whereas our analysis includes them. } }

\begin{figure}
    \centering
    \includegraphics[width=1\linewidth]{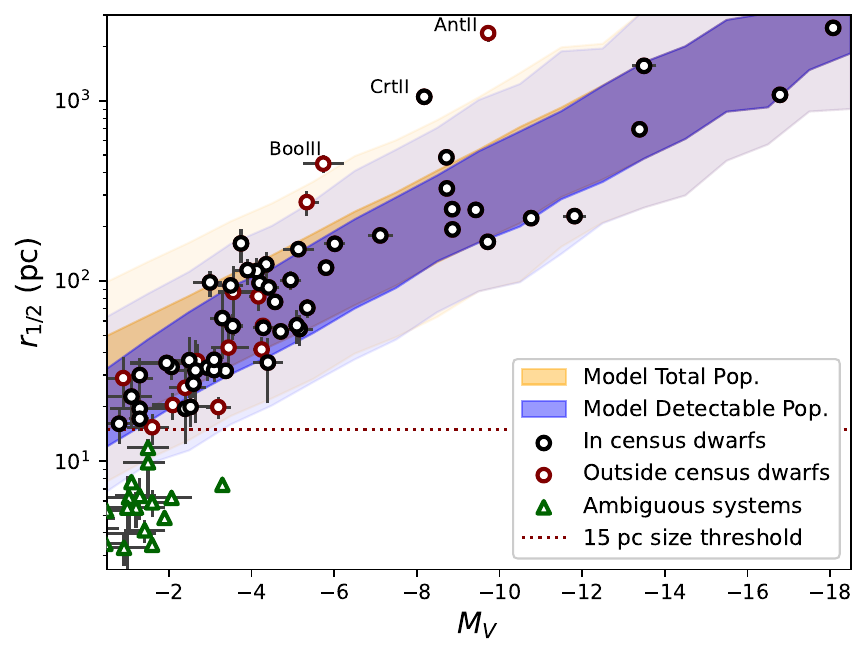}
    \caption{The inferred size--luminosity relation for both the total and detectable populations of Milky Way satellite galaxies. The dark (light)  bands correspond to 68\% (95\%) credible intervals. The inferred relations for the total (yellow) and detectable (purple) satellite populations are largely overlapping, except at the faintest magnitudes, where the detectable population contains fewer diffuse systems. Black circles indicate satellites recovered in our census, while maroon circles represent those that were not. \cyfive{Grey error bars are added for systems that have reported measurement uncertainties.} Green triangles denote ambiguous compact systems with $r_{1/2} < 15$\,pc (red dotted line), which are excluded from our census \cyfive{(see Appendix \ref{Appendix:ambigous})}. We also label the ultra-diffuse satellites that are significant outliers in the size--luminosity relation.} \label{fig:size_dist}
\end{figure}

\cyone{In Figure \ref{fig:radial_dist}, we show the  radial distribution of the total and detectable Milky Way satellite population based on our empirical model assuming a cored-NFW profile. We also show the binned radial distribution of satellites recovered in our census, along with a completeness-corrected version obtained using the same method as for the luminosity function. As shown in the figure, there is good agreement between our  cored-NFW-based empirical model and the observed satellite population. \cyfour{We note a slight downturn in the number of observed satellites at large radii compared to model predictions, which may be a statistical fluctuation due to high incompleteness at large distance.}
%This discrepancy could arise from significant incompleteness at the largest distances, or from source blending of satellite member stars not accounted for in our catalog-injection methods, which disproportionately affects distant, compact systems \citep{Zhang:2025}.}

Finally, in Figure \ref{fig:size_dist}, we present the size–luminosity relation of Milky Way satellites as inferred from our census data. As expected, we find that the detectable population of faint galaxies is on average more compact than the total population, since more compact galaxies are generally easier to detect. For comparison, we also include Milky Way dwarf galaxies both within and outside our census sample. We find again good agreement between our model and the observed galaxies, with the exception of outliers such as Antlia~II and Crater~II.}

\cyone{In addition to the best-fit values of the free parameters in our empirical model for Milky Way satellites, Table~\ref{table:model_variables} also lists the corresponding values from a similar model for M31 satellites by \citet{Doliva-Dolinsky:2023}. Using their parameter space, defined by $-17 \leq M_V \leq -5.5 $, $ 63 \leq r_{1/2}\,(\mathrm{pc}) \leq 10{,}000 $, and $30 \leq D_{GC}\,(\mathrm{kpc}) \leq 300$, our Milky Way model predicts $29^{+9}_{-7}$ galaxies, in contrast to the $92^{+19}_{-26}$ satellites predicted for M31. Comparing the two measurements, we find that the M31-to-Milky-Way satellite count ratio is $3.2\pm1.2$, which is within $1\sigma$ of the measured M31-to-Milky-Way mass ratio of $2–3$ \citep{2009MNRAS.397.1990B, 2018ApJ...857...78P, 2023ApJ...948..104P}. This agreement supports the expectation that the number of subhaloes, and thus satellites, scales approximately linearly with host halo mass \citep{2012MNRAS.424.2715W}. Furthermore, the shallower slope of the M31 size--luminosity relation relative to the Milky Way is consistent with M31 having a noisier assembly history and possibly more tidal stripping of satellites.}

\begin{figure*}
    \centering
    \includegraphics[width=\linewidth]{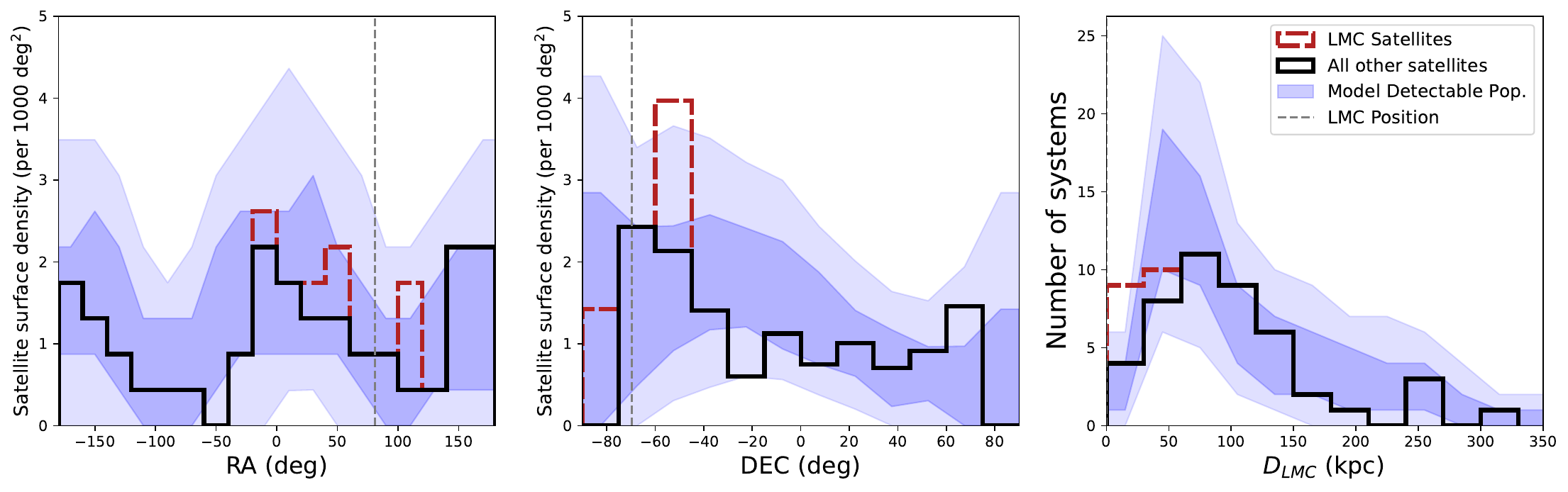}
    \caption{\cyone{{Testing the apparent anisotropy in the Milky Way satellites}. Left \& Middle: We compared the observed right ascension (RA) and  declination (Dec) of galaxies distributed with the one expected from an isotropic distribution (corrected for the detection selection function). \cyfour{An unusually high concentration of satellites is seen near Dec $\sim-60^\circ$, driven by satellites associated with the LMC.} Right: We also compare the observed distribution of satellites as a function of the distance to the LMC with their expected prediction from an isotropic distribution. }\label{fig:anisotropy_test}}
\end{figure*}

\subsection{Anisotropy in the Milky Way satellite distribution}
\label{sec:anisotropy}

$\Lambda$CDM simulations predict anisotropies in the spatial distribution of the satellite populations of Milky-Way-mass halos \citep[e.g.,][]{D'Onghia:2008, 2009MNRAS.399..550L, 2017MNRAS.466.3119A, 2018MNRAS.476.1796S, MeziniL2024, Buch:2024}. These anisotropies have been further studied for bright satellites of Milky-Way-mass galaxies \citep[e.g.,][]{Brainerd:2020, Samuels:2023}. In this section, we examine the on-sky angular distribution of Milky Way satellites to test for consistency with the isotropic assumption, when accounting for the anisotropic detection limits of our satellite census.

Figure~\ref{fig:anisotropy_test} shows the distribution in RA and Dec of satellites recovered in the census, along with predictions from an isotropic model. As expected, the isotropic model predicts a higher number of detectable galaxies in the southern hemisphere due to the greater sensitivity of DES and DELVE in that region.  However, even after accounting for selection bias, the number of galaxies observed near ${\rm Dec} \sim -60^\circ$ remains unusually high. 

We perform a Kolmogorov–Smirnov (KS) test for the Dec distribution of the satellites and find that the isotropic hypothesis is mildly disfavored, with a $p$-value of $p=2.2\times10^{-2}$, corresponding to a Gaussian significance of $2.3\sigma$. Note that this is smaller than the $p=4.3\times10^{-4}$ $(3.5\sigma)$ reported by \citet{Drlica-Wagner:2015} since their analysis only considered galaxies in the DES footprint while we consider most of the high-Galactic-latitude sky. In contrast, the RA distribution is more consistent with isotropy, with a KS test for the isotropic hypothesis yielding $p=2.3\times10^{-1}$ $(1.2\sigma)$.

\cyone{Most of the satellites with Dec $\sim-60^\circ$ are close to the LMC. Thus, it has been suggested that the anisotropy can be explained by a group infall scenario whereby the LMC brought its own population of smaller satellites \citep[e.g.,][]{D'Onghia:2008, 2008MNRAS.385.1365L, 2015ApJ...807...49W, Jethwa:2016, MWCensus2, 2021MNRAS.504.4551S}. Using detailed phase-space measurements and orbital modeling, seven dwarf galaxies (Carina~II, Carina~III, Horologium~I, Hydrus~I, Phoenix~II, Pictor~II, and Reticulum~II) have been identified as LMC satellites \citep{Kallivayalil:2018, Erkal:2020,  Patel:2020, 2022MNRAS.511.2610C, 2024MNRAS.527..437V,  Pace:2025}. As shown in Figure \ref{fig:anisotropy_test}, these satellites are located closer to the LMC than expected for an isotropic distribution and are concentrated around Dec $\sim-60^\circ$.}

If we remove these seven LMC satellites from the sample, the significance of the anisotropy decreases substantially. The KS test $p$-value for the Dec distribution increases to $p=1.4\times10^{-1}$ $(1.5\sigma)$ and the RA distribution to $p=4.6\times10^{-1}$ $(0.7\sigma)$.
\cyone{Therefore, we conclude that the presence of a mild anisotropy ($2.3\sigma$) in the angular distribution of the Milky Way satellites is driven primarily by the satellites associated with the LMC. 
This is in contrast to M31, which exhibits a stronger anisotropy in its satellite distribution with many more M31 satellites residing in the hemisphere facing the Milky Way \citep{Savino:2022, Doliva-Dolinsky:2023}.}

\section{Conclusion}
\label{sec:discussion}
We use imaging data from DES Y6, DELVE DR3, and PS1 DR1 to construct a stellar catalog spanning \CHECK{$\sim$27,700}\,deg$^2$ (see Figure \ref{Figure:mask}) and use this catalog to perform a systematic census of the  Milky Way satellite dwarf galaxies.  Our detection pipeline identifies thousands of hotspots with the highest-significance detections corresponding purely to known systems, while many lower-significance detections are likely false positives. By imposing a strict detection threshold, we successfully recovered a pure galaxy sample for our census. This sample consists of \CHECK{49} of the \CHECK{62} known Milky Way satellites found in our census footprint (Tables \ref{table:des_search}, \ref{table:delve_search}, and \ref{table:ps1_search} in Appendix \ref{appendix:detection_efficiencies}). This is the largest Milky Way satellite galaxy sample assembled from a uniform census to date. While we did not include any newly discovered systems in our census due to our conservative detection threshold, we identified several promising lower-significance candidates that will be presented in an upcoming paper.

We estimated the detection efficiency of our census by injecting simulated satellites with a wide range of properties and running the same detection algorithms \cytwo{and thresholds} to attempt to recover them. We express our satellite detection efficiency as a function of physical properties and sky location using two different methods. The first analytical method is based on the 50\% detectability contour (Figures \ref{fig:des_effiencency} and \ref{fig:delve_effiencency}), which delineates the parameters of simulated satellites detected within the 50\% detection efficiency.  In addition, we use the simulated satellites to build \texttt{XGBoost}-based machine learning models, which estimate the detection probability of Milky Way satellites as a function of their absolute magnitude, $M_V$, physical size, $r_{1/2}$, heliocentric distance, $D_\odot$, and sky position (RA, Dec). These models and other aspects of this analysis are publicly available online,$^{\ref{github_link}}$ enabling the community to apply the same selection function to other Milky Way satellite models and/or simulations. Our intent is to facilitate direct comparisons between observations and models of the Milky Way satellite galaxy population.

By adopting an empirical model for the Milky Way satellite population and combining it with our recovered sample and detection efficiency models, we estimate the completeness-corrected total number of Milky Way satellites with 
$-20\leq M_V\leq0$,  $15\leq r_{1/2} (\rm pc) \leq 3000$, and $10\leq D_{GC} (\rm kpc)\leq 300$ to be \CHECK{$265_{-47}^{+79}$}, consistent with several other estimates in the literature \citep[e.g.,][]{MWCensus2, Manwadkar:2022, Jethwa:2018}. We also construct an empirical model to estimate the luminosity function, radial distribution, and size--luminosity relation of the full satellite population, and we find good agreement with the observed satellites in our census (Figures~\ref{fig:lum_func}, \ref{fig:radial_dist}, and~\ref{fig:size_dist}). We compare our results to the empirical model of M31 from \citet{Doliva-Dolinsky:2023} and found that the  M31-to-Milky Way satellite count ratio is $3.2\pm1.2$, suggesting that M31 is much more massive than the Milky Way. Furthermore,
we  examined the apparent anisotropy in the spatial distribution of Milky Way satellites and detected the presence of a mild anisotropy in declination (2.3$\sigma$ significance) that is primarily driven by satellites associated with the LMC.

This deeper dataset and expanded footprint provide new leverage on the faint end of the galaxy--halo connection, enabling us to probe the suppression of galaxy formation by reionization and to place constraints on alternative dark matter scenarios. The census can also serve as an empirical anchor for low-mass semi-analytical models and abundance-matching methods (see \citealt{Wechsler:2018} for a review). In future work, we will present a more detailed comparison between the observed data and theoretical  predictions from numerical simulations with galaxy--halo connection framework.

While our survey covers $\sim${\CHECK{13,600}}\,deg$^2$ of the sky at a minimum depth of $g \sim 24.0$, we are still unable to recover some of the faintest galaxies discovered by deeper surveys such as Cetus~III and Virgo~I. However, the upcoming Rubin LSST is expected to cover an unmasked area of $\sim18,300\,$deg$^2$ at a depth of $r \sim  27.5$ mag (S/N = 5, point-like sources), which will allow a much more sensitive census of the Milky Way and nearby Local Volume satellites in the near future \citep{Mutlu-Pakdil:2021, Tsiane:2025}. In fact, it is expected that LSST will discover dozens to hundreds of ultra-faint satellites around the Milky Way \citep{Tollerud:2008, Hargis:2014, Jethwa:2018, Newton:2018, MWCensus2, Manwadkar:2022, Tsiane:2025}. Furthermore, other current and upcoming surveys such as UNIONS \citep{Gwyn:2025}, Euclid \citep{Euclid:2022}, and the Roman Space Telescope \citep{Spergel:2015} are also expected to have the ability to discover faint and distant systems \citep{Nadler:2024}. Maximizing the overlap between ground- and space-based observations will be critical to fully leverage the available data \citep{Han:2023}. In addition, large spectroscopic surveys may be able to detect large low surface brightness objects that are elusive in photometric surveys \citep{Chandra:2022, Aganze:2025}.  We expect that the growing population of Milky Way satellite galaxies will enable new insights into reionization, galaxy formation, and the nature of dark matter.

\facilities{Blanco (DECam), PS1, Gaia, Subaru (HSC)} 

\software{
\texttt{astropy} \citep{2013A&A...558A..33A, 2018AJ....156..123A},
\texttt{emcee} \citep{Foreman_Mackey:2013},
\texttt{fitsio},\footnote{\url{https://github.com/esheldon/fitsio}}
\texttt{healpix} \citep{Gorski:2005},\footnote{\url{http://healpix.sourceforge.net}}
\texttt{healpy} \citep{Zonca:2019},\footnote{\url{https://github.com/healpy/healpy}}
\texttt{healsparse},\footnote{\url{https://healsparse.readthedocs.io/en/latest/}}
\texttt{matplotlib} \citep{Hunter:2007},
\texttt{numpy} \citep{NumPy:2020},
\texttt{simple} \citep{Bechtol:2015},
\texttt{scipy} \citep{Scipy:2020},
\texttt{skymap},\footnote{\url{https://github.com/kadrlica/skymap}}
\texttt{ugali} \citep{Bechtol:2015, MWCensus1},
}

\section*{Acknowledgments}
\input{ack.tex}

\section*{Author Contributions}
CYT performed the main dwarf search, conducted the corresponding analysis, produced all plots and tables in the paper, and led the writing. ADW provided direct supervision for the research and guided the data processing and search algorithms. ABP and WC  guided major analysis decisions regarding building Table \ref{table:all_dwarfs} and  \ref{table:ambi_dwarfs} as well as thresholds for including satellites in the census. EON and ADD  advised on modeling choices for the empirical model. \cyfive{DA help run \texttt{balrog} for blending simulations.}  TSL, JDS,  AKV, and ARW internally reviewed the paper. The authors from MA to RHW contributed to producing and characterizing one or more of the following data products used in the paper: DES Y6 and DELVE DR3 source catalogs, known dwarf galaxy catalogs, and/or provided valuable comments that improved the paper’s clarity and quality. Builders: The remaining authors contributed to this work through the construction of DECam and other aspects of data collection; data processing and calibration; developing widely used methods, codes, and simulations; running pipelines and validation tests; and promoting the science analysis.


\bibliography{main}{}
\bibliographystyle{aasjournal}

%Panstarrs
%\begin{figure*}[ht]
%\centering
%\includegraphics[width=0.9\linewidth]{figures/PS1.png}
%%\vspace{-3cm}
%\caption{PS1 Detection effiency  }
%\end{figure*}

\appendix

\input{appendix}

%\section{Skymaps}
%interesting skymaps ugali 

%\begin{figure}[h]
%\centering
%\includegraphics[width=\linewidth]{figures/mod_16.5.png}
%%\vspace{-3cm}
%\caption{ugali search map for the entire sky (labeled)}
%\end{figure}



\end{document}

%% file: authors.tex
\correspondingauthor{Chin Yi Tan}
\email{chinyi@uchicago.edu}
\author[0000-0003-0478-0473]{C.~Y.~Tan}
\email{chinyi@uchicago.edu}
\affiliation{Kavli Institute for Cosmological Physics, University of Chicago, Chicago, IL 60637, USA}
\affiliation{Department of Physics, University of Chicago, Chicago, IL 60637, USA}
\affiliation{NSF-Simons AI Institute for the Sky (SkAI),172 E. Chestnut St., Chicago, IL 60611, USA}

\author[0000-0001-8251-933X]{A.~Drlica-Wagner}
\email{kadrlica@fnal.gov}
\affiliation{Fermi National Accelerator Laboratory, P.O.\ Box 500, Batavia, IL 60510, USA}
\affiliation{Kavli Institute for Cosmological Physics, University of Chicago, Chicago, IL 60637, USA}
\affiliation{Department of Astronomy and Astrophysics, University of Chicago, Chicago, IL 60637, USA}
\affiliation{NSF-Simons AI Institute for the Sky (SkAI),172 E. Chestnut St., Chicago, IL 60611, USA}

\author[0000-0002-6021-8760]
{A.~B.~Pace}
\affiliation{Department of Astronomy, University of Virginia, 530 McCormick Road, Charlottesville, VA 22904, USA}
\email{pvpace1@gmail.com}

\author[0000-0003-1697-7062]{W.~Cerny}
\affiliation{Department of Astronomy, Yale University, New Haven, CT 06520, USA}
\email{william.cerny@yale.edu}

\author[0000-0002-1182-3825]{E.~O.~Nadler}
\affiliation{
Department of Astronomy \& Astrophysics, University of California, San Diego, La Jolla, CA 92093, USA}
\email{enadler@ucsd.edu}

\author[0000-0001-9775-9029]{A.~Doliva-Dolinsky}
\affiliation{Department of Physics, University of Surrey, Guildford GU2 7XH, UK}
\email{amandine.doliva-dolinsky@dartmouth.edu}

 \author[0000-0003-3312-909X]{D.~Anbajagane}
 \affiliation{Department of Astronomy and Astrophysics, University of Chicago, Chicago, IL 60637, USA}
 \affiliation{Kavli Institute for Cosmological Physics, University of Chicago, Chicago, IL 60637, USA}
\affiliation{NSF-Simons AI Institute for the Sky (SkAI),172 E. Chestnut St., Chicago, IL 60611, USA}
\email{dhayaa@uchicago.edu}

\author[0000-0002-9110-6163]{T.~S.~Li}
\affiliation{Department of Astronomy and Astrophysics, University of Toronto, 50 St. George Street, Toronto ON, M5S 3H4, Canada}
\email{tingli@carnegiescience.edu}

\author[0000-0002-4733-4994]{J.~D.~Simon}
 \affiliation{Observatories of the Carnegie Institution for Science, 813 Santa Barbara St., Pasadena, CA 91101, USA}
\email{jsimon@carnegiescience.edu}

\author[0000-0003-4341-6172]{A.~K.~Vivas}
\affiliation{Cerro Tololo Inter-American Observatory/NSF NOIRLab, Casilla 603, La Serena, Chile}
\email{kathy.vivas@noirlab.edu}

\author[0000-0002-7123-8943, gname='Alistair', sname='Walker']{A.~R.~Walker}
\affiliation{Cerro Tololo Inter-American Observatory/NSF NOIRLab, Casilla 603, La Serena, Chile}
\email{alistair.walker@noirlab.edu}

%\author{et. al.}

% \author[0000-0002-7007-9725]{M.~Geha}
% \affiliation{Department of Astronomy, Yale University, New Haven, CT 06520, USA}

 \author[0000-0002-6904-359X]{M.~Adam\'ow}
 \affiliation{Center for Astrophysical Surveys, National Center for Supercomputing Applications, 1205 West Clark St., Urbana, IL 61801, USA}
  \email{madamow@illinois.edu}

% \author[0000-0003-4383-2969]{C.~R.~Bom}
% \affiliation{Centro Brasileiro de Pesquisas F\'isicas, Rua Dr. Xavier Sigaud 150, 22290-180 Rio de Janeiro, RJ, Brazil}

 \author[0000-0001-8156-0429]
 {K.~Bechtol}
 \affiliation{Physics Department, 2320 Chamberlin Hall, University of Wisconsin-Madison, 1150 University Avenue Madison, WI  53706-1390}
  \email{kbechtol@wisc.edu}

 \author[0000-0002-3936-9628]
 {J.~L.~Carlin}
 \affiliation{Rubin Observatory/AURA, 950 North Cherry Avenue, Tucson, AZ, 85719, USA}
  \email{jeffreylcarlin@gmail.com}
  
 \author[0009-0005-9002-4800]{Q.~O.~Casey}
 \affiliation{Department of Physics and Astronomy, Dartmouth College, Hanover, NH 03755, USA}
  \email{Quinn.O.Casey.GR@dartmouth.edu}
 \author[0000-0002-7887-0896]{C.~Chang}
 \affiliation{Department of Astronomy and Astrophysics, University of Chicago, Chicago, IL 60637, USA}
 \affiliation{Kavli Institute for Cosmological Physics, University of Chicago, Chicago, IL 60637, USA}
 \affiliation{NSF-Simons AI Institute for the Sky (SkAI),172 E. Chestnut St., Chicago, IL 60611, USA}
  \email{chihway@kicp.uchicago.edu}

 \author[0000-0001-5143-1255]{A.~Chaturvedi}
 \affiliation{Department of Physics, University of Surrey, Guildford GU2 7XH, UK}
  \email{aa07223@surrey.ac.uk}
\author[0000-0001-8670-4495, gname='Ting-Yun', sname='Cheng']{T.-Y.~Cheng}
\affiliation{Kapteyn Astronomical Institute, University of Groningen, Landleven 12 (Kapteynborg, 5419), 9747 AD Groningen, The Netherlands}
\email{tycheng.sunny@gmail.com}
  
 \author[0000-0002-7155-679X]{A.~Chiti}
 \affiliation{Department of Astronomy and Astrophysics, University of Chicago, Chicago, IL 60637, USA}
 \affiliation{Kavli Institute for Cosmological Physics, University of Chicago, Chicago, IL 60637, USA}
 \email{achiti@stanford.edu}

 \author[0000-0003-1680-1884]{Y.~Choi}
 \affiliation{NSF NOIRLab, 950 N. Cherry Ave., Tucson, AZ 85719, USA}
 \email{yumi.choi@noirlab.edu}

\author[0000-0002-1763-4128]{D.~Crnojevi\'c}
\affiliation{Department of Physics \& Astronomy, University of Tampa, 401 West Kennedy Boulevard, Tampa, FL 33606, USA}
\email{dcrnojevic@ut.edu}
 
% \author[0000-0002-1693-3265]{M.~L.~M.~Collins}
% \affiliation{Department of Physics, University of Surrey, Guildford GU2 7XH, UK}

 \author[0000-0001-6957-1627]{P.~S.~Ferguson}
 \affiliation{DiRAC Institute, Department of Astronomy, University of Washington, 3910 15th Ave NE, Seattle, WA, 98195, USA}
 \affiliation{Department of Physics, University of Wisconsin-Madison, Madison, WI 53706, USA}
 \email{pferguso@uw.edu}
 \author[0000-0002-4588-6517]{R.~A.~Gruendl}
 \affiliation{Department of Astronomy, University of Illinois, 1002 W. Green Street, Urbana, IL 61801, USA}
 \affiliation{Center for Astrophysical Surveys, National Center for Supercomputing Applications, 1205 West Clark St., Urbana, IL 61801, USA}
  \email{gruendl@illinois.edu}

\author[0000-0002-4863-8842]{A.~P.~Ji}
\affiliation{Kavli Institute for Cosmological Physics, University of Chicago, Chicago, IL 60637, USA}
\affiliation{Department of Astronomy and Astrophysics, University of Chicago, Chicago, IL 60637, USA}
\email{alexji@uchicago.edu}

 \author[0000-0002-9269-8287]{G.~Limberg}
 \affiliation{Kavli Institute for Cosmological Physics, University of Chicago, Chicago, IL 60637, USA}
 \email{limberg@uchicago.edu}
% \author[0000-0001-9438-5228]{M.~Navabi}
% \affiliation{Department of Physics, University of Surrey, Guildford GU2 7XH, UK}
% \author[0000-0003-3835-2231]{D.~Mart\'{i}nez-Delgado}
% \affiliation{Instituto de Astrof\'{i}sica de Andaluc\'{i}a, CSIC, E-18080 Granada, Spain}

 \author[0000-0003-0105-9576]{G.~E.~Medina}
 \affiliation{Department of Astronomy and Astrophysics, University of Toronto, 50 St. George Street, Toronto ON, M5S 3H4, Canada}
 \email{gustavo.medina@utoronto.ca}
 \author[0000-0001-9649-4815]{B.~Mutlu-Pakdil}
 \affiliation{Department of Physics and Astronomy, Dartmouth College, Hanover, NH 03755, USA}
\email{Burcin.Mutlu-Pakdil@dartmouth.edu}
% \author[0000-0002-1793-3689]{D.~L.~Nidever}
% \affiliation{Department of Physics, Montana State University, P.O. Box 173840, Bozeman, MT 59717-3840}
% \affiliation{NSF NOIRLab, 950 N. Cherry Ave., Tucson, AZ 85719, USA}

 \author[0000-0002-8282-469X]{N.~E.~D.~No\"el}
 \affiliation{Department of Physics, University of Surrey, Guildford GU2 7XH, UK}
 \email{n.noel@surrey.ac.uk}

\author[0009-0008-0959-0162]{K.~Overdeck}
\affiliation{Kavli Institute for Cosmological Physics, University of Chicago, Chicago, IL 60637, USA}
\affiliation{Department of Astronomy and Astrophysics, University of Chicago, Chicago, IL 60637, USA}
\affiliation{NSF-Simons AI Institute for the Sky (SkAI),172 E. Chestnut St., Chicago, IL 60611, USA}
 \email{koverdeck@uchicago.edu}
\author[0000-0003-4479-1265]{V.~M.~Placco}
 \affiliation{NSF NOIRLab, 950 N. Cherry Ave., Tucson, AZ 85719, USA}
 \email{vinicius.placco@noirlab.edu}

 \author[0000-0001-5805-5766]{A.~H.~Riley}
 \affiliation{Institute for Computational Cosmology, Department of Physics, Durham University, South Road, Durham DH1 3LE, UK}
 \affiliation{ Lund Observatory, Division of Astrophysics, Department of Physics, Lund University, SE-221 00 Lund, Sweden}
\email{alexander.riley@fysik.lu.se}

% \author[0000-0002-1594-1466]{J.~D.~Sakowska}
% \affiliation{Department of Physics, University of Surrey, Guildford GU2 7XH, UK}
 \author[0000-0003-4102-380X]{D.~J.~Sand}
 \affiliation{Department of Astronomy/Steward Observatory, 933 North Cherry Avenue, Room N204, Tucson, AZ 85721-0065, USA}
 \email{dsand@arizona.edu}
 \author[0009-0001-1133-5047]{J.~Sharp}
 \affiliation{Kavli Institute for Cosmological Physics, University of Chicago, Chicago, IL 60637, USA}
 \affiliation{Department of Astronomy and Astrophysics, University of Chicago, Chicago, IL 60637, USA}
  \email{sharpj@uchicago.edu}

\author[0000-0001-5399-0114]{N.~F.~Sherman}
\affiliation{Institute for Astrophysical Research, Boston University, 725 Commonwealth Avenue, Boston, MA 02215, USA}
\email{norafs@bu.edu}

 \author[0000-0003-1479-3059]{G.~S.~Stringfellow}
 \affiliation{Center for Astrophysics and Space Astronomy, University of Colorado, 389 UCB, Boulder, CO 80309-0389, USA}
 \email{Guy.Stringfellow@colorado.edu}

\author[0000-0003-2229-011X]{R.~H.~Wechsler}
 \affiliation{Department of Physics, Stanford University, 382 Via Pueblo Mall, Stanford, CA 94305, USA}
  \affiliation{Kavli Institute for Particle Astrophysics \& Cosmology, P.O.\ Box 2450, Stanford University, Stanford, CA 94305, USA}
   \affiliation{SLAC National Accelerator Laboratory, Menlo Park, CA 94025, USA}
  \email{rwechsler@stanford.edu}
  
\author[0000-0001-5679-6747, gname='Michel', sname='Aguena']{M.~Aguena}
\affiliation{INAF-Osservatorio Astronomico di Trieste, via G. B. Tiepolo 11, I-34143 Trieste, Italy}
\affiliation{Laborat\'orio Interinstitucional de e-Astronomia - LIneA, Av. Pastor Martin Luther King Jr, 126 Del Castilho, Nova Am\'erica Offices, Torre 3000/sala 817 CEP: 20765-000, Brazil}
\email{aguena@linea.org.br}

\author[0000-0002-7069-7857, gname='Sahar', sname='Allam']{S.~Allam}
\affiliation{Fermi National Accelerator Laboratory, P. O. Box 500, Batavia, IL 60510, USA}
\email{sallamatfnalgov@gmail.com}

\author[0000-0002-7394-9466,gname='Otavio', sname='Alves']{O.~Alves}
\affiliation{Department of Physics, University of Michigan, Ann Arbor, MI 48109, USA}
\email{oalves@umich.edu}

\author[0000-0002-2562-8537,gname='David', sname='Bacon']{D.~Bacon}
\affiliation{Institute of Cosmology and Gravitation, University of Portsmouth, Portsmouth, PO1 3FX, UK}
\email{bacond@fnal.gov}

\author[0000-0002-8458-5047, gname='David', sname='Brooks']{D.~Brooks}
\affiliation{Department of Physics \& Astronomy, University College London, Gower Street, London, WC1E 6BT, UK}
\email{david.brooks@ucl.ac.uk}

\author[0000-0003-1866-1950, gname='David', sname='Burke']{D.~L.~Burke}
\affiliation{Kavli Institute for Particle Astrophysics \& Cosmology, P. O. Box 2450, Stanford University, Stanford, CA 94305, USA}
\affiliation{SLAC National Accelerator Laboratory, Menlo Park, CA 94025, USA}
\email{daveb@slac.stanford.edu}

\author[0000-0002-7436-3950,gname='Ryan', sname='Camilleri']{R.~Camilleri}
\affiliation{School of Mathematics and Physics, University of Queensland,  Brisbane, QLD 4072, Australia}
\email{r.camilleri@uq.net.au}
 \author[0000-0002-3690-105X]{J.~A.~Carballo-Bello}
 \affiliation{Instituto de Alta Investigaci\'on, Universidad de Tarapac\'a, Casilla 7D, Arica, Chile}
  \email{jcarballo@academicos.uta.cl}

\author[0000-0003-3044-5150, gname='Aurelio', sname='Carnero Rosell']{A.~Carnero~Rosell}
\affiliation{Instituto de Astrofisica de Canarias, E-38205 La Laguna, Tenerife, Spain}
\affiliation{Laborat\'orio Interinstitucional de e-Astronomia - LIneA, Av. Pastor Martin Luther King Jr, 126 Del Castilho, Nova Am\'erica Offices, Torre 3000/sala 817 CEP: 20765-000, Brazil}
\affiliation{Universidad de La Laguna, Dpto. Astrofísica, E-38206 La Laguna, Tenerife, Spain}
\email{aurelio.crosell@gmail.com}

\author[0000-0002-3130-0204, gname='Jorge', sname='Carretero']{J.~Carretero}
\affiliation{Institut de F\'{\i}sica d'Altes Energies (IFAE), The Barcelona Institute of Science and Technology, Campus UAB, 08193 Bellaterra (Barcelona) Spain}
\email{jorgecarreteropalacios@gmail.com}

\author[0000-0002-7731-277X, gname='Luiz', sname='da Costa']{L.~N.~da Costa}
\affiliation{Laborat\'orio Interinstitucional de e-Astronomia - LIneA, Av. Pastor Martin Luther King Jr, 126 Del Castilho, Nova Am\'erica Offices, Torre 3000/sala 817 CEP: 20765-000, Brazil}
\email{ldacosta@fnal.gov}

\author[0000-0002-7131-7684,gname='Maria Elidaiana', sname='da Silva Pereira']{M.~E.~da Silva Pereira}
\affiliation{Hamburger Sternwarte, Universit\"{a}t Hamburg, Gojenbergsweg 112, 21029 Hamburg, Germany}
\email{elidaiana.sp@gmail.com}

\author[0000-0002-4213-8783, gname='Tamara', sname='Davis']{T.~M.~Davis}
\affiliation{School of Mathematics and Physics, University of Queensland,  Brisbane, QLD 4072, Australia}
\email{tamarad@physics.uq.edu.au }

\author[0000-0001-8318-6813, gname='Juan', sname='De Vicente']{J.~De~Vicente}
\affiliation{Centro de Investigaciones Energ\'eticas, Medioambientales y Tecnol\'ogicas (CIEMAT), Madrid, Spain}
\email{juan.vicente@ciemat.es}

\author[0000-0002-0466-3288, gname='Shantanu', sname='Desai']{S.~Desai}
\affiliation{Department of Physics, IIT Hyderabad, Kandi, Telangana 502285, India}
\email{shntn05@gmail.com}

\author[0000-0002-3745-2882,gname='Spencer', sname='Everett']{S.~Everett}
\affiliation{California Institute of Technology, 1200 East California Blvd, MC 249-17, Pasadena, CA 91125, USA}
\email{severett@caltech.edu}

\author[0000-0002-2367-5049, gname='Brenna', sname='Flaugher']{B.~Flaugher}
\affiliation{Fermi National Accelerator Laboratory, P. O. Box 500, Batavia, IL 60510, USA}
\email{brenna@fnal.gov}

\author[0000-0003-4079-3263, gname='Josh', sname='Frieman']{J.~Frieman}
\affiliation{Department of Astronomy and Astrophysics, University of Chicago, Chicago, IL 60637, USA}
\affiliation{Fermi National Accelerator Laboratory, P. O. Box 500, Batavia, IL 60510, USA}
\affiliation{Kavli Institute for Cosmological Physics, University of Chicago, Chicago, IL 60637, USA}
\email{frieman@fnal.gov}

\author[0000-0002-9370-8360, gname='Juan', sname='Garcia-Bellido']{J.~Garc\'ia-Bellido}
\affiliation{Instituto de Fisica Teorica UAM/CSIC, Universidad Autonoma de Madrid, 28049 Madrid, Spain}
\email{juan.garciabellido@uam.es}

\author[0000-0003-3270-7644, gname='Daniel', sname='Gruen']{D.~Gruen}
\affiliation{University Observatory, LMU Faculty of Physics, Scheinerstr. 1, 81679 Munich, Germany}
\email{dgruen@usm.uni-muenchen.de}

\author[0000-0003-0825-0517, gname='Gaston', sname='Gutierrez']{G.~Gutierrez}
\affiliation{Fermi National Accelerator Laboratory, P. O. Box 500, Batavia, IL 60510, USA}
\email{gaston@fnal.gov}

\author[0000-0001-6718-2978, gname='Kenneth', sname='Herner']{K.~Herner}
\affiliation{Fermi National Accelerator Laboratory, P. O. Box 500, Batavia, IL 60510, USA}
\email{kherner@fnal.gov}

\author[0000-0003-2071-9349, gname='Samuel', sname='Hinton']{S.~R.~Hinton}
\affiliation{School of Mathematics and Physics, University of Queensland,  Brisbane, QLD 4072, Australia}
\email{samuelreay@gmail.com}

\author[0000-0002-9369-4157, gname='Devon L.', sname='Hollowood']{D.~L.~Hollowood}
\affiliation{Santa Cruz Institute for Particle Physics, Santa Cruz, CA 95064, USA}
\email{dhollowo@ucsc.edu}

 \author[0000-0001-5160-4486]{D.~J.~James}
 \affiliation{ASTRAVEO LLC, PO Box 1668, Gloucester, MA 01931}
 \affiliation{Applied Materials Inc., 35 Dory Road, Gloucester, MA 01930}
\email{djames44@gmail.com}

\author[0000-0003-0120-0808, gname='Kyler', sname='Kuehn']{K.~Kuehn}
\affiliation{Australian Astronomical Optics, Macquarie University, North Ryde, NSW 2113, Australia}
\affiliation{Lowell Observatory, 1400 Mars Hill Rd, Flagstaff, AZ 86001, USA}
\email{kkuehn@lowell.edu}

\author[0000-0002-1134-9035, gname='Ofer', sname='Lahav']{O.~Lahav}
\affiliation{Department of Physics \& Astronomy, University College London, Gower Street, London, WC1E 6BT, UK}
\email{o.lahav@ucl.ac.uk}

\author[0000-0002-8289-740X, gname='Sujeong', sname='Lee']{S.~Lee}
\affiliation{Jet Propulsion Laboratory, California Institute of Technology, 4800 Oak Grove Dr., Pasadena, CA 91109, USA}
\email{sujeong.lee@jpl.nasa.gov}

\author[0000-0003-0710-9474, gname='Jennifer', sname='Marshall']{J.~L.~Marshall}
\affiliation{George P. and Cynthia Woods Mitchell Institute for Fundamental Physics and Astronomy, and Department of Physics and Astronomy, Texas A\&M University, College Station, TX 77843,  USA}
\email{jlm076@tamu.edu}

 \author[0000-0002-9144-7726]{C.~E.~Mart\'inez-V\'azquez}
 \affiliation{International Gemini Observatory/NSF NOIRLab, 670 N. A'ohoku Place, Hilo, Hawai'i, 96720, USA}
 \email{clara.martinez@noirlab.edu}

 \author[0000-0002-8093-7471]{P.~Massana}
 \affiliation{NSF NOIRLab, Casilla 603, La Serena, Chile}
 \email{pol.massana@noirlab.edu}

\author[0000-0001-9497-7266, gname='Juan', sname='Mena-Fernández']{J. Mena-Fern{\'a}ndez}
\affiliation{Universit\'e Grenoble Alpes, CNRS, LPSC-IN2P3, 38000 Grenoble, France}
\email{juan.menafernandez@lpsc.in2p3.fr}

\author[0000-0002-6610-4836, gname='Ramon', sname='Miquel']{R.~Miquel}
\affiliation{Instituci\'o Catalana de Recerca i Estudis Avan\c{c}ats, E-08010 Barcelona, Spain}
\affiliation{Institut de F\'{\i}sica d'Altes Energies (IFAE), The Barcelona Institute of Science and Technology, Campus UAB, 08193 Bellaterra (Barcelona) Spain}
\email{rmiquel@ifae.es}

\author[0000-0002-7579-770X, gname='Jessica', sname='Muir']{J.~Muir}
\affiliation{Department of Physics, University of Cincinnati, Cincinnati, Ohio 45221, USA}
\affiliation{Perimeter Institute for Theoretical Physics, 31 Caroline St. North, Waterloo, ON N2L 2Y5, Canada}
\email{muirjc.ucmail.uc.edu}

\author[0000-0001-6145-5859, gname='Justin', sname='Myles']{J.~Myles}
\affiliation{Department of Astrophysical Sciences, Princeton University, Peyton Hall, Princeton, NJ 08544, USA}
\email{jmyles@astro.princeton.edu}

\author[0000-0003-2120-1154, gname='Ricardo', sname='Ogando']{R.~L.~C.~Ogando}
\affiliation{Centro de Tecnologia da Informa\c{c}\~ao Renato Archer, Campinas, SP - 13069-901, Brazil }
\affiliation{Observat\'orio Nacional, Rua Gal. Jos\'e Cristino 77, Rio de Janeiro, RJ - 20921-400, Brazil}
\email{ricardo.ogando@cti.gov.br}

\author[0000-0002-2598-0514, gname='Andrés', sname='Plazas Malagón']{A.~A.~Plazas~Malag\'on}
\affiliation{Kavli Institute for Particle Astrophysics \& Cosmology, P. O. Box 2450, Stanford University, Stanford, CA 94305, USA}
\affiliation{SLAC National Accelerator Laboratory, Menlo Park, CA 94025, USA}
\email{plazasmalagon@gmail.com}

\author[0000-0002-2762-2024, gname='Anna', sname='Porredon']{A.~Porredon}
\affiliation{Centro de Investigaciones Energ\'eticas, Medioambientales y Tecnol\'ogicas (CIEMAT), Madrid, Spain}
\affiliation{Ruhr University Bochum, Faculty of Physics and Astronomy, Astronomical Institute, German Centre for Cosmological Lensing, 44780 Bochum, Germany}
\email{annam.porredon@gmail.com}

\author[0000-0002-9646-8198, gname='Eusebio', sname='Sanchez']{E.~Sanchez}
\affiliation{Centro de Investigaciones Energ\'eticas, Medioambientales y Tecnol\'ogicas (CIEMAT), Madrid, Spain}
\email{eusebio.sanchez@ciemat.es}

\author[0000-0003-3054-7907, gname='David', sname='Sanchez Cid']{D.~Sanchez Cid}
\affiliation{Centro de Investigaciones Energ\'eticas, Medioambientales y Tecnol\'ogicas (CIEMAT), Madrid, Spain}
\affiliation{Physik-Institut, University of Zürich, Winterthurerstrasse 190, CH-8057 Zürich, Switzerland}
\email{david.sanchezcid@physik.uzh.ch}

\author[0000-0002-1831-1953, gname='Ignacio', sname='Sevilla']{I.~Sevilla-Noarbe}
\affiliation{Centro de Investigaciones Energ\'eticas, Medioambientales y Tecnol\'ogicas (CIEMAT), Madrid, Spain}
\email{ignacio.sevilla@ciemat.es}

\author[0000-0002-3321-1432, gname='Mathew', sname='Smith']{M.~Smith}
\affiliation{Physics Department, Lancaster University, Lancaster, LA1 4YB, UK}
\email{mat.smith@soton.ac.uk}

\author[0000-0002-7047-9358, gname='Eric', sname='Suchyta']{E.~Suchyta}
\affiliation{Computer Science and Mathematics Division, Oak Ridge National Laboratory, Oak Ridge, TN 37831}
\email{eric.d.suchyta@gmail.com}

\author[0000-0002-1488-8552,gname='Molly', sname='Swanson']{M.~E.~C.~Swanson}
\affiliation{Center for Astrophysical Surveys, National Center for Supercomputing Applications, 1205 West Clark St., Urbana, IL 61801, USA}
\email{molly.swanson@gmail.com}

\author[0000-0001-7836-2261, gname='Chun-Hao', sname='To']{C.~To}
\affiliation{Department of Astronomy and Astrophysics, University of Chicago, Chicago, IL 60637, USA}
\email{chunhaoto@gmail.com}

 \author[0000-0002-9599-310X]{E.~J.~Tollerud}
 \affiliation{Space Telescope Science Institute, 3700 San Martin Drive, Baltimore, MD 21218, USA}
  \email{etollerud@stsci.edu}

\author[0000-0001-7211-5729, gname='Douglas', sname='Tucker']{D.~L.~Tucker}
\affiliation{Fermi National Accelerator Laboratory, P. O. Box 500, Batavia, IL 60510, USA}
\email{dtucker@fnal.gov}

\author[gname='Vinu', sname='Vikram']{V.~Vikram}
\affiliation{Department of Physics, Central University of Kerala, 93VR+RWF, CUK Rd, Kerala 671316, India}
\email{vvinuv@gmail.com}

\author[0000-0001-9382-5199, gname='Noah', sname='Weaverdyck']{N.~Weaverdyck}
\affiliation{Department of Astronomy, University of California, Berkeley,  501 Campbell Hall, Berkeley, CA 94720, USA}
\affiliation{Lawrence Berkeley National Laboratory, 1 Cyclotron Road, Berkeley, CA 94720, USA}
\email{nweaverd@umich.edu}

\author[0000-0003-1585-997X, gname='Masaya', sname='Yamamoto']{M.~Yamamoto}
\affiliation{Department of Astrophysical Sciences, Princeton University, Peyton Hall, Princeton, NJ 08544, USA}
\affiliation{Department of Physics, Duke University Durham, NC 27708, USA}
\email{masaya.yamamoto@princeton.edu}

\author[0000-0001-6455-9135]{A.~Zenteno}
\affiliation{Cerro Tololo Inter-American Observatory/NSF NOIRLab, Casilla 603, La Serena, Chile}
\email{alfredo.zenrteno@noirlab.edu}

\collaboration{87}{(DELVE and DES Collaborations)}

%% file: table1.tex
\startlongtable
\begin{deluxetable*}{lcccccccccccc}
\tabletypesize{\scriptsize}
\tablewidth{\textwidth} 
%\tablenum{1}
\tablecaption{{Confirmed and candidate Milky Way satellite galaxies with azimuthally averaged projected physical half-light radius $r_{1/2} > 15$ pc within a Galactocentric distance of $D <$ 400 kpc. We report on-sky location (RA, Dec), distance modulus ($(m-M)_0$), heliocentric distance ($D_\odot$), azimuthally averaged physical half-light radius ($r_{1/2}$) and absolute magnitude ($M_V$). We also note the survey that is used to analyze each system, its probability of inclusion in our census as predicted by the \texttt{XGBoost}–based classifier based on its properties (Section \ref{sec:selfunc_ML}), and whether it actually passes the detection threshold defined for our census. Bright satellites  that lie outside the footprint ($^\dagger$) can be incorporated into the census under the additional assumption that all bright satellites ($M_V < -12.5$) are detected regardless of their location or distance (see Section  \ref{sec:results} for more details). \label{table:all_dwarfs} }}
\tablehead{
Name & Class. & Abbrev. &  RA & Dec & $(m-M)_0$  & $D_\odot$ &$r_{1/2}$ & $M_V$ & Survey/ &  $P_{\rm det}$  & Pass Census & Ref.\\ & & & (deg) & (deg) & (mag)  &  (kpc) & (pc)  &  (mag)& Region &  & Threshold &  } 
\startdata 
%%%%%%%%%%%
Antlia II &  D  & AntII & 143.8 & $-$36.7 & 20.5 & 124$^{+5}_{-5}$ & 2379$^{+248}_{-248}$  &  $-$9.7$^{+0.1}_{-0.1}$  & Gal. Plane &   - & No & 1, 2 \\ 
Aquarius II &  D  & AqrII & 338.5 & $-$9.3 & 20.2 & 108$^{+4}_{-3}$ & 124$^{+22}_{-22}$  &  $-$4.4$^{+0.1}_{-0.1}$  & DELVE &   0.69 & Yes & 3 \\ 
Aquarius III &  D  & AqrIII & 357.2 & $-$3.5 & 19.7 & 86$^{+4}_{-4}$ & 36$^{+14}_{-13}$  &  $-$2.5$^{+0.3}_{-0.5}$  & DELVE &   0.17 & Yes & 4 \\ 
Bo\"{o}tes I &  D  & BooI & 210.0 & 14.5 & 19.1 & 66$^{+2}_{-2}$ & 161$^{+8}_{-8}$  &  $-$6.0$^{+0.2}_{-0.2}$  & DELVE &   1.00 & Yes & 5, 6 \\ 
Bo\"{o}tes II &  D  & BooII & 209.5 & 12.9 & 18.1 & 42$^{+1}_{-1}$ & 33$^{+5}_{-5}$  &  $-$2.9$^{+0.2}_{-0.2}$  & DELVE &   0.96 & Yes & 5, 7 \\ 
Bo\"{o}tes III &  D  & BooIII & 209.3 & 26.8 & 18.3 & 47$^{+1}_{-1}$ & 448$^{+52}_{-52}$  &  $-$5.7$^{+0.5}_{-0.5}$  & PS1 &   0.01 & No & 8, 9, 10 \\ 
Bo\"{o}tes IV &  DC  & BooIV & 233.7 & 43.7 & 21.6 & 209$^{+20}_{-18}$ & 274$^{+45}_{-42}$  &  $-$5.3$^{+0.3}_{-0.2}$  & PS1 &   0.01 & No & 11, 12 \\ 
Bo\"{o}tes V &  DC  & BooV & 213.9 & 32.9 & 20.0 & 102$^{+7}_{-7}$ & 20$^{+3}_{-3}$  &  $-$3.2$^{+0.3}_{-0.3}$  & PS1 &   0.06 & No & 13 \\ 
Canes Venatici I &  D  & CVnI & 202.0 & 33.6 & 21.6 & 211$^{+6}_{-6}$ & 326$^{+16}_{-16}$  &  $-$8.7$^{+0.1}_{-0.1}$  & PS1 &   1.00 & Yes & 5, 14 \\ 
Canes Venatici II &  D  & CVnII & 194.3 & 34.3 & 21.0 & 160$^{+4}_{-4}$ & 54$^{+11}_{-10}$  &  $-$5.2$^{+0.3}_{-0.3}$  & PS1 &   0.93 & Yes & 5, 15 \\ 
Carina &  D  & Car & 100.4 & $-$51.0 & 20.1 & 106$^{+6}_{-5}$ & 248$^{+13}_{-13}$  &  $-$9.4$^{+0.1}_{-0.1}$  & DELVE &   1.00 & Yes & 5, 16 \\ 
Carina II &  D  & CarII & 114.1 & $-$58.0 & 17.9 & 37$^{+1}_{-1}$ & 77$^{+8}_{-8}$  &  $-$4.6$^{+0.1}_{-0.1}$  & DELVE &   0.99 & Yes & 17 \\ 
Carina III &  D  & CarIII & 114.6 & $-$57.9 & 17.2 & 28$^{+1}_{-1}$ & 20$^{+7}_{-7}$  &  $-$2.4$^{+0.2}_{-0.2}$  & DELVE &   0.91 & Yes & 17 \\ 
Centaurus I &  D  & CenI & 189.6 & $-$40.9 & 20.4 & 118$^{+4}_{-4}$ & 71$^{+9}_{-9}$  &  $-$5.4$^{+0.2}_{-0.2}$  & DELVE &   0.98 & Yes & 18, 19 \\ 
Cetus II &  DC  & CetII & 19.5 & $-$17.4 & 17.4 & 30$^{+3}_{-3}$ & 16$^{+7}_{-6}$  &  0.0$^{+0.7}_{-0.7}$  & DES &   0.79 & Yes & 20 \\ 
Cetus III &  DC  & CetIII & 31.3 & $-$4.3 & 22.0 & 251$^{+24}_{-22}$ & 43$^{+14}_{-13}$  &  $-$3.5$^{+0.5}_{-0.4}$  & DES &   0.70 & No & 12, 21 \\ 
Columba I &  D  & ColI & 82.9 & $-$28.0 & 21.3 & 183$^{+9}_{-9}$ & 97$^{+12}_{-12}$  &  $-$4.2$^{+0.2}_{-0.2}$  & DES &   0.76 & Yes & 22 \\ 
Coma Berenices &  D  & Com & 186.7 & 23.9 & 18.1 & 42$^{+2}_{-2}$ & 55$^{+4}_{-4}$  &  $-$4.3$^{+0.2}_{-0.2}$  & DELVE &   0.99 & Yes & 5, 23 \\ 
Crater II &  D  & CrtII & 177.3 & $-$18.4 & 20.3 & 117$^{+4}_{-4}$ & 1054$^{+93}_{-89}$  &  $-$8.2$^{+0.1}_{-0.1}$  & DELVE &   0.95 & Yes & 24 \\ 
DELVE 2 &  DC  & DEL2 & 28.8 & $-$68.3 & 19.3 & 71$^{+3}_{-3}$ & 20$^{+4}_{-4}$  &  $-$2.1$^{+0.4}_{-0.5}$  & DELVE &   0.31 & No & 25 \\ 
Draco &  D  & Dra & 260.1 & 57.9 & 19.6 & 82$^{+2}_{-1}$ & 193$^{+4}_{-4}$  &  $-$8.9$^{+0.1}_{-0.1}$  & PS1 &   1.00 & Yes & 5, 26 \\ 
Draco II &  DC  & DraII & 238.2 & 64.6 & 16.7 & 22$^{+1}_{-1}$ & 16$^{+4}_{-4}$  &  $-$0.8$^{+0.4}_{-1.0}$  & PS1 &   0.24 & Yes & 27 \\ 
Eridanus II &  D  & EriII & 56.1 & $-$43.5 & 22.8 & 370$^{+9}_{-8}$ & 179$^{+12}_{-12}$  &  $-$7.1$^{+0.3}_{-0.3}$  & DES &   1.00 & Yes & 28, 29 \\ 
Eridanus IV &  D  & EriIV & 76.4 & $-$9.5 & 19.2 & 70$^{+4}_{-3}$ & 56$^{+9}_{-9}$  &  $-$3.5$^{+0.2}_{-0.2}$  & DELVE &   0.93 & Yes & 18, 30 \\ 
Fornax &  D  & For & 40.0 & $-$34.5 & 20.8 & 143$^{+3}_{-3}$ & 695$^{+16}_{-16}$  &  $-$13.4$^{+0.2}_{-0.2}$  & DES &   1.00 & Yes & 5, 31, 32 \\ 
Grus I &  D  & GruI & 344.2 & $-$50.2 & 20.5 & 126$^{+6}_{-6}$ & 113$^{+21}_{-20}$  &  $-$4.1$^{+0.3}_{-0.3}$  & DES &   0.96 & Yes & 33, 34 \\ 
Grus II &  D  & GruII & 331.0 & $-$46.4 & 18.7 & 55$^{+3}_{-2}$ & 94$^{+9}_{-9}$  &  $-$3.5$^{+0.3}_{-0.3}$  & DES &   0.98 & Yes & 34, 35 \\ 
Hercules &  D  & Her & 247.8 & 12.8 & 20.6 & 131$^{+6}_{-6}$ & 119$^{+12}_{-12}$  &  $-$5.8$^{+0.2}_{-0.2}$  & PS1 &   0.42 & Yes & 5, 36 \\ 
Horologium I &  D  & HorI & 43.9 & $-$54.1 & 19.5 & 79$^{+4}_{-4}$ & 32$^{+3}_{-3}$  &  $-$3.4$^{+0.1}_{-0.1}$  & DES &   0.99 & Yes & 37, 38 \\ 
Horologium II &  DC  & HorII & 49.1 & $-$50.0 & 19.5 & 78$^{+8}_{-7}$ & 33$^{+5}_{-5}$  &  $-$2.1$^{+0.1}_{-0.1}$  & DES &   0.86 & Yes & 37, 39 \\ 
Hydra II &  D  & HyaII & 185.4 & $-$32.0 & 20.9 & 151$^{+8}_{-7}$ & 57$^{+12}_{-12}$  &  $-$5.1$^{+0.1}_{-0.2}$  & DELVE &   0.99 & Yes & 5, 37, 40 \\ 
Hydrus I &  D  & HyiI & 37.4 & $-$79.3 & 17.2 & 28$^{+1}_{-1}$ & 52$^{+5}_{-6}$  &  $-$4.7$^{+0.1}_{-0.1}$  & DELVE &   1.00 & Yes & 41 \\ 
Leo I &  D  & LeoI & 152.1 & 12.3 & 22.1 & 258$^{+10}_{-9}$ & 229$^{+18}_{-19}$  &  $-$11.8$^{+0.3}_{-0.3}$  & PS1 &   1.00 & Yes & 5, 42 \\ 
Leo II &  D  & LeoII & 168.4 & 22.2 & 21.8 & 233$^{+14}_{-14}$ & 165$^{+10}_{-10}$  &  $-$9.7$^{+0.1}_{-0.1}$  & DELVE &   1.00 & Yes & 5, 43 \\ 
Leo IV &  D  & LeoIV & 173.2 & $-$0.5 & 20.9 & 151$^{+4}_{-4}$ & 101$^{+12}_{-12}$  &  $-$4.9$^{+0.3}_{-0.3}$  & DELVE &   0.93 & Yes & 5, 44 \\ 
Leo V &  D  & LeoV & 172.8 & 2.2 & 21.1 & 169$^{+5}_{-5}$ & 35$^{+14}_{-13}$  &  $-$4.4$^{+0.4}_{-0.4}$  & DELVE &   0.90 & Yes & 5, 44 \\ 
Leo VI &  D  & LeoVI & 171.1 & 24.9 & 20.2 & 111$^{+4}_{-6}$ & 87$^{+37}_{-34}$  &  $-$3.6$^{+0.5}_{-0.4}$  & DELVE &   0.08 & No & 45 \\ 
Leo Minor I &  DC  & LMiI & 164.3 & 28.9 & 19.6 & 82$^{+4}_{-7}$ & 26$^{+9}_{-8}$  &  $-$2.4$^{+0.5}_{-0.4}$  & DELVE &   0.36 & No & 13 \\ 
LMC &  D  & LMC & 78.8 & $-$69.2 & 18.5 & 50$^{+1}_{-1}$ & 2544$^{+29}_{-28}$  &  $-$18.1$^{+0.1}_{-0.1}$  & MC &   - & No$^\dagger$ & 46, 47, 48 \\ 
Pegasus III &  D  & PegIII & 336.1 & 5.4 & 21.7 & 215$^{+12}_{-12}$ & 82$^{+14}_{-13}$  &  $-$4.2$^{+0.2}_{-0.2}$  & PS1 &   0.00 & No & 49, 50 \\ 
Pegasus IV &  D  & PegIV & 328.5 & 26.6 & 19.8 & 90$^{+1}_{-1}$ & 42$^{+7}_{-7}$  &  $-$4.2$^{+0.2}_{-0.2}$  & PS1 &   0.03 & No & 51 \\ 
Phoenix II &  D  & PheII & 355.0 & $-$54.4 & 19.6 & 83$^{+8}_{-7}$ & 27$^{+5}_{-4}$  &  $-$2.6$^{+0.1}_{-0.1}$  & DES &   0.89 & Yes & 37, 52 \\ 
Pictor I &  DC  & PicI & 70.9 & $-$50.3 & 20.3 & 115$^{+5}_{-5}$ & 32$^{+5}_{-5}$  &  $-$3.1$^{+0.3}_{-0.3}$  & DES &   0.95 & Yes & 8, 38 \\ 
Pictor II &  D  & PicII & 101.2 & $-$59.9 & 18.3 & 45$^{+1}_{-1}$ & 32$^{+4}_{-4}$  &  $-$2.6$^{+0.7}_{-0.5}$  & DELVE &   0.88 & Yes & 53 \\ 
Pisces II &  D  & PscII & 344.6 & 6.0 & 21.3 & 183$^{+15}_{-14}$ & 56$^{+6}_{-5}$  &  $-$4.3$^{+0.2}_{-0.2}$  & PS1 &   0.04 & No & 49, 54 \\ 
Reticulum II &  D  & RetII & 53.9 & $-$54.1 & 17.5 & 32$^{+1}_{-1}$ & 36$^{+5}_{-6}$  &  $-$3.1$^{+0.1}_{-0.1}$  & DES &   0.99 & Yes & 52 \\ 
Reticulum III &  D  & RetIII & 56.4 & $-$60.5 & 19.8 & 92$^{+14}_{-12}$ & 62$^{+25}_{-23}$  &  $-$3.3$^{+0.3}_{-0.3}$  & DES &   0.94 & Yes & 20 \\ 
Sagittarius &  D  & Sgr & 284.1 & $-$30.5 & 17.1 & 26$^{+2}_{-2}$ & 1566$^{+133}_{-126}$  &  $-$13.5$^{+0.3}_{-0.3}$  & Gal. Plane &   - & No$^\dagger$ & 55 \\ 
Sculptor &  D  & Scl & 15.0 & $-$33.7 & 19.6 & 84$^{+2}_{-2}$ & 223$^{+5}_{-5}$  &  $-$10.8$^{+0.1}_{-0.1}$  & DES &   1.00 & Yes & 5, 56 \\ 
Segue 1 &  D  & Seg1 & 151.8 & 16.1 & 16.8 & 23$^{+2}_{-2}$ & 20$^{+3}_{-3}$  &  $-$1.3$^{+0.7}_{-0.7}$  & DELVE &   0.89 & Yes & 5, 57 \\ 
Segue 2 &  D  & Seg2 & 34.8 & 20.2 & 17.8 & 36$^{+2}_{-2}$ & 35$^{+4}_{-4}$  &  $-$1.9$^{+0.9}_{-0.9}$  & DELVE &   0.77 & Yes & 5, 58 \\ 
Sextans &  D  & Sex & 153.3 & $-$1.6 & 19.7 & 86$^{+4}_{-4}$ & 486$^{+29}_{-28}$  &  $-$8.7$^{+0.1}_{-0.1}$  & DELVE &   0.99 & Yes & 5, 59, 60 \\ 
Sextans II &  DC  & SexII & 156.4 & $-$0.6 & 20.5 & 126$^{+12}_{-11}$ & 115$^{+20}_{-18}$  &  $-$3.9$^{+0.4}_{-0.4}$  & DELVE &   0.20 & Yes & 12 \\ 
SMC &  D  & SMC & 16.2 & $-$72.4 & 19.0 & 63$^{+3}_{-3}$ & 1081$^{+93}_{-87}$  &  $-$16.8$^{+0.2}_{-0.2}$  & MC &   - & No$^\dagger$ & 5, 48, 61 \\ 
Triangulum II &  D  & TriII & 33.3 & 36.2 & 17.3 & 28$^{+1}_{-1}$ & 17$^{+3}_{-3}$  &  $-$1.3$^{+0.2}_{-0.2}$  & PS1 &   0.08 & Yes & 22, 37 \\ 
Tucana II &  D  & TucII & 343.0 & $-$58.6 & 18.8 & 56$^{+5}_{-5}$ & 162$^{+35}_{-34}$  &  $-$3.8$^{+0.1}_{-0.1}$  & DES &   0.95 & Yes & 38, 62 \\ 
Tucana III &  DC  & TucIII & 359.1 & $-$59.6 & 16.8 & 23$^{+1}_{-1}$ & 30$^{+7}_{-7}$  &  $-$1.3$^{+0.2}_{-0.2}$  & DES &   0.92 & Yes & 52 \\ 
Tucana IV &  D  & TucIV & 0.7 & $-$60.8 & 18.4 & 47$^{+4}_{-4}$ & 98$^{+17}_{-16}$  &  $-$3.0$^{+0.3}_{-0.4}$  & DES &   0.93 & Yes & 35 \\ 
Tucana V &  D  & TucV & 354.3 & $-$63.3 & 18.7 & 55$^{+6}_{-5}$ & 23$^{+7}_{-7}$  &  $-$1.1$^{+0.5}_{-0.6}$  & DES &   0.85 & Yes & 35 \\ 
Ursa Major I &  D  & UMaI & 158.8 & 51.9 & 19.9 & 97$^{+6}_{-6}$ & 150$^{+13}_{-12}$  &  $-$5.1$^{+0.4}_{-0.4}$  & PS1 &   0.24 & Yes & 5, 63 \\ 
Ursa Major II &  D  & UMaII & 132.9 & 63.1 & 17.7 & 35$^{+2}_{-2}$ & 92$^{+7}_{-7}$  &  $-$4.4$^{+0.3}_{-0.3}$  & PS1 &   0.90 & Yes & 5, 64 \\ 
Ursa Minor &  D  & UMi & 227.2 & 67.2 & 19.2 & 70$^{+4}_{-3}$ & 250$^{+13}_{-13}$  &  $-$8.9$^{+0.1}_{-0.1}$  & PS1 &   1.00 & Yes & 5, 65 \\ 
Virgo I &  DC  & VirI & 180.0 & $-$0.7 & 19.8 & 91$^{+9}_{-8}$ & 29$^{+10}_{-9}$  &  $-$0.9$^{+0.7}_{-0.7}$  & DELVE &   0.01 & No & 12, 21 \\ 
Virgo II &  DC  & VirII & 225.1 & 5.9 & 19.3 & 72$^{+8}_{-7}$ & 15$^{+3}_{-3}$  &  $-$1.6$^{+0.4}_{-0.6}$  & DELVE &   0.15 & No & 13 \\ 
Virgo III &  DC  & VirIII & 186.3 & 4.4 & 20.9 & 151$^{+15}_{-13}$ & 36$^{+9}_{-9}$  &  $-$2.7$^{+0.5}_{-0.6}$  & DELVE &   0.01 & No & 12 \\ 
Willman 1 &  DC  & Wil1 & 162.3 & 51.1 & 17.9 & 38$^{+8}_{-6}$ & 20$^{+5}_{-4}$  &  $-$2.5$^{+0.7}_{-0.7}$  & PS1 &   0.54 & Yes & 5, 66 \\ 
%%%%%%%%%%%%%%
\enddata
{\tiny \tablecomments{ Classification Status: D: \cytwo{Spectroscopically} Confirmed Dwarf Galaxy, DC: Dwarf Candidate. \cythree{We adopt the classifications from the LVDB \citep{Pace:2024}, which confirms galaxies based on the presence of a large velocity dispersion or metallicity spread. The only exception is Willman 1, which we treat as a dwarf candidate given its uncertain kinematics \citep{Willman:2011}.}}}
{\tiny \tablerefs{Literature references for size, distance, and magnitude: (1) \cite{Ji:2021}, (2) \cite{Vivas:2022}, (3) \cite{Torrealba:2016}, (4) \cite{Cerny:2024}, (5) \cite{Munoz:2018}, (6) \cite{Dallora:2006}, (7) \cite{Walsh:2008}, (8) \cite{Moskowitz:2020}, (9) \cite{Carlin:2018}, (10) \cite{Correnti:2009}, 
(11) \cite{Homma:2019},
(12) \cite{Homma:2024},
(13) \cite{Cerny:2023c},
(14) \cite{Kuehn:2008},
(15) \cite{Greco:2008},
(16) \cite{Karczmarek:2015},
(17) \cite{Torrealba:2018},
(18) \cite{Casey:2025},
(19) \cite{Martinez-Vazquez:2021},
(20) \cite{Drlica-Wagner:2015},
(21) \cite{Homma:2018},
(22) \cite{Carlin:2017},
(23) \cite{Musella:2009},
(24) \cite{Torrealba:2016b},
(25) \cite{Cerny:2021b},
(26) \cite{Bhardwaj:2024},
(27) \cite{Longeard:2018},
(28) \cite{Crnojevic:2016},
(29) \cite{Martinez-Vazquez:2021b},
(30) \cite{Cerny:2021},
(31) \cite{Wang:2019},
(32) \cite{Oakes:2022},
(33) \cite{Cantu:2021},
(34) \cite{Martinez-Vazquez:2019},
(35) \cite{Simon:2020},
(36) \cite{MutluPakdil:2020},
(37) \cite{Richstein:2024},
(38) \cite{Koposov15a},
(39) \cite{Kim:2015b},
(40) \cite{Vivas:2016},
(41) \cite{Koposov:2018},
(42) \cite{Stetson:2014},
(43) \cite{Bellazzini:2005},
(44) \cite{Medina:2018},
(45) \cite{Tan:2025},
(46) \cite{Choi:2018},
(47) \cite{Pietrzynski:2019},
(48) \cite{deVaucouleurs:1991},
(49) \cite{Richstein:2022},
(50) \cite{Kim:2016},
(51) \cite{Cerny:2023},
(52) \cite{MutluPakdil:2018},
(53) \cite{Pace:2025},
(54) \cite{Sand:2012},
(55) \cite{McConnachie:2012},
(56) \cite{Martinez-Vazquez:2015},
(57) \cite{Belokurov07},
(58) \cite{Boettcher:2013},
(59) \cite{Cicuendez:2018},
(60) \cite{Lee:2009},
(61) \cite{Cioni:2000},
(62) \cite{Vivas:2020},
(63) \cite{Garofalo:2013},
(64) \cite{DallOra:2012},
(65) \cite{Garofalo:2025},
(66) \cite{Willman:2006}.}}
\end{deluxetable*}

%% file: ack.tex
\cyfive{We thank the anonymous referee for the many useful comments that helped us improve this manuscript.} CYT was supported by the U.S. National Science Foundation (NSF) through the grants AST-2108168 and AST-2307126. WC gratefully acknowledges support from a Gruber Science Fellowship at Yale University.  This material is based upon work supported by the National Science Foundation Graduate
Research Fellowship Program under Grant No. DGE2139841. Any opinions, findings, and conclusions or recommendations expressed in this material are those of the author(s) and do not necessarily reflect the views of the National Science Foundation. \cytwo{ADD acknowledges support from STFC grants ST/Y002857/1.} DJS acknowledges support from NSF grant AST-2205863.
\cyfive{This research was supported in part by grant NSF PHY-2309135 to the Kavli Institute for Theoretical Physics (KITP).}

The DELVE project is partially supported by the NASA Fermi Guest Investigator Program Cycle 9 No.\ 91201 and by Fermilab LDRD project L2019-011. This material is based upon work supported by the National Science Foundation under Grant No.\ AST-2108168, AST-2108169, AST-2307126, and AST-2407526. This research award is partially funded by a generous gift of Charles Simonyi to the NSF Division of Astronomical Sciences.  The award is made in recognition of significant contributions to Rubin Observatory’s Legacy Survey of Space and Time.

Funding for the DES Projects has been provided by the U.S. Department of Energy, the U.S. National Science Foundation, the Ministry of Science and Education of Spain, 
the Science and Technology Facilities Council of the United Kingdom, the Higher Education Funding Council for England, the National Center for Supercomputing 
Applications at the University of Illinois at Urbana-Champaign, the Kavli Institute of Cosmological Physics at the University of Chicago, 
the Center for Cosmology and Astro-Particle Physics at the Ohio State University,
the Mitchell Institute for Fundamental Physics and Astronomy at Texas A\&M University, Financiadora de Estudos e Projetos, 
Funda{\c c}{\~a}o Carlos Chagas Filho de Amparo {\`a} Pesquisa do Estado do Rio de Janeiro, Conselho Nacional de Desenvolvimento Cient{\'i}fico e Tecnol{\'o}gico and 
the Minist{\'e}rio da Ci{\^e}ncia, Tecnologia e Inova{\c c}{\~a}o, the Deutsche Forschungsgemeinschaft and the Collaborating Institutions in the Dark Energy Survey. 

The Collaborating Institutions are Argonne National Laboratory, the University of California at Santa Cruz, the University of Cambridge, Centro de Investigaciones Energ{\'e}ticas, 
Medioambientales y Tecnol{\'o}gicas-Madrid, the University of Chicago, University College London, the DES-Brazil Consortium, the University of Edinburgh, 
the Eidgen{\"o}ssische Technische Hochschule (ETH) Z{\"u}rich, 
Fermi National Accelerator Laboratory, the University of Illinois at Urbana-Champaign, the Institut de Ci{\`e}ncies de l'Espai (IEEC/CSIC), 
the Institut de F{\'i}sica d'Altes Energies, Lawrence Berkeley National Laboratory, the Ludwig-Maximilians Universit{\"a}t M{\"u}nchen and the associated Excellence Cluster Universe, 
the University of Michigan, NSF NOIRLab, the University of Nottingham, The Ohio State University, the University of Pennsylvania, the University of Portsmouth, 
SLAC National Accelerator Laboratory, Stanford University, the University of Sussex, Texas A\&M University, and the OzDES Membership Consortium.

Based in part on observations at NSF Cerro Tololo Inter-American Observatory at NSF NOIRLab (NOIRLab Prop. ID 2012B-0001; PI: J. Frieman), which is managed by the Association of Universities for Research in Astronomy (AURA) under a cooperative agreement with the National Science Foundation.

The DES data management system is supported by the National Science Foundation under Grant Numbers AST-1138766 and AST-1536171.
The DES participants from Spanish institutions are partially supported by MICINN under grants PID2021-123012, PID2021-128989 PID2022-141079, SEV-2016-0588, CEX2020-001058-M and CEX2020-001007-S, some of which include ERDF funds from the European Union. IFAE is partially funded by the CERCA program of the Generalitat de Catalunya.

We  acknowledge support from the Brazilian Instituto Nacional de Ci\^encia
e Tecnologia (INCT) do e-Universo (CNPq grant 465376/2014-2).

This document was prepared by the DES Collaboration using the resources of the Fermi National Accelerator Laboratory (Fermilab), a U.S. Department of Energy, Office of Science, Office of High Energy Physics HEP User Facility. Fermilab is managed by Fermi Forward Discovery Group, LLC, acting under Contract No.\ 89243024CSC000002.

%% file: appendix.tex
\section{Ambiguous Compact Systems }
\label{Appendix:ambigous}
\cytwo{For our main analysis, we have masked  compact Milky Way satellite systems with $r_{1/2} < 15$\,pc. The classification of these systems as star clusters or dwarf galaxies is still undetermined. Our selection is motivated by the desire to avoid including misclassified star clusters into our census of dwarf galaxies. However, these compact systems are interesting objects, and follow-up observations may reveal that some of them are dwarf galaxies. \cyfive{Thus, we calculate the detection significance for ambiguous compact satellite systems with $r_{1/2} < 15$ pc listed in the LVDB and report the results in } Table~\ref{table:ambi_dwarfs}}. Furthermore, we include some confirmed star clusters, which were previously classified as possible dwarf galaxies. We note that four systems (DES~4, DES~5, Gaia~3 and To~1) fall within the LMC mask footprint. Even when the mask is removed, we find poor performance for our search algorithms for the first three aforementioned systems due to crowding from LMC stars. We note when running our search algorithms on To 1 using DELVE data, we get a detection significance of $\sqrt{\mathrm{TS}_{gr}}=7.0$,  and  SIG$_{gr}=8.5$, despite its location near the LMC. %\cyfive{In addition, we did not recover in the deeper DES Y6 data the candidate DES J0111−1341, originally reported by \citet{Luque:2017} ($\sqrt{\mathrm{TS}{gr}}=5.9$ and SIG${gr}=3.5$), even though it has a detection probability of $P_{\rm det}=0.89$. However, the system is considered a false positive in LVDB due to the lack of an expected Gaia proper motion signal in the search performed by \citep{Pace:2019b}. }

\section{\cyfive{Quantifying Source Blending Using Image-Level Dwarf Injection}}
\label{appendix:balrog}
\cyfive{In Section \ref{sec:sims}, we quantified the detection efficiency of our census by injecting stars from a large population of simulated satellites into our catalogs and applying our satellite detection algorithms. Full image-level simulations provide an alternative and more realistic analysis approach; however, their computational cost was prohibitively high for the large set of satellites that we simulated. As reported by \cite{Zhang:2025}, image-level dwarf galaxy simulations show a greater loss in detection efficiency particularly for distant, compact satellites than catalog-level simulations, due to observational effects that are not accounted for in catalog-level simulations, such as source blending. In this appendix, we use a small set of image-level injection simulations to assess the impact of these systematics on our catalog-level detection efficiency.  }

\begin{figure}
    \centering
    \includegraphics[width=1\linewidth]{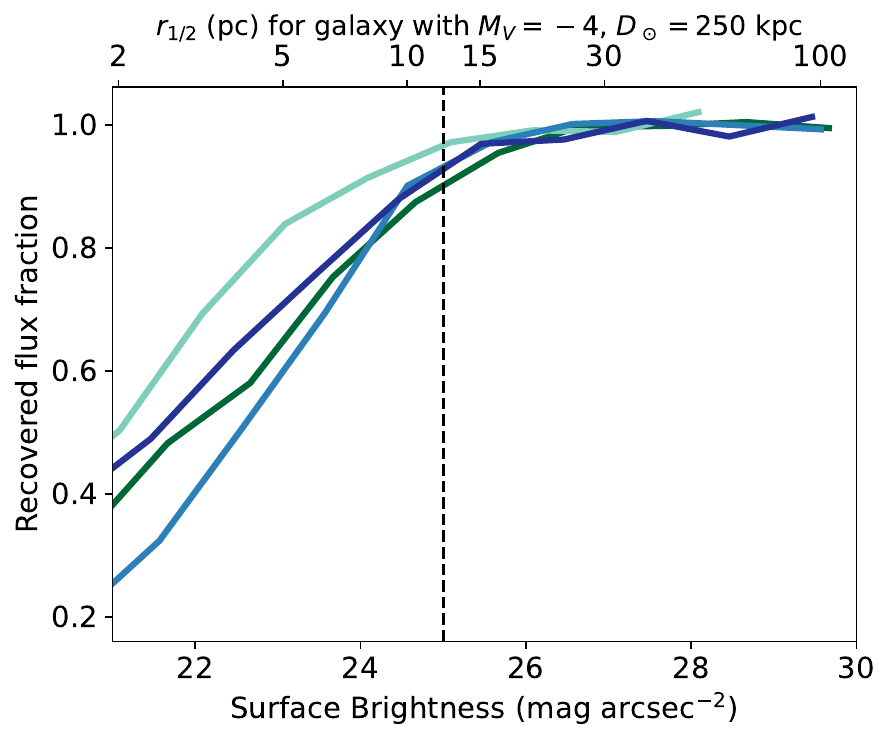}
    \caption{\cyfive{The median recovered flux fraction of image-level simulated satellites as a function of effective surface brightness, averaged within the half-light radius. The different colors represent the  four different simulated satellite stellar populations with $M_V=-4$ and $D_\odot = 250$\,kpc. The top x-axis indicates the corresponding physical half-light radius of the simulated satellites. In our analysis, which considers only satellites with $r_h \geq 15$ pc and $D_\odot \leq 400$ kpc, satellites lying on the 50\%  detectability contour have surface brightnesses of $\mu_V \gtrsim 25\,\mathrm{mag\ arcsec^{-2}}$ (dotted line) and are therefore not significantly affected by blending.}\label{fig:blending}}
\end{figure}

\cyfive{We injected simulated dwarfs at image level using a Synthetic Source Injection (SSI) pipeline built for DELVE DR3 \citep{Anbajagane:2025b, Doliva-Dolinsky:2025b}, which follows on the \texttt{balrog} pipeline from DES \citep{Suchyta:2016, Everett:2022, Anbajagane:2025OJAp....8E..65A}. We generate stellar populations for four satellites with $M_V = - 4$ and $D_\odot = 250$\,kpc drawing the age and metallicity of each randomly from the distributions described in Table~\ref{table:galsims_params}. We spatially distribute each stellar population to sample physical half-light radii ranging from 1 to 100\,pc. 
We calculate their effective surface brightness, averaged within the half-light radius using  $\mu_V = M_V + 36.57 + 2.5\log[2\pi(r_{1/2}/\rm{kpc})]^2$.
The distance and range of sizes were selected to explore a regime susceptible to blending effects, and the absolute magnitude places these systems near the 50\% satellite detectability contour at that distance.}

\cyfive{To facilitate image simulation, we model each satellite as the sum of resolved and diffuse components, both following an exponential profile. For the resolved component, we use sources from the mock catalogs with magnitudes $g < 25.5$, while the flux from fainter sources is represented by the diffuse component. Image simulations for both the resolved and diffuse components are generated using the \texttt{Galsim} package \citep{Rowe:2015}, and the resulting source-injected images are processed through the same pipeline described in Section \ref{sec:data} to produce object catalogs. These simulated images contain the same properties as the data images, i.e. the same PSF, mask, background, and noise.}

%We begin with the mock catalogs of satellites described in Section \ref{sec:sims}. For our SSI pipeline, we model each dwarf galaxy as the sum of resolved and diffuse components, both following an exponential profile. For the resolved component, we use sources from the mock catalogs with magnitudes $g < 25.5$, while the flux from fainter sources is represented by the diffuse component. Image simulations for both the resolved and diffuse components are generated using the \texttt{Galsim} package \citep{Rowe:2015}, and the resulting source-injected images are processed through the same pipeline described in Section \ref{sec:data} to produce object catalogs. These simulated images contain the same properties as the data images, i.e. the same PSF, mask, background, and noise.}

\cyfive{
%We selected four independent realizations of the isochrone for galaxies with $M_V\sim-4$ and $D_\odot = 250$\,kpc, and injected each realization multiple times into the images. For each injection, we varied the positions of the stars to sample physical half-light radii ranging from 1 to 100\,pc.
%We analyze stars generated through the image simulations to assess the properties of the satellite.
We match the stars recovered in our SSI catalogs with the original injected galaxies and calculated the total flux lost due to unrecovered stars.
We calculate the recovered flux fraction as the ratio of the flux of all stars that are recovered in the SSI catalog to the flux of all stars injected into the image.
We show the recovered flux fraction as a function of the galaxy surface brightness in Figure~\ref{fig:blending}.
The flux loss for our simulated systems is substantial at high surface brightness due to source blending.  However, when we impose our lower limit on galaxy sizes of 15\,pc, we find that galaxies lying on the 50\% satellite detectability contour (Section \ref{sec:selfunc_analytical}) have a surface brightness of $\mu_V \gtrsim 25\,\mathrm{mag\ arcsec^{-2}}$, which implies flux losses due to blending of at most $\sim10\%$. This indicates that our satellite detection sensitivity are not significantly affected by blending for galaxies with $r_h \geq 15$ pc and $D_\odot \leq 400$ kpc.
This result is consistent with the findings of \citet{Zhang:2025}, who found minimal differences for satellites with $r_h \gtrsim 15$\,pc, and that the magnitude required to reach 50\% detectability is significantly brighter in the image-level simulations only once systems reach $r_h \lesssim 5$\,pc and $D_\odot \gtrsim 150$\,kpc, which corresponds to a surface brightnesses of $\sim 24\,\mathrm{mag\ arcsec^{-2}}$.}

%However, at sufficiently large radii the flux loss becomes largely independent of galaxy size, indicating that blending does not significantly affect more extended systems

\section{Empirical Model with Measurement uncertainties}
\label{appendix:uncertainties}
\begin{figure*}
    \centering
    \includegraphics[width=0.9\linewidth]{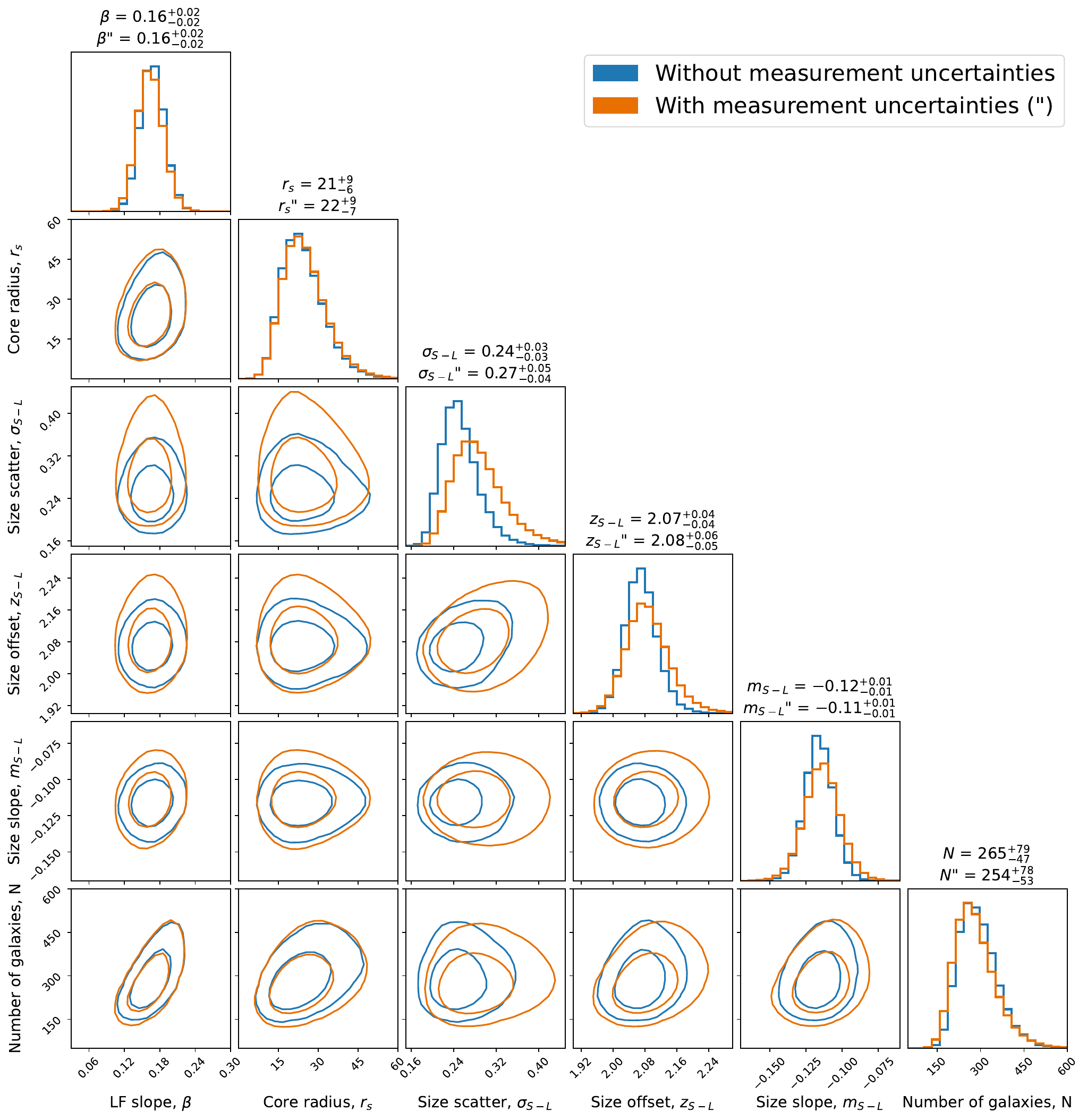}
    \caption{{Posterior distributions of Milky Way satellite population model parameters}. \cytwo{We show parameter estimates from both our main analysis and the alternative analysis which incorporates measurement uncertainties (denoted by a double prime symbol). The contour represent 68\% (95\%) credible intervals. $\beta$ is the slope of the luminosity function, $r_s$ is the scale radius of the cored-NFW radial profile, $z_{S-L}$ and $m_{S-L}$ are the offset and slope of the median size--luminosity relations while $\sigma_{S-L}$ is its scatter. Finally, $N$ is the total number of Milky Way satellites.} \label{fig:mcmc_chains}}
\end{figure*}

\cytwo{In our main analysis (Section~\ref{sec:mwpop}), we assumed the median measured values for the observed galaxy parameters. Here, we incorporate their measurement uncertainties, obtained from Table \ref{table:all_dwarfs}. To do so, we perform multiple MCMC runs in which the properties of the observed satellite galaxies are randomly sampled from Gaussian distributions defined by their respective uncertainties. We then combine the resulting posteriors and compare them to those from our main analysis, as shown in Figure~\ref{fig:mcmc_chains}. We find that incorporating uncertainties leads to minimal differences, aside from a modest increase in the scatter of the size--luminosity relation.}

\section{Detection Efficiencies}
\label{appendix:detection_efficiencies}
Tables  \ref{table:delve_search}, \ref{table:des_search}  and \ref{table:ps1_search} show the detection significance for each of the three search methods for candidates located in the  DELVE, DES, and PS1 regions, respectively. We also present the detection significance of confirmed dwarf galaxies that fall below our detection threshold.

\begin{deluxetable*}{lccccccccccccccc}
\tabletypesize{\scriptsize}
\tablewidth{\textwidth} 
%\tablenum{1}
\tablecaption{\cyone{Compact Milky Way satellite systems and known star clusters previously classified as possible dwarf galaxies. We include their location and structural parameters as described in Table~\ref{table:all_dwarfs}.  We also show the survey used to analyze each system, its \texttt{XGBoost}-predicted detection probability, its detection significance, and whether it passes the detection threshold defined in our census.\label{table:ambi_dwarfs} }}
\tablehead{
Name & Class.  &  RA & DEC & $(m-M)_0$  & $D$ &$r_{1/2}$ & $M_V$ & Survey/ & $P_{\rm det}$  & $\sqrt{\mathrm{TS}_{gr}}$    & $\rm{SIG}_{gr}$   &   Pass Census & Ref. \\ & & (deg) & (deg) &   &  (kpc) & (pc)  &  (mag)& Region & &&& Threshold  &  } 
\startdata 
%%%%%%%%%%% 
Alice &  A & 220.8 & $-$28.0 & 19.4 & 74    &  $-$  & $-$ & DELVE & $-$ & 3.8 & 2.1 & No & 1  \\ 
Balbinot 1 &  A & 332.7 & 14.9 & 17.5 & 32    &  6  & $-$1.2 & PS1 & 0.50 & 13.0 & 8.4 & Yes & 2  \\ 
BLISS 1 &  A & 177.5 & $-$41.8 & 16.9 & 24    &  4  & 0.0 & DELVE & 0.77 & 14.9 & 11.9 & Yes & 3  \\ 
Crater/Laevens 1 &  SC & 174.1 & $-$10.9 & 20.8 & 146    &  20  & $-$5.3 & DELVE & 1.00 & 13.2 & 15.7 & Yes & 4  \\ 
DELVE 1 &  A & 247.7 & $-$1.0 & 16.7 & 22    &  5  & $-$0.5 & DELVE & 0.86 & 10.5 & 7.9 & Yes & 5, 6  \\ 
DELVE 3 &  A & 290.4 & $-$60.8 & 18.7 & 56    &  6  & $-$1.3 & DELVE & 0.81 & 11.1 & 8.1 & Yes & 7  \\ 
DELVE 4 &  A & 230.8 & 27.4 & 18.3 & 45    &  5  & $-$0.2 & PS1 & 0.00 & 4.1 & 3.9 & No & 7  \\ 
DELVE 5 &  A & 222.1 & 17.5 & 18.0 & 39    &  5  & 0.4 & PS1 & 0.01 & 5.1 & 4.7 & No & 7  \\ 
DELVE 6 &  A & 33.1 & $-$66.1 & 19.5 & 80    &  10  & $-$1.5 & DELVE & 0.39 & 6.9 & 5.8 & No & 8  \\ 
DES 1 &  A & 8.5 & $-$49.0 & 19.4 & 76    &  4  & $-$1.4 & DES & 0.90 & 9.2 & 6.9 & Yes & 9  \\ 
DES 3 &  A & 325.1 & $-$52.5 & 19.4 & 76    &  6  & $-$1.6 & DES & 0.91 & 8.6 & 6.9 & Yes & 10  \\ 
DES 4 &  A & 82.1 & $-$61.7 & 17.5 & 32    &  8  & $-$1.1 & MC & $-$ & $-$ & $-$ & No & 11   \\ 
DES 5 &  A & 77.5 & $-$62.6 & 17.0 & 25    &  1  & 0.3 & MC & $-$ & $-$ & $-$ & No & 11   \\ 
Eridanus III &  A & 35.7 & $-$52.3 & 19.8 & 91    &  6  & $-$2.1 & DES & 0.95 & 12.8 & 10.4 & Yes & 9  \\ 
Gaia 3 &  A & 95.1 & $-$73.4 & 18.4 & 48    &  7  & $-$3.3 & MC & $-$ & $-$ & $-$ & No & 11   \\ 
HSC 1 &  A & 334.3 & 3.5 & 18.3 & 46    &  4  & $-$0.2 & PS1 & 0.00 & 4.0 & 3.4 & No & 12  \\ 
Kim 1 &  A & 332.9 & 7.0 & 16.5 & 20    &  5  & 0.3 & PS1 & 0.21 & 7.6 & 6.4 & Yes & 13  \\ 
Kim 2 &  A & 317.2 & $-$51.2 & 20.1 & 105    &  12  & $-$1.5 & DES & 0.54 & 14.2 & 12.3 & Yes & 14  \\ 
Kim 3 &  A & 200.7 & $-$30.6 & 15.9 & 15    &  2  & 0.7 & DELVE & 0.83 & 8.9 & 8.4 & Yes & 15  \\ 
Koposov 1 &  A & 179.8 & 12.3 & 18.4 & 48    &  6  & $-$1.0 & DELVE & 0.74 & 11.0 & 9.0 & Yes & 16  \\ 
Koposov 2 &  A & 119.6 & 26.3 & 17.7 & 35    &  3  & $-$0.9 & DELVE & 0.86 & 10.7 & 9.1 & Yes & 16  \\ 
Laevens 3 &  SC & 316.7 & 15.0 & 18.9 & 61    &  11  & $-$2.8 & PS1 & 0.02 & 7.0 & 4.8 & No & 17  \\ 
PS1 1 &  A & 289.2 & $-$27.8 & 17.4 & 30    &  5  & $-$1.9 & No Cov. & - &  $-$ & $-$ & No & 11   \\ 
Sagittarius II &  SC & 298.2 & $-$22.1 & 19.0 & 64    &  34  & $-$5.2 & DELVE & 1.00 & 54.2 & 37.5 & Yes & 18, 19  \\ 
Segue 3 &  SC & 320.4 & 19.1 & 16.1 & 17    &  2  & $-$0.1 & PS1 & 0.58 & 11.6 & 7.2 & Yes & 20  \\ 
SMASH 1 &  A & 95.2 & $-$80.4 & 18.8 & 58    &  6  & $-$1.0 & DELVE & 0.67 & 4.3 & 3.3 & No & 21  \\ 
To 1 &  A & 56.1 & $-$69.4 & 18.2 & 44    &  3  & $-$1.6 & MC & $-$ & $-$ & $-$ & No & 11   \\ 
Ursa Major III &  A & 174.7 & 31.1 & 15.0 & 10    &  2  & 2.2 & PS1 & 0.01 & 4.2 & 4.0 & No & 22  \\ 
YMCA-1 &  A & 110.8 & $-$64.8 & 18.7 & 55    &  3  & $-$0.5 & DELVE & 0.31 & 3.3 & 4.1 & No & 23  \\ 
%%%%%%%%%%%%%%
\enddata
{\tiny \tablecomments{ Classification Status: A: Ambiguous Systems, SC: Confirmed Star Clusters. Literature references for the size, distance, and magnitude measurements are:
(1) \cite{Popova:2025}, 
(2) \cite{Balbinot:2013}, 
(3) \cite{Mau:2019},
(4) \cite{Weisz:2016}, 
(5) \cite{Mau:2020},
(6) \cite{Simon:2024},
(7) \cite{Cerny:2023c},
(8) \cite{Cerny:2023b},
(9) \cite{Conn:2018},
(10) \cite{Luque:2018},
(11) \cite{Torrealba:2019},
(12) \cite{Homma:2019},
(13) \cite{Kim:2015a},
(14) \cite{Kim:2015},
(15) \cite{Kim:2016b},
(16) \cite{Munoz:2018},
(17) \cite{Longeard:2019},
(18) \cite{Richstein:2024},
(19) \cite{Joo:2019},
(20) \cite{Fadely:2011},
(21) \cite{Martin:2016},
(22) \cite{Smith:2024},
(23) \cite{Gatto:2022}
}}
\end{deluxetable*}

\input{table2}

%% file: table2.tex
\begin{deluxetable}{lccccc}
\tabletypesize{\scriptsize}
\tablewidth{0pt} 
%\tablenum{1}
\tablecaption{Milky Way satellites in the DELVE DR3 region. We separate systems above and below the census detection threshold into the top and bottom sections of the table. $\sqrt{\mathrm{TS}_{g\{r,i\}}}$ is the \texttt{ugali} detection significance using $g$ and \{$r$ or $i$\} bands, SIG$_{gr}$ is the \texttt{simple} detection significance, and $P_{\rm det}$ is the estimated detection probability of the system based on the \texttt{XGBoost} classifier (Section \ref{sec:selfunc_ML}).  \cytwo{We also include the apparent magnitude, $m_V$, of each system as a rough proxy for its detectability.}  \label{table:delve_search} }
\tablehead{
Name & $\sqrt{\mathrm{TS}_{gr}}$  & $\sqrt{\mathrm{TS}_{gi}}$  & $\rm{SIG}_{gr}$  &  $m_V$  & $P_{\rm det}$  } 
\startdata 
%%%%%%%%%%%%%%%%%%%%%%%%%%%%%%%%%%%
Carina & 192.6 & 192.1 & 37.5 & 10.7  & 1.00 \\ 
Sextans & 149.6 & 180.5 & 37.5 & 10.9  & 0.99 \\ 
Leo II & 103.7 & 100.5 & 37.5 & 12.1  & 1.00 \\ 
Hydrus I & 55.5 & 55.1 & 37.5 & 12.5  & 1.00 \\ 
Bo\"{o}tes I & 54.1 & 64.6 & 37.0 & 13.1  & 1.00 \\ 
Carina II & 34.4 & 33.6 & 17.3 & 13.3  & 0.99 \\ 
Coma Berenices & 30.1 & 22.3 & 17.7 & 13.8  & 0.99 \\ 
Crater II & 28.2 & 48.8 & 19.2 & 12.2  & 0.95 \\ 
Segue 2 & 22.6 & 16.5 & 17.4 & 15.9  & 0.77 \\ 
Eridanus IV & 22.4 & 16.4 & 15.1 & 15.7  & 0.93 \\ 
Carina III\tablenotemark{\tiny a} & 21.6 & 19.5 & 7.1 & 14.8  & 0.91 \\ 
Pictor II & 17.4 & 16.5 & 14.2 & 15.6  & 0.88 \\ 
Centaurus I & 17.3 & 17.5 & 20.8 & 15.0  & 0.98 \\ 
Segue 1 & 15.3 & 14.4 & 12.9 & 15.5  & 0.89 \\ 
Leo IV & 14.7 & 13.2 & 13.2 & 15.9  & 0.93 \\ 
Bo\"{o}tes II & 14.6 & 17.9 & 15.1 & 15.2  & 0.96 \\ 
Aquarius III & 13.6 & 9.1 & 9.8 & 17.2  & 0.17 \\ 
Aquarius II & 12.8 & 15.7 & 14.8 & 15.8  & 0.69 \\ 
Hydra II & 9.8 & 9.7 & 10.9 & 15.8  & 0.99 \\ 
Leo V & 8.4 & 8.6 & 6.9 & 16.7  & 0.90 \\ 
Sextans II & 8.3 & 8.9 & 9.0 & 16.6  & 0.20 \\ 
\hline
\textbf{Threshold} & \textbf{8.0} & \textbf{7.5} & \textbf{6.5}  & - &  - \\
\hline
Virgo II & 9.8 & 4.8 & 7.3 & 17.7  & 0.15 \\ 
Leo Minor I & 7.6 & 3.9 & 6.1 & 17.2  & 0.36 \\ 
Virgo III & 6.8 & 6.8 & 5.1 & 18.2  & 0.01 \\ 
Leo VI & 6.4 & 4.9 & 6.5 & 16.7  & 0.08 \\ 
DELVE 2 & 6.1 & 5.8 & 10.1 & 17.2  & 0.31 \\ 
Virgo I & 5.7 & 4.7 & 5.6 & 18.9  & 0.01 \\ 
%%%%%%%%%%%%%%%%%%%%%%%%%%%%%%%%%%%%%%%%%%%%%%%
\enddata
\tablenotetext{\tiny a}{\cyone{Due to its on-sky proximity to Carina II, Carina III was not identified as an independent hotspot by \texttt{ugali} but was identified by \texttt{simple}}. }
\end{deluxetable}

\begin{deluxetable}{lccccc}
\tabletypesize{\scriptsize}
\tablewidth{0pt} 
%\tablenum{1}
\tablecaption{Table of Milky Way satellites in the DES Y6  region. The table description is the same as Table \ref{table:delve_search} \label{table:des_search} }
\tablehead{
Name & $\sqrt{\mathrm{TS}_{gr}}$  & $\sqrt{\mathrm{TS}_{gi}}$  & $\rm{SIG}_{gr}$  &  $m_V$  & $P_{\rm det}$  } 
\startdata 
%%%%%%%%%%%%%%%%%%%%%%%%%%%%%%%%%%%
Sculptor & 609.5 & 522.9 & 37.5 & 8.9  & 1.00 \\ 
Fornax & 586.2 & 467.2 & 37.5 & 7.4  & 1.00 \\ 
Reticulum II & 62.5 & 59.8 & 37.5 & 14.4  & 0.99 \\ 
Eridanus II & 40.3 & 36.0 & 31.1 & 15.7  & 1.00 \\ 
Horologium I & 31.3 & 26.4 & 24.9 & 16.1  & 0.99 \\ 
Tucana II & 30.9 & 29.6 & 14.6 & 15.0  & 0.95 \\ 
Grus II & 25.7 & 25.5 & 12.0 & 15.2  & 0.98 \\ 
Tucana IV & 22.3 & 21.2 & 14.5 & 15.4  & 0.93 \\ 
Tucana III & 19.6 & 19.0 & 12.7 & 15.5  & 0.92 \\ 
Phoenix II & 17.1 & 15.0 & 14.2 & 17.0  & 0.89 \\ 
Tucana V & 14.9 & 14.2 & 12.2 & 17.6  & 0.85 \\ 
Horologium II & 14.5 & 12.3 & 11.7 & 17.4  & 0.86 \\ 
Pictor I & 13.7 & 12.8 & 12.1 & 17.2  & 0.95 \\ 
Cetus II & 12.4 & 11.4 & 8.3 & 17.4  & 0.79 \\ 
Reticulum III & 12.4 & 10.1 & 11.6 & 16.5  & 0.94 \\ 
Grus I & 12.1 & 10.3 & 9.9 & 16.4  & 0.96 \\ 
Columba I & 10.9 & 9.9 & 10.4 & 17.1  & 0.76 \\ 
\hline
\textbf{Threshold} & \textbf{5.0} & \textbf{5.0} & \textbf{5.5} & -  & - \\ 
\hline
Cetus III & 4.8 & 4.1 & 4.9 & 18.6  & 0.70 \\ 
%%%%%%%%%%%%%%%%%%%%%%%%%%%%%%%%%%%%%%%%%%%%%%%
\enddata

\end{deluxetable}

\begin{deluxetable}{lcccc}
\tabletypesize{\scriptsize}
\tablewidth{0pt} 
%\tablenum{1}
\tablecaption{Table of Milky Way satellites in the the PS1 region from the analysis of \citet{MWCensus1}. The table descriptions are the same as Table \ref{table:delve_search}. We include the detection significances for Pegasus IV and Bo\"{o}tes V, which were discovered after the publication of \citet{MWCensus1}.  \label{table:ps1_search} } %
\tablehead{
Name & $\sqrt{\mathrm{TS}_{gr}}$  & $\rm{SIG}_{gr}$  &  $m_V$  & $P_{\rm det}$  } 
\startdata 
%%%%%%%%%%%%%%%%%%%%%%%%%%%%%%%%%%%
Leo I & 157.6 & 37.5 & 10.2   & 1.00 \\ 
Draco & 96.9 & 37.5 & 10.7   & 1.00 \\ 
Ursa Minor & 83.1 & 37.5 & 10.4   & 1.00 \\ 
Canes Venatici I & 36.0 & 25.3 & 12.9   & 1.00 \\ 
Ursa Major II & 18.7 & 8.9 & 13.3   & 0.90 \\ 
Willman 1 & 15.0 & 12.5 & 15.4   & 0.54 \\ 
Canes Venatici II & 11.7 & 8.8 & 15.8   & 0.93 \\ 
Ursa Major I & 10.2 & 6.0 & 14.8   & 0.24 \\ 
Draco II & 9.8 & 7.9 & 15.9   & 0.24 \\ 
Triangulum II & 9.5 & 6.8 & 16.0   & 0.08 \\ 
Hercules & 9.1 & 6.4 & 14.8   & 0.42 \\ 
\hline
\textbf{Threshold} &\textbf{6.0} & \textbf{6.0}  & - &  - \\ 
\hline
Pegasus IV & \CHECK{6.7} & \CHECK{3.9} & 15.5   & 0.03 \\ 
Bo\"{o}tes V & \CHECK{6.4} & \CHECK{5.1} & 16.8   & 0.06 \\ 
Pisces II & 6.2 & 4.4 & 17.0   & 0.04 \\ 
Bo\"{o}tes IV & 5.4 & 4.7 & 16.3   & 0.01 \\ 
Pegasus III & 4.8 & 3.7 & 17.5   & 0.00 \\ 
Bo\"{o}tes III & 4.0 & 4.3 & 12.6   & 0.01 \\ 
%%%%%%%%%%%%%%%%%%%%%%%%%%%%%%%%%%%%%%%%%%%%%%%
\enddata
\end{deluxetable}